\documentclass[traditabstract]{aa}  

\pdfoutput=1
\usepackage{blindtext}
\usepackage{subfigure}
\usepackage{graphicx}
\usepackage{multicol}
\usepackage{tabularx}
\usepackage{rotating, booktabs}
\usepackage[varg]{txfonts}
\usepackage{filecontents}
\usepackage{threeparttable}

\usepackage{natbib}
\usepackage[toc,page]{appendix}

\bibpunct{(}{)}{;}{a}{}{,}

\usepackage{gensymb}

\def\aox{$\rm {\alpha_{OX}}$}

\def\ltsima{$\; \buildrel < \over \sim \;$}
\def\simlt{\lower.5ex\hbox{\ltsima}}
\def\gtsima{$\; \buildrel > \over \sim \;$}
\def\simgt{\lower.5ex\hbox{\gtsima}}

\def\ergs{{erg s$^{-1}$}}
\def\cm2{{cm$^{-2}$}}
\def\kms{{km s$^{-1}$}}

\def\lbol{L$\rm_{Bol}$}
\def\edd{$\rm\lambda_{Edd}$}
\def\vciv{v$\rm_{CIV}^{peak}$}
\def\vout{v$\rm_{CIV}^{out}$}
\def\ekin{$\rm\dot{E}_{kin}$}
\def\oiii{{\it{[OIII]}}}
\def\weak{{\it{Weak [OIII]}}}
\def\vmax{v$_{CIV}^{max}$}
\def\mpc{Mpc$^{-1}$}
\def\mbh{M$\rm_{BH}$}
\def\mion{M$\rm_{ion}$}
\def\mdot{$\rm\dot{M}_{ion}$}

\begin{document}

   \title{The WISSH Quasars Project}

   \subtitle{IV. BLR versus kpc-scale winds}
   \author{G. Vietri\inst{1, 2,3,4}
 \and E. Piconcelli \inst{1}
   \and M. Bischetti \inst{1, 5}
   \and F. Duras \inst{6,1}
   \and S. Martocchia \inst{7}
	\and A. Bongiorno\inst{1}
   \and A. Marconi \inst{8,9}
    \and L. Zappacosta \inst{1}
    \and S. Bisogni \inst{10}
    \and G. Bruni \inst{11}
   \and M. Brusa \inst{12, 13}
   \and A. Comastri \inst{13}
   \and G. Cresci \inst{9}
   \and C. Feruglio \inst{14, 1}
   \and E. Giallongo \inst{1}
   \and F. La Franca \inst{6}
  \and V. Mainieri \inst{4}
   \and F. Mannucci \inst{9}
  \and F. Ricci \inst{6}
   \and E. Sani \inst{15}
   \and V. Testa \inst{1}
  \and F. Tombesi \inst{5,16,17}
   \and C. Vignali \inst{12, 13}
 \and F. Fiore \inst{1}
          }

   \institute{INAF - Osservatorio Astronomico di Roma, via Frascati 33, 00078  Monteporzio Catone, Italy
   \and Universit\`a degli Studi di Roma "La Sapienza", Piazzale Aldo Moro 5, I--00185 Roma, Italy
   \and Excellence Cluster Universe, Technische Universit\"{a}t M\"{u}nchen, Boltzmannstr. 2, D-85748, Garching, Germany 
      \and European Southern Observatory, Karl-Schwarzschild-str. 2, 85748 Garching bei M\"{u}nchen, Germany
    \and Dipartimento di Fisica, Universit\`a degli Studi di Roma "Tor Vergata", Via della Ricerca Scientifica 1, I-00133 Roma, Italy
   \and Dipartimento di Matematica e Fisica, Universit\`a degli Studi Roma Tre, via della Vasca Navale 84, I--00146, Roma, Italy
   \and Astrophysics Research Institute, Liverpool John Moores University, 146 Brownlow Hill, Liverpool L3 5RF, UK
      \and Dipartimento di Fisica e Astronomia, Universit\`a di Firenze, Via G. Sansone 1, I--50019, Sesto Fiorentino (Firenze), Italy
   \and INAF - Osservatorio Astrofisico di Arcetri, Largo E. Fermi 5, I--50125, Firenze, Italy
   \and Harvard-Smithsonian Center for Astrophysics, 60 Garden St., Cambridge, MA 02138, USA
      \and INAF - Istituto di Astrofisica e Planetologia Spaziali, via Fosso del Cavaliere 100, I-00133 Rome, Italy
\and Dipartimento di Fisica e Astronomia, Universit\`a degli Studi di Bologna, Via Gobetti 93/2, 40129 Bologna, Italy
\and INAF -- Osservatorio Astronomico di Bologna, Via Gobetti 93/3, 40129 Bologna, Italy
   \and INAF - Osservatorio Astronomico di Trieste, via G.B. Tiepolo 11, 34143 Trieste, Italy
  \and European Southern Observatory (ESO), Alonso de Cordova 3107, Vitacura (Santiago), Chile
   \and Department of Astronomy, University of Maryland, College Park, MD 20742, USA
   \and NASA / Goddard Space Flight Center, Code 662, Greenbelt, MD 20771, USA}
   \date{}

 
  \abstract
 {AGN-driven winds are invoked in the most successful models of galaxy evolution to explain the observed physical and evolutionary properties of massive galaxies. Winds are expected to deposit energy and momentum into the interstellar medium (ISM), thus regulating both star formation and supermassive black hole (SMBH) growth. We have undertaken a multi-band observing program aimed at obtaining a complete census of winds in a sample of WISE/SDSS selected hyper-luminous (WISSH) QSOs at $z$ $\approx$ 2--4. 
We have analyzed the rest-frame optical (i.e. LBT/LUCI and VLT/SINFONI) and UV (i.e. SDSS) spectra of 18 randomly selected WISSH QSOs to measure the SMBH mass and study the properties of winds both in the narrow line region (NLR) and broad line region (BLR) traced by blueshifted/skewed [OIII] and CIV emission lines, respectively. These WISSH QSOs are powered by SMBH with masses $\simgt$ 10$^9$ M$_{\odot}$  accreting at 0.4 $<$ \edd\ $< 3.1$.
We have found the existence of two sub-populations of hyper-luminous QSOs characterized by the presence of outflows at different distances from the SMBH. One population (i.e. \oiii\ sources) exhibits powerful [OIII] outflows, rest-frame equivalent width (REW) of the CIV emission REW$\rm_{CIV}$ $\approx$ 20-40 \AA\  and modest CIV velocity shift (\vciv) with respect to the systemic redshift (\vciv\ $\simlt$ 2,000 \kms). The second population (i.e. \weak\ sources), representing $\sim$70\% of the analyzed WISSH QSOs, shows weak/absent [OIII] emission and an extremely large blueshifted CIV emission (\vciv\ up to $\sim$ 8,000 \kms\ and REW$\rm_{CIV} \simlt$ 20 \AA ). We propose two explanations for the observed behavior of the strength of the [OIII] emission in terms of orientation effects of the line of sight and ionization cone. The dichotomy in the presence of BLR and NLR winds could be likely due to inclination effects considering a polar geometry scenario for the BLR winds. In a few cases these winds are remarkably as powerful as those revealed in the NLR in the \oiii\ QSOs (\ekin\ $\sim$ 10$^{44-45}$ \ergs).
We have also investigated the dependence of these CIV winds on fundamental AGN parameters such as bolometric luminosity (\lbol), Eddington ratio (\edd) and UV-to-X-ray continuum slope (\aox). We find a strong correlation with \lbol\ and an anti-correlation with \aox\, whereby the higher the luminosity, the steeper the ionizing continuum described by means of \aox\ and the larger is the blueshift of the CIV emission line. Finally, the observed dependence \vciv $\propto$ L$\rm_{Bol}^{0.28\pm0.04}$ is consistent with radiatively driven winds scenario, where strong UV continuum is necessary to launch the wind and a weakness of the X-ray emission is fundamental to prevent overionization of the wind itself.
}

   \maketitle
%
%

\section{Introduction}\label{sec:intro}
Supermassive Black Holes (SMBHs) at the center of galaxies grow through accretion of nearby matter and a fraction of the gravitational potential of the accreted mass is converted into radiation. In this case the SMBH is called an active galactic nucleus (AGN). The idea of a possible connection between the growth of the SMBH and the evolution of its host galaxy has been put forward, once a correlation between black hole mass (\mbh) and the bulge luminosity was discovered (\citealt{Kormendy1995}; \citealt{Silk1998}; \citealt{Granato2004}; \citealt{DiMatteo2005}; \citealt{Menci2008}). 
Since the energy released by the growth of the BH can be larger than the binding energy of the galaxy bulge, the AGN can have an important effect on the evolution of its host galaxy.  
Indeed, fast winds can be launched from the accretion disk, shock against the interstellar medium (ISM) and drive powerful kpc-scale outflows which may sweep out the gas in the galaxy and then may be responsible for the regulation of the star formation and BH accretion (\citealt{Zubovas2012}; \citealt{Faucher2012}; \citealt{Fabian2012}, see \citealt{Fiore2017} for a detailed discussion). 

Over the past few years efforts have been made to search for AGN outflows in different ISM phases. Evidence of radiatively-driven winds are observed at sub-pc scale, in the innermost regions of the AGN through the detection of blueshifted highly ionized Fe K-shell transitions, i.e. Ultra Fast Outflows (UFOs), with velocities $\sim$ 0.2c (\citealt{King2015}; \citealt{Tombesi2010,Tombesi2014}; \citealt{Gofford2013});  at pc scale via warm absorbers and broad absorption lines (BAL), moving at velocities up to 10,000 \kms\ (\citealt{King2015}; \citealt{TombesiCappi2013}); at kpc-scale through different gas phases such as ionized gas, i.e. broad asymmetric blueward [OIII] emission lines (\citealt{Harrison2012, Harrison2014}; \citealt{Cano-Diaz2012}; \citealt{Brusa2015}; \citealt{Carniani2015}); neutral atomic as NaID (\citealt{Rupke2015}), and molecular gas with velocities up to a few thousands \kms\ (\citealt{Feruglio2010}; \citealt{Cicone2014}; \citealt{Spoon2013}) . The investigation of possible correlations of AGN properties over a wide range of spatial scales can give us more information about the nature of AGNs and the impact on their surroundings.

The most popular theoretical models of AGN-driven outflows (\citealt{Faucher2012}) suggest that the kinetic energy of the nuclear fast outflows is transferred to the ISM and drive the kpc-scale flows, in a likely energy-conserving scenario. This has been reflected in the observations of both UFO and molecular winds in two sources, i.e. Markarian 231 (\citealt{Feruglio2015}) and IRASF11119+13257 (\citealt{Tombesi2015}, but see {\citealt{Veilleux2017}) 

The high ionization lines, such as CIV$\lambda$1549 \AA, are known to exhibit asymmetric profiles towards the blue-side, with the peak of the emission line blueshifted with respect to the low ionization lines, such as MgII or H$\beta$ (\citealt{Gaskell1982}; \citealt{Sulentic2000} and \citealt{Richards2011}). This behavior cannot be ascribed to virialized motion but it can be interpreted in terms of outflowing gas, which possibly leads to biased BH mass estimates derived from the entire CIV emission line (\citealt{Denney2012}; \citealt{Coatman2017}). 

The blueshift of the CIV emission line can be a valuable tool to trace the winds in the BLR.
The CIV emission line typically shows a blueshift of $\sim$ 600 \kms\ up to 2,000 \kms\ (\citealt{Shen2011}; \citealt{Richards2002}). The largest blueshifts have been discovered in an extreme quasar population at 2.2 $\simlt$ z $\simlt$ 5.9 the so-called Weak Line QSOs (hereafter WLQs), which exhibit weak/undetectable UV emission lines, i.e. they are defined as QSOs with rest-frame equivalent width (REW) of the CIV emission line of less than 10 \AA\ (\citealt{Fan1999}; \citealt{Diamond2009}; \citealt{Plotkin2010}; \citealt{Wu2011}).
The velocity shift of the CIV emission line correlates with several parameters of optical and ultraviolet emission lines. In particular it is part of 4D Eigenvector 1 parameter space (\citealt{Boroson1992}; \citealt{Sulentic2000}), which also involves the broad Fe II emission, the equivalent width of the narrow [OIII] component, the width of H$\beta$ emission line and the X-ray photon index.
The physical driver of the 4D Eigenvector 1 is thought to be the Eddington ratio (\edd), since QSOs showing large blueshifts are accreting at high rate with \edd\ $\geq$ 0.2 (\citealt{Marziani2001}; \citealt{Boroson2002}; \citealt{Shen2014}). 

As found by \cite{Marziani2016}, these winds occurring in the BLR may affect the host galaxy. They studied a sample of QSOs with \lbol\ = 10$^{45-48.1}$ \ergs\  at 0.9 $\simlt$ z $\simlt$ 3 and found that the outflow kinetic power of these winds traced by CIV emission line is comparable to the binding energy of the gas in a massive spheroid, underlying the importance of considering also these winds in the feedback scenario.
It is therefore crucial to study such AGN-driven outflows at the golden epoch of AGN activity, e.g. 1.5 $<$ z $<$ 3.5.  Furthermore, from theory and observations we know that the efficiency in driving energetic winds increases with AGN luminosity (\citealt{Menci2008}; \citealt{Faucher2012}), which brings out the need to investigate the properties of the most luminous QSOs.


We have undertaken a multiwavelength survey to investigate a sample of 86 Type-1 WISE/SDSS selected hyper-luminous AGNs (\lbol\ $>$ 2 $\times$ 10$^{47}$ \ergs ; Duras et al in prep.), i.e. the WISSH sample (see \citealt{Bischetti2017} for further details, Paper I hereafter; \citealt{Duras2017}).
In Paper I we discussed about the discovery of extreme kpc-scale outflows in five WISSH QSOs, traced via the blueshift of the [OIII]$\lambda$5007 \AA\ emission line.
In this paper we report on the discovery of fast BLR winds (with velocities up to 8,000 \kms) traced by the blueshift of the CIV with respect to the systemic redshift and their possible relation with NLR winds. We concentrate the analysis on the first 18 WISSH QSOs with  rest-frame optical observations and ultraviolet (UV) archival data at z $\sim$ 2.2--3.6. In Sect. \ref{obs}, we present details about observations and data reduction of proprietary data, such as LBT/LUCI1 and VLT/SINFONI spectra, and archival data from SDSS DR10.  In Sect. \ref{sec:analysis}, we outline the models used in our spectral analysis and present the results of the spectral fit, both for optical and UV spectra. In Sect. \ref{oiii_hbeta}, we discuss the properties of the [OIII] and H$\beta$ emission lines in the context of a scenario based on the orientation of our line of sight. H$\beta$-based SMBH masses and Eddington ratios are presented in Sect. \ref{sec:mbh}. We discuss the properties of the CIV emission line in Sect. \ref{sec:shift}, with particular emphasis on the blueshifted component and the mass outflow rate and kinetic power of the associated BLR winds in the WISSH QSOs, and their relation with NLR winds traced by [OIII]. In Sect. \ref{driver} we investigate the dependence of the CIV velocity shifts with fundamental AGN parameters such as \lbol , \edd\ and the spectral index \aox\ defined as:
 \begin{equation}
  \alpha\rm_{OX} = \frac{\rm Log(L_{2keV}/L_{2500 \AA})}{\rm Log(\nu_{2keV}/\nu_{2500 \AA})} 
  \end{equation}
 and representing the slope of a power law defined by the rest-frame monochromatic luminosities at 2 keV (L$\rm_{2keV}$)  and 2500 \AA\ (L$\rm_{2500 \AA}$).

 Finally we summarize our findings in Sect. \ref{summary}.
Throughout this paper, we assume H$_{0}$ = 71 \kms \mpc , $\Omega_{\Lambda}$ = 0.73 and $\Omega_{m}$ = 0.27.

\begin{table*}[t]
	\centering
	\begin{threeparttable}
	\caption{Properties of the WISSH QSOs considered in this paper.}\label{tab:data1}
		\begin{tabular}{lcccccccccccc}
			\hline
			\hline
			SDSS name& RA & Dec  & z$_{SDSS}$ & \it{u} & \it{g} & \it{r} & \it{i} & \it{z} & J & H & K &  E(B--V) \\ 
			(1) & (2) & (3) & (4) & (5) & (6) &(7) & (8) & (9) & (10) & (11) & (12)& (13) \\
			\hline
			J0801$+$5210 & 08:01:17.79&  $+$52:10:34.5 &  3.217 &  19.73	&17.26	&16.90	&16.76	&16.69& 15.71  &15.34 &  14.61               & 0.0 \\
			J0958$+$2827 & 09:58:41.21 & $+$28:27:29.5 &  2.120 &  20.95&	18.50&	17.84&	17.71&	17.51&16.31  &16.04 &  15.41   					&0 .0 \\
			J1106$+$6400 & 11:06:10.72 & $+$64:00:09.6  &  2.205 &  16.44&	16.11&	16.01&	15.98&	15.81 & 	15.06&  14.55   &13.95			& 0.0 \\
			J1111$+$1336 & 11:11:19.10 & $+$13:36:03.9 &  3.462 &  22.26&	18.11&	17.30&	17.18&	17.05 &	15.89 &15.51 &  15.03 & 0.0 \\ 
			J1157$+$2724 & 11:57:47.99 &$+$ 27:24:59.6 &  2.214 &  19.98&	19.11&	18.25&	17.77&	17.31 &16.14&  15.43  & 14.48	 & 0.2 \\ 
			J1201$+$0116 & 12:01:44.36 & $+$01:16:11.6 &  3.240 &  19.50&	17.72&	17.39&	17.32&	17.27 &	15.90 & 15.88 &  14.84  & 0.0 \\ 
		    J1236$+$6554 & 12:36:41.45 & $+$65:54 :42.1  &  3.358 &  20.52	&17.81&	17.29&	17.19&	16.96 &	16.00 & 15.47&   15.31& 0.0 \\ 
			J1421$+$4633 & 14:21:23.97 & $+$46:33:18.0 &  3.404 &  21.96&	17.97&	17.38&	17.22&	17.09 &	16.28 & 15.49 & 14.89  	 & 0.0 \\ 
			J1422$+$4417 & 14:22:43.02 & $+$44:17:21.2 &  3.605 &  23.45&	19.36&	18.15&	17.57&	17.11 &15.81 & 15.21 &  14.46		 & 0.1 \\ 
			J1521$+$5202 & 15:21:56.48 & $+$52:02:38.5 &  2.195 &  16.44&	15.90&	15.60&	15.44&	15.33& 14.64 & 14.06  & 13.35			& 0.0\\ 
			J1538$+$0855 & 15:38:30.55 & $+$08:55:17.0 &  3.551 &  24.14&	18.17&	17.17&	17.00&	16.84& 	15.67&  15.48 &  14.72  	 & 0.0 \\ 
			J2123$-$0050 & 21:23:29.46 & $-$00:50:52.9 &  2.269&  17.16&	16.63&	16.42&	16.34&	16.10&	   15.18 &14.62   &13.90			  & 0.0 \\ 
			J2346$-$0016 & 23:46:25.66 & $-$00:16:00.4 &  3.489&  21.24&18.69&	17.91&	17.75&	17.56 & 	16.87  &16.13  & 15.30   			&0.0 \\ 
			\hline
			\hline
			J0745$+$4734\tnote{a} & 07:45:21.78 & $+$47:34:36.1 &  3.214 &  19.56 & 16.63 & 16.35 & 16.29 & 16.19 & 15.08 & 14.61 & 13.95&0.0 \\
			J0900$+$4215\tnote{a} & 09:00:33.50 & $+$42:15:47.0 & 3.297 &  22.97 & 17.11 & 16.74 & 16.69 & 16.58 & 15.37 & 14.68 & 14.00 &0.0\\
			J1201$+$1206\tnote{a}& 12:01:47.90 & $+$12:06:30.3 &  3.495 &  20.77 & 18.31 & 17.41 & 17.31 & 17.18 & 15.87 & 15.25 & 14.61&0.0 \\
			J1326$-$0005\tnote{a}& 13:26:54.95 & $-$00:05:30.1 &  3.307 &  22.83 & 20.90 & 20.54 & 20.02 & 19.28 & 17.39 & 16.75 & 15.31 &0.3\\ 
			J1549$+$1245\tnote{a} & 15:49:38.71 & $+$12:45:09.1 &  2.386 &  20.23 & 18.67 & 17.84 & 17.38 & 16.90 & 15.86 & 14.56 & 13.52 &0.1\\

			\hline 
		\end{tabular} 
	 \begin{tablenotes}[para,flushleft]
	 \item {\bf{Notes}}. Columns give the following information: (1) SDSS name,  (2--3) celestial coordinates, (4) redshift from SDSS DR10, (5--9) SDSS DR10 photometric data, (10--12) 2MASS photometric data, and  (13) colour excess E(B$-$V) (Duras et al. in prep).

 \item[a] See Paper I for the analysis of the LBT/LUCI spectra of these objects.
     				  \end{tablenotes}
     				       \end{threeparttable}
	
\end{table*}

\begin{table*}[]
	\centering
\begin{threeparttable}
	\caption{Journal of the LBT/LUCI1 observations.}\label{tab:data2}
		\begin{tabular}{lccccc}
			\hline
			\hline
			SDSS name &  Obs. date& t$_{exp}$ & Grating & R & Seeing  \\
			(1) & (2) & (3) & (4) & (5)&(6)  \\
			\hline
			J0801+5210& Feb 10\textsuperscript{th} 2015& 2400& 150 K$_s$ & 4150 & 1.05 \\
			J0958+2827 &Mar 21\textsuperscript{th} 2015&3600&150 K$_s$& 4150& 0.75\\
			J1106+6400 & Apr 17\textsuperscript{th} 2015&1800&210 zJHK & 7838 & 0.69 \\
			J1111+1336 & Apr 20\textsuperscript{th} 2014&2400&150 K$_s$ & 4150 & 0.55 \\
			J1157+2724 &Mar 22\textsuperscript{th} 2015&1200& 210 zJHK & 7838  & 0.9\\
			J1201+0116 &Apr 27\textsuperscript{th} 2015&1440& 150 K$_s$ & 4150 & 0.78\\
			J1236+6554 &Apr 20\textsuperscript{th} 2014&2400 & 150 K$_s$ & 4150 & 0.66\\
			J1421+4633 & Apr 20\textsuperscript{th} 2014& 1500&150 K$_s$ & 4150 & 0.52 \\
			J1422+4417 & Apr 20\textsuperscript{th} 2014& 800& 150 K$_s$ & 4150& 0.55\\
			J1521+5202 & Apr 20\textsuperscript{th} 2014& 1200&210 zJHK & 7838  & 1.01\\
			\hline 
		\end{tabular} 
		\begin{tablenotes}[para,flushleft]
	\item {\bf{Notes.}} Columns give the following information: (1) SDSS name  (2) observation date, (3) exposure time (in units of s), (4) grating, (3) resolving power R, and (5) on-site average seeing (in units of arcsec).
\end{tablenotes}
\end{threeparttable}
	
\end{table*}

\begin{table*}[]
	\centering
	\begin{threeparttable}
	\caption{Journal of the SINFONI observations.}\label{tab:data3}
	
		\begin{tabular}{lccccc}
			\hline
			\hline
			SDSS name&Obs. date&t$_{exp}$ &  Grating & R  & Spatial resolution\\
			(1) & (2) & (3) & (4) &(5)&(6) \\
				\hline
			J1538+0855 & Jul 19\textsuperscript{th} 2014&2400&K & 4000 & 0.3\\
			J2123-0050 & Jul 20\textsuperscript{th} 2014& 600& H& 3000&0.2\\
			J2346-0016 & Aug18\textsuperscript{th} - Sept 1\textsuperscript{st} 2014& 4800& K& 4000&0.3\\
			\hline 
		\end{tabular} 
\begin{tablenotes}[para,flushleft]
\item {\bf{Notes.}} Columns give the following information: (1) SDSS name, (2) observation date, (3) exposure time (in units of s), (4)  grating, (5) resolving power R, and (6) the spatial resolution obtained with AO technique (in units of arcsec).
\end{tablenotes}
\end{threeparttable}

\end{table*}

\section{Observations and data reduction}\label{obs}
We have collected near-infrared (NIR) spectroscopic observations for 18 WISSH QSOs. Five of them have been discussed in Paper I, while the properties of additional 13 QSOs, with LUCI1/LBT (10 objects) and VLT/SINFONI (3 objects) observations are presented in the present paper.  The coordinates, SDSS 10th data release (DR10) redshift, optical photometric data (\citealt{Ahn2014}),  2MASS NIR photometric data (\citealt{Skrutskie2006}) and colour excess E(B$-$V) derived from broad-band SED fitting (Duras et al 2018, in prep) of the WISSH QSOs analyzed here are listed in Table \ref{tab:data1}.
 

\subsection{LUCI/LBT observations}

The observations of ten WISSH QSOs were carried out with LUCI1,  NIR spectrograph and imager at the Large Binocular Telescope (LBT) located in Mount Graham, Arizona. 
We required medium resolution spectroscopic observations in longslit mode with a slit of 1 arcsec width, using the N1.8 camera. The gratings 150\_K$_s$ (R = 4140) in the K band and 210\_zJHK (R = 7838) in H band were used for objects at redshift bins 2.1 $<$ z $<$ 2.4 and 3.1 $<$ z $<$ 3.6, respectively. The wavelength ranges covered are 1.50-1.75 $\mu$m and 1.95-2.40 $\mu$m for the H and K bands, respectively. The observations were performed from April 2014 to March 2015, in two different Cycles. The average on-site seeing of both cycles is comparable, i.e. $\sim$0.8 arcsec (See Table \ref{tab:data2} for further details).
Observations of stars with known spectral type were taken during the observing nights, to account for telluric absorption and flux calibration.

The data reduction was performed using IRAF tasks and IDL routines on both targets and standard stars. The reduced final frames are sky lines and telluric absorption free, wavelength and flux calibrated.
 We used Argon and Neon arc-lamps for the wavelength calibration of J0801$+$5210, J0958$+$2827, J1106$+$6400, J1157$+$2724 and J1201$+$0116; OH sky emission lines for the J1111$+$1336, J1236$+$6554, J1421$+$4633 and J1422$+$4417; finally, Argon and Xenon arc-lamps for J1521$+$5202. For the lines identification, a cubic spline was used to fit the pixel coordinates to the wavelengths provided. Analysis of the night sky lines indicated an uncertainty in the wavelength calibration $\simlt$ 20 \kms . We performed an unweighted extraction of the 1D spectrum, using different apertures according to the spatial profile of each 2D spectrum, which was traced using a Chebyshev function of the 3rd order.
 
 The sky subtraction was done by subtracting frames corresponding to different telescope pointings, along the slit (ABBA method). To remove the atmospheric absorption features we used the IDL routine XTELLCOR$\_$GENERAL (\citealt{Vacca2003}). This routine makes use of (i) the observed standard stars spectra, which are affected by telluric absorption as the targets, and (ii) a Vega model spectrum, useful to build up an atmospheric absorption free spectrum. As the magnitude of the standard stars are also known, we calibrated the target spectra in absolute flux, with the same IDL routine.

\subsection{SINFONI observations}
Three QSOs were observed with Spectrograph for INtegral Field Observations in the Near Infrared (SINFONI), at the VLT installed at the Cassegrain focus of Unit Telescope 4 (UT4). The observations were performed in service mode during the period 093.A, with the adaptive optics (AO) - Laser Guide Star mode, by which the atmospheric turbulence can be partially corrected.
J1538+0855 and J2123-0050 were observed during the nights of Jul 19\textsuperscript{th}-Jul 20\textsuperscript{th} 2014, while J2346$-$0016 was observed during Aug 18\textsuperscript{th} and Sept 1\textsuperscript{st} 2014, as part of the program 093.A-0175.
All the targets were observed within a field of view of 3''$\times$3'', with a spaxel scale of 0.05''$\times$0.1''. The final spatial resolution obtained thanks to the LGS correction is $\sim$ 0.3'' for J1538$+$0855 and J2346$-$0016, and $\sim$ 0.2'' for J2123-0050, as derived by the H$\beta$ broad emission line.
Given the redshift range of the QSOs, we used K and H filters to target the H$\beta$ - [OIII]$\lambda$5007\AA\ spectral region with a resolving power of R$\sim$4000 and 3000, respectively.
The data reduction was performed using ESO pipelines,  in Reflex environment (\citealt{reflex}). The background sky emission was removed with the IDL routine SKYSUB, using object frames taken at different telescope pointings (see \citealt{Davies2007}).
After the cube reconstruction, the flux calibration was performed with our own IDL routines, by means of telluric spectra acquired during the same nights as the targets. As a final step, we extracted a 1D spectrum for each source with a circular aperture of 1'' diameter. This procedure does not allow to have the spatial information, but we are interested in the integrated emission from different AGN components (i.e. BLR and NLR).  The analysis of the spatially-resolved kinematics of the [OIII]$\lambda$5007 emission line will be presented in a forthcoming paper.
Detailed information about SINFONI observations are listed in Table \ref{tab:data3}.

\subsection{SDSS archival data}
We retrieved the rest-frame UV spectrum of the 18 WISSH QSOs with NIR spectroscopic data, from the Sloan Digital Sky Survey (SDSS) DR10 public archive, in order to derive the properties of the CIV emission line. The SDSS survey uses a dedicated wide-field 2.5m telescope (\citealt{Gunn2006}) at Apache Point Observatory in Southern New Mexico. The spectra were obtained from February 2001 to February 2012 with two spectrographs, SDSS-I and BOSS, with a resolving power of 1500 at 3800 \AA\ and 2500 at 9000 \AA .  The optical brightness of our objects, i.e. 15.6 $< r <$ 18.25, ensures data with an excellent S/N ratio ($>$15) of the analyzed spectra.



\section{Spectroscopic analysis}\label{sec:analysis}
\subsection{Modeling of the H$\beta$  and [OIII] emission}\label{sec:OIII_analysis}

  \begin{table*}[]
	\centering
	 \small
\begin{threeparttable}
	\caption{Properties of core and BLR components of the H$\beta$ emission line derived from parametric model fits. Objects in boldface belong to the \oiii\ sample (see Sect. \ref{sec:OIII_analysis} for details).}\label{tab:beta}

			\begin{tabular}{lcccccccc}
				\hline
				\hline	
				SDSS Name & z\tnote{a}& FWHM$\rm ^{core}_{H\beta}$&REW$\rm ^{core}_{H\beta}$& $\rm \lambda^{BLR}_{H\beta}$  & FWHM$\rm ^{BLR}_{H\beta}$  &REW$\rm ^{BLR}_{H\beta}$ & Flux H$\rm \beta^{BLR}$/ Flux H$\rm \beta^{Tot}$&Model$\rm ^{BLR}_{H\beta}$\\
				&&\footnotesize{(km s$^{-1}$)}&  \footnotesize{(\AA)} & \footnotesize{(\AA)}&\footnotesize{(km s$^{-1}$)} & \footnotesize{(\AA)} & \\
				\hline
			J0801+5210& 3.257$\pm0.002$&1000$^{b}$&2.4\footnotesize{$_{-0.3}^{+0.2}$}&4862$\pm1$&6080\footnotesize{$_{-160}^{+130}$}&43.8\footnotesize{$_{-0.8}^{+0.5}$}&0.95\footnotesize{$_{-0.09}^{+0.05}$}&Gaussian\\
			J0958+2827 &3.434$\pm0.001$&1000$^{b}$&2.1\footnotesize{$_{-0.6}^{+0.2}$}&4859$\pm1$&5060\footnotesize{$_{-160}^{+140}$}&39.4\footnotesize{$_{-0.5}^{+0.8}$}&0.95$\pm0.05$&Gaussian\\
			J1106+6400&2.221\footnotesize{$_{-0.003}^{+0.002}$}&1000$^{b}$&0.6$\pm0.2$&4850\footnotesize{$_{-2}^{+1}$}&7920\footnotesize{$_{-180}^{+300}$}&92.0$\pm3.0$&0.99$\pm0.01$&BPL\\
			J1111+1336&3.490\footnotesize{$_{-0.005}^{+0.004}$}&1000$^{b}$&0.9\footnotesize{$_{-0.1}^{+0.3}$}&4871\footnotesize{$_{-2}^{+1}$}&7390$\pm110$&71.0\footnotesize{$_{-1.0}^{+2.0}$}&0.99\footnotesize{$_{-0.02}^{+0.01}$}&BPL\\
			J1157+2724&2.217$\pm0.001$&1000$^{b}$&3.0\footnotesize{$_{-0.4}^{+0.1}$}&4849$\pm1$&4200\footnotesize{$_{-100}^{+140}$}&56.0$\pm1.0$&0.95$\pm0.03$&BPL\\
			J1201+0116&3.248\footnotesize{$_{-0.001}^{+0.002}$}&1000$^{b}$&6.8\footnotesize{$_{-1.0}^{+0.3}$}&4855$\pm2$&4670\footnotesize{$_{-330}^{+130}$}&55.0$\pm2.0$&0.89\footnotesize{$_{-0.04}^{+0.05}$}&BPL\\
			J1236+6554&3.424$\pm0.001$&1000$^{b}$&1.2$\pm0.2$&4848$\pm2$&5330\footnotesize{$_{-220}^{+150}$}&64.0$\pm2.0$&0.98\footnotesize{$_{-0.03}^{+0.02}$}&BPL\\
			J1421+4633&3.454$\pm0.002$&1000$^{b}$&2.2$\pm0.3$&4838$\pm3$&6480$\pm250$&55.0\footnotesize{$_{-1.0}^{+2.0}$}&0.96\footnotesize{$_{-0.03}^{+0.04}$}&BPL\\
			J1422+4417&3.648$\pm0.001$&1000$^{b}$&3.0\footnotesize{$_{-0.4}^{+0.2}$}&4851$\pm2$&4430\footnotesize{$_{-240}^{+270}$}&21.0$\pm1.0$&0.89$\pm0.20$&Gaussian\\
			J1521+5202&2.218$\pm0.001$&1000$^{b}$&1.2$\pm0.1$&4861$\pm1$&7240\footnotesize{$_{-130}^{+140}$}&25.9$\pm0.5$&0.96\footnotesize{$_{-0.06}^{+0.04}$}&Gaussian\\
			{\bf{J1538+0855}}&3.567\footnotesize{$_{-0.002}^{+0.003}$}&1000$^{b}$&2.1\footnotesize{$_{-0.4}^{+0.2}$}&4843$\pm2$&5490\footnotesize{$_{-210}^{+400}$}&104.0\footnotesize{$_{-3.0}^{+4.0}$}&0.98$\pm0.02$&BPL\\
			J2123-0050&2.282$\pm0.001$&1000$^{b}$&1.2$\pm0.2$&4850$\pm2$&4900\footnotesize{$_{-160}^{+150}$}&41.1$\pm0.5$&0.97\footnotesize{$_{-0.02}^{+0.01}$}&BPL\\
			J2346-0016&3.511\footnotesize{$_{-0.002}^{+0.001}$}&1000$^{b}$&4.2$\pm$0.9&4864\footnotesize{$_{-7}^{+4}$}&5820\footnotesize{$_{-560}^{+310}$}&66.0\footnotesize{$_{-5.0}^{+6.0}$}&0.94\footnotesize{$_{-0.09}^{+0.06}$}&BPL\\
				\\
				\hline
				\hline
			{\bf{J0745+473}}4\tnote{c}&3.225$\pm0.001$&470$\pm$170&-&4862$\pm1$&8600\footnotesize{$_{-200}^{+230}$}&85.5\footnotesize{$_{-1.8}^{+2.7}$}&0.98\footnotesize{$_{-0.08}^{+0.02}$}&BPL\\
			{\bf{J0900+4215}}\tnote{c}&3.294$\pm0.001$&300$\pm$130 &0.2\footnotesize{$_{-0.2}^{+1.0}$}&4861$\pm1$&3210$\pm190$&54.9\footnotesize{$_{-0.5}^{+0.9}$}&0.98$\pm0.02$&BPL\\
			{\bf{J1201+1206}}\tnote{c}&3.512$\pm0.002$&700$\pm$200&10.2$\pm$1.5&4865$\pm2$&6160$\pm250$&68.4\footnotesize{$_{-1.3}^{+1.2}$}&0.86$\pm0.07$&Gaussian\\
			{\bf{J1326-0005}}\tnote{c}&3.303$\pm0.001$&710$\pm$170&0.4\footnotesize{$_{-0.4}^{+1.2}$}&4862$\pm1$&3700$\pm160$&96.3\footnotesize{$_{-0.5}^{+0.7}$}&0.93\footnotesize{$_{-0.09}^{+0.07}$}&BPL\\
			{\bf{J1549+1245}}\tnote{c}&2.365$\pm0.001$&270$\pm$160&0.3\footnotesize{$_{-0.3}^{+0.5}$}&4871$\pm2$&8340$\pm280$&63.9\footnotesize{$_{-2.1}^{+1.8}$}&0.98\footnotesize{$_{-0.09}^{+0.02}$}&BPL\\

\hline
					\end{tabular}
		 \begin{tablenotes}[para,flushleft]
		 			\item[a] Redshifts are based on the core component of H$\beta$ emission line from NIR spectroscopy.
		 			
		 			\item[b] Best fit value corresponding to the maximum allowed value of FWHM (see Sect. \ref{sec:OIII_analysis}). For these objects we only report the FWHM$\rm_{[H\beta]}^{core}$ corresponding to the best-fit values.

       			 \item[c] WISSH QSOs with LUCI1 spectra analyzed in \cite{Bischetti2017}. For these objects the redshifts are based on the core component of the [OIII] emission line.
     				  \end{tablenotes}
       \end{threeparttable}
    
			\end{table*} 
			
The visual inspection of the rest-frame optical spectra of the 13 WISSH QSOs taken with LBT/LUCI or VLT/SINFONI  revealed the weakness of the [OIII]$\lambda\lambda$4959,5007 \AA\ doublet emission lines and the presence of a strong and complex FeII emission (see Appendix \ref{sec:app_lbt}). This combination makes the determination of [OIII] spectral parameters an extremely challenging task. In one case, i.e. J1422+4417, the LBT/LUCI1 bandwidth does not cover the [OIII] spectral region.

We developed a model based on the IDL package MPFIT (\citealt{Markwardt2009}), in order to accurately infer the properties of the [OIII] and H$\beta$ emission in these hyper-luminous QSOs.
The NIR spectra were fitted with a spectral model which consists on:

\begin{itemize}

\item {\bf{H$\beta$/[OIII] core components associated with the narrow line region (NLR)}}. Specifically, one Gaussian component for fitting the core of H$\beta$ 4861 \AA\ line with FWHM$\rm_{H\beta}^{core}$ $\leq$ 1,000\footnote{The limit of 1,000 \kms\ for the  FWHM has been set in order to account for a possible subtle H$\beta$ outflow component.} \kms\ plus one Gaussian profile for each component of the [OIII]$\lambda\lambda$4959,5007 \AA\ doublet. The FWHM of the [OIII] doublet was forced to be the same as that of the H$\beta$ core component. Moreover, we assumed the theoretical flux ratio for [OIII]$\lambda\lambda$4959,5007 \AA\ equal to 1:3 and the separation between [OIII]$\lambda\lambda$4959--5007 \AA\ and H$\beta$ fixed to 98--146 \AA\ in the rest-frame, respectively.

\item {\bf{[OIII] outflow components}}. Specifically, one Gaussian profile with the same FWHM for each component of the [OIII] in order to accurately constrain the possible presence of subtle broad/shifted components associated with outflows.
Furthermore, we left free to vary the values of the FWHM (FWHM$\rm_{[OIII]}^{broad}$) and centroids ($\lambda\rm_{[OIII]}^{broad}$)  in the range 500--2,500 \kms\ and 4980--5034 \AA , respectively, which are based on the values derived for these spectral parameters from the analysis of the [OIII] broad features in Paper I.

\item {\bf{H$\beta$ broad component associated with the broad line region (BLR)}}. Specifically, one Gaussian profile with width free to vary or a broken power-law (BPL) component convolved with a Gaussian curve (e.g. \citealt{Nagao2006}, \citealt{Carniani2015}, Paper I).

\end{itemize}  
 Similarly to Paper I, in order to fit the prominent FeII emission, we included one or two Fe II templates depending on the complexity of the iron emission. Specifically, this was done by adding to the fit one or two templates chosen through the $\chi^2$ minimization process among a library consisting of the three observational templates by \cite{Boroson1992}, \cite{Veron2004}, \cite{Tsuzuki2006} and twenty synthetic FeII templates obtained with the CLOUDY plasma simulation code (\citealt{Ferland2013}), which takes into account different hydrogen densities, ionizing photon densities and turbulence. The FeII template was then convolved with a Gaussian function with a velocity dispersion $\sigma$ in the range 500-4,000 \kms .
  The continuum was fitted by a power law component with the normalization left free to vary and the slope derived from the best-fit SED in the wavelength range of the observations. This parametrization worked well for all but two sources, i.e. J1538+0855 and J2123-0050, for which we used a slope left free to vary. We estimated the statistical noise from the line-free continuum emission and assumed to be constant over the entire wavelength bandwidth. 
Finally, the uncertainty of each parameter has been evaluated considering the variation of $\chi^2$ around the best fit value, with a grid step. From the corresponding $\chi^2$ curve we measured the $\Delta \chi^2$ with respect to the minimum best-fit value. Hereafter, the reported errors refer to the one standard deviation confidence level, corresponding to a $\Delta \chi^2$ = 1 for each parameter of interest.

    \begin{table}[]  
\centering
\small
\makebox[1\columnwidth]{
\begin{threeparttable}
		\caption{Properties of the core (C) and broad (B) components  of the [OIII]$\lambda$5007 emission line derived from parametric model fits (see Sect. \ref{sec:OIII_analysis}). Objects in boldface belong to the \oiii\ sample.}\label{tab:oiii}
	\setlength\tabcolsep{2pt}

			\begin{tabular}{cccccc}
				\hline
				\hline	
				SDSS Name &Component&FWHM$\rm_{[OIII]}$ & $\rm\lambda_{[OIII]}$&REW$\rm_{[OIII]}$&L$\rm_{[OIII]}$\\
				& &\footnotesize{(km s$^{-1})$}&\footnotesize{(\AA)}&\footnotesize{(\AA)}& \footnotesize{(10$^{42}$ erg s$^{-1}$)} \\
				\hline
			J0801+5210 &C&1000\tnote{a}&5007$\pm1$&$<$0.1&$<$2.3\\
								&B&1210\footnotesize{$_{-340}^{+320}$} &5026$\pm5$ &0.7\footnotesize{$_{-0.3}^{+0.2}$}&15.0$\pm7.0$\\

			J0958+2827 &C&1000\tnote{a}&5007$\pm2$&1.0$\pm0.2$&13.0$\pm3.0$\\
									&B&2500\tnote{a}  &4980\tnote{a}&3.5\footnotesize{$_{-0.3}^{+0.4}$}&44.0$\pm5.0$\\
			J1106+6400&C&1000\tnote{a}&5007$\pm4$&0.6$\pm0.2$&10.0\footnotesize{$_{-5.0}^{+4.0}$}\\ 
							    	&B&500\tnote{a}&5022$\pm2$&0.3$\pm0.1$&5.0\footnotesize{$_{-2.0}^{+2.0}$}\\
			J1111+1336&C&1000\tnote{a}&5007\footnotesize{$_{-2}^{+3}$}&$<$0.04&$<$0.6\\
								&B&1020\footnotesize{$_{-170}^{+200}$}&5029$\pm2$&0.9$\pm0.2$&16.0$\pm4.0$\\

			J1157+2724&C&1000\tnote{a}&5007$\pm1$&$<$0.1&$<$1.5\\
								 &B&500\tnote{a}&4995$\pm2$&0.5$\pm0.1$&6.0$\pm2.0$\\

			J1201+0116&C&1000\tnote{a}&5006$\pm1$&$<$0.2&$<$3.0\\
								&B&500\tnote{a}&5022\footnotesize{$_{-35}^{+7}$}&0.2$\pm0.2$&4.7\footnotesize{$_{-4.7}^{+18.6}$}\\

			J1236+6554&C&1000\tnote{a}&5007$\pm2$&1.3$\pm0.2$&16.0$\pm$3.0\\
								 &B&820\footnotesize{$_{-210}^{+270}$} &5022$\pm2$&1.0\footnotesize{$_{-0.3}^{+0.4}$}&12.0$\pm4.0$\\

			J1421+4633&C&1000\tnote{a}&5007$\pm2$&$<$0.01&$<$1.4\\
								&B&700$\pm140$&5027$\pm2$&1.4$\pm0.2$&26.0$\pm6.0$\\

			J1521+5202&C&1000\tnote{a}&5007$\pm1$&$<$0.1&$<$0.1\\
								&B&1270$\pm140$&5033$\pm1$&1.5$\pm0.1$&46.0$\pm4.0$\\

			{\bf{J1538+0855}}&C&1000\tnote{a}&5007$\pm3$&$<$0.6&$<$16.0\\
												&B&1980\footnotesize{$_{-300}^{+320}$} &4987\footnotesize{$_{-2}^{+3}$}&7.7$\pm0.9$&210.0\footnotesize{$_{-30.0}^{+40.0}$}\\

			J2123-0050&C&1000\tnote{a}&5007$\pm2$&$<$0.01&$<$0.3\\
												&B&2300&4980\tnote{a}&2.8\footnotesize{$_{-0.2}^{+0.3}$}&50.0$\pm10$\\

			J2346-0016&C&1000\tnote{a}&5007$\pm2$&$<$0.2&$<$3.5\\
											&B&1480$\pm350$&5030\footnotesize{$_{-3}^{+4}$}&3.2$\pm0.5$&50.0$\pm20.0$\\

			\hline
			\hline

			{\bf{J0745+4734}}\tnote{b}&C&470$\pm170$&5007$\pm1$&3.6$\pm0.6$&170$\pm80$ \\
		   										&B&1630$\pm180$ &4999$\pm1$ &24.8$\pm0.8$ &1140$\pm290$ \\
		
			{\bf{J0900+4215}}\tnote{b}&C&300$\pm130$ &5007$\pm1$ &1.1$\pm0.6$ &50\footnotesize{$_{-30}^{+100}$} \\
												&B&2240$\pm170$ &4999$\pm1$ &12.5$\pm1.0$ &530$\pm100$ \\
			
			{\bf{J1201+1206}}\tnote{b}&C&1670$\pm190$ &5007$\pm2$ &18.8$\pm1.1$ &500$\pm100$ \\
												&B&940$\pm$210 &4990$\pm2$ &5.6$\pm1.0$ &150$\pm80$ \\
			
			{\bf{J1326-0005}}\tnote{b}&C&710$\pm170$ &5007$\pm1$ &26.5$\pm0.6$ &600$\pm310$ \\
												&B&1870$\pm170$ &4997$\pm1$ &50.4$\pm0.8$ &1140$\pm260$ \\
			
			{\bf{J1549+1245}}\tnote{b}&C&270$\pm160$ &5007$\pm1$ &0.6$\pm0.5$ &20\footnotesize{$_{-10}^{+40}$} \\
												&B&1240$\pm210$ &5012$\pm1$ &13.6$\pm0.7$ &470$\pm90$ \\
				\hline

					\end{tabular}

		 \begin{tablenotes}[para,flushleft]
       				\item[a] Best fit value corresponding to the minimum or maximum allowed value of FWHM or centroid (see Sect. \ref{sec:OIII_analysis}).
       				
	       			 \item[b] WISSH QSOs analyzed in Paper I.
       				
     				  \end{tablenotes}
       \end{threeparttable}
 }
  
			\end{table}

This fit yielded a good description of the observed NIR spectrum of the WISSH QSOs. The results of our spectral analysis for the H$\beta$ and [OIII]$\lambda$5007 emission lines are listed in Table \ref{tab:beta} and Table \ref{tab:oiii}, respectively. Fig. \ref{fig:esempio_hbeta} shows an example of the H$\beta$ and [OIII] spectral line fitting decomposition.

\begin{figure}
\includegraphics[width=1\columnwidth]{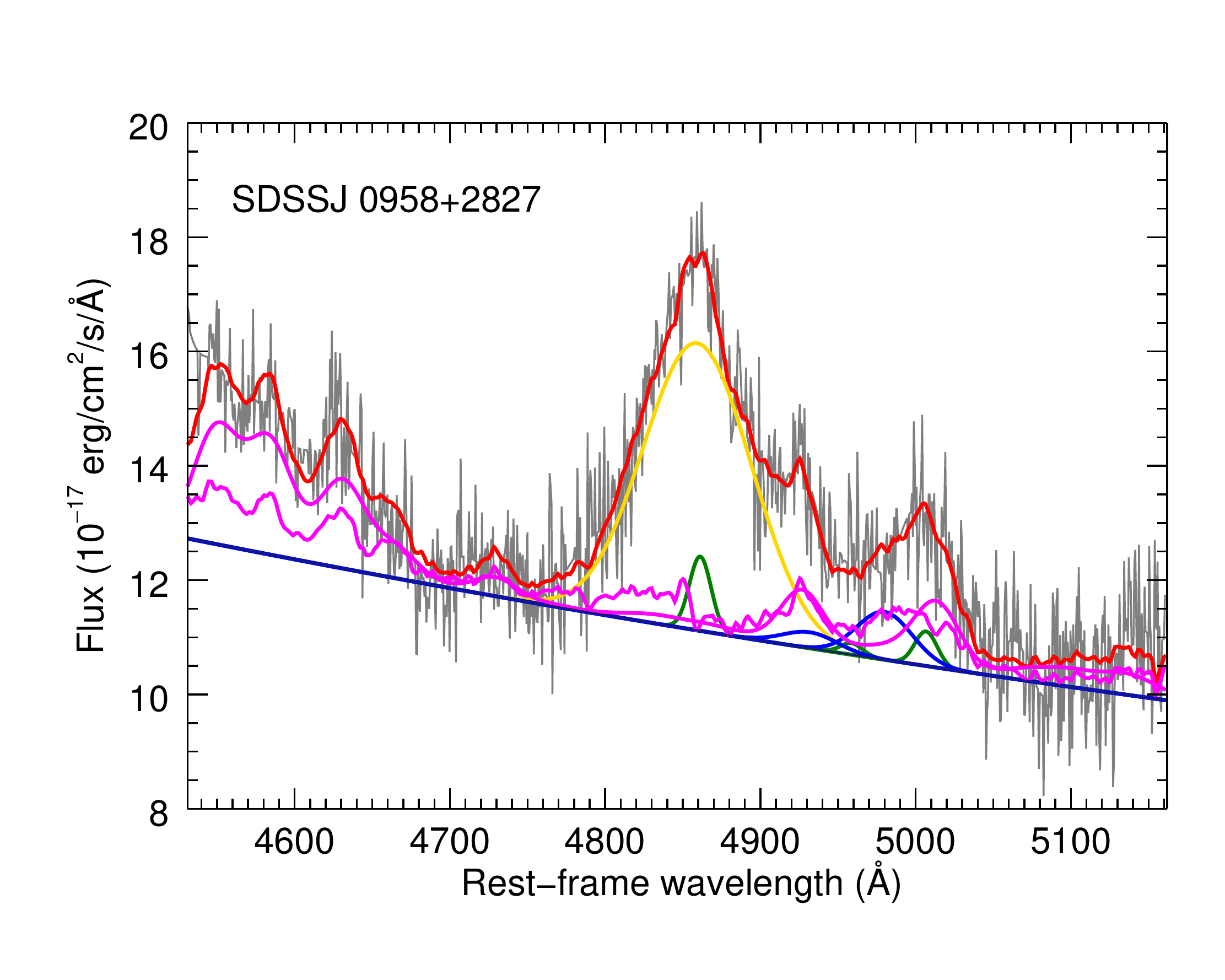}
\caption{Parametrization of the H$\beta$-[OIII] region of the WISSH QSO J0958+2827. The red line shows the best-fit to the data. Green curve refers to the core component associated with the NLR emission of H$\beta$ and [OIII] emission lines. The blue curve refers to the broad blueshifted emission of [OIII]$\lambda\lambda$4959,5007 \AA , indicative of outflow (but see Sect. \ref{sec:OIII_analysis} for more details).
Gold curve indicates the broad component of H$\beta$ associated with BLR emission. FeII emission is marked in magenta.}\label{fig:esempio_hbeta}
\end{figure}

We estimated the spectroscopic redshift from the H$\beta$ core component.
The bulk of the H$\beta$ emission is provided by the BLR component, which is well fitted by a broken power-law for all but four cases (for J0801+5210, J0958+2827, J1422+4417, J1521+5202 a Gaussian component was preferred).
The FWHM of H$\beta$ associated with the BLR has a range of 4,000-8,000 \kms, with a rest-frame equivalent width of 20 -- 100 \AA\ . 

The normalization of the [OIII] emission line associated with the NLR is consistent with zero in all but three QSOs (namely J0958+2827, J1106+6400 and J1236+6554) with a REW$\rm_{[OIII]}^{core}$  $\approx$ 1 \AA ,  confirming the weakness of this feature in the spectra considered here. 
Regarding the spectral components associated with a wind, for all but one (J1538+0855) QSOs the best value of their FWHMs or centroids is equal to the maximum or minimum value allowed by our model (500--2,500 km/s and 4980\AA , respectively), or the centroids fall extremely redward to 5007 \AA\ (i.e. $>$ 5020 \AA). The combination of these findings strongly suggests that this weak (REW$\rm_{[OIII]}^{broad}$ $\leq$ 3 \AA) Gaussian component is likely accounting for additional FeII emission not properly fitted by the templates, instead of being due to a wind. In the case of J1538+0855, we measure a broad (FWHM$\rm_{[OIII]}^{broad}$ = 1980 \kms), blueshifted (v$\rm_{[OIII]}^{broad}$ = 1,200 \kms) component with a REW$\rm_{[OIII]}^{broad}$ = 7.7 \AA , making us highly confident to associate this spectral feature with blueshifted [OIII] emission.
Hereafter, we refer to "\oiii\ sample" as the group of WISSH QSOs showing a REW of the [OIII] emission line  $\geq$ 5 \AA , while QSOs with REW$\rm_{[OIII]}^{broad}$ $<$ 5 \AA\ are included in the "\weak\ sample". Despite the fact that its NIR spectrum not covers the [OIII] region, we also included the WISSH QSO J1422+4417 as a \weak\ source, as it shows distinguishing features of this subclass (i.e. CIV velocity shift $\sim$5,000 \kms , see Sect. \ref{sec:SDSS} and Table \ref{tab:CIV}).



	\subsection{Modeling of the CIV emission}\label{sec:SDSS}
\begin{figure}[]
    \includegraphics[width=1\columnwidth]{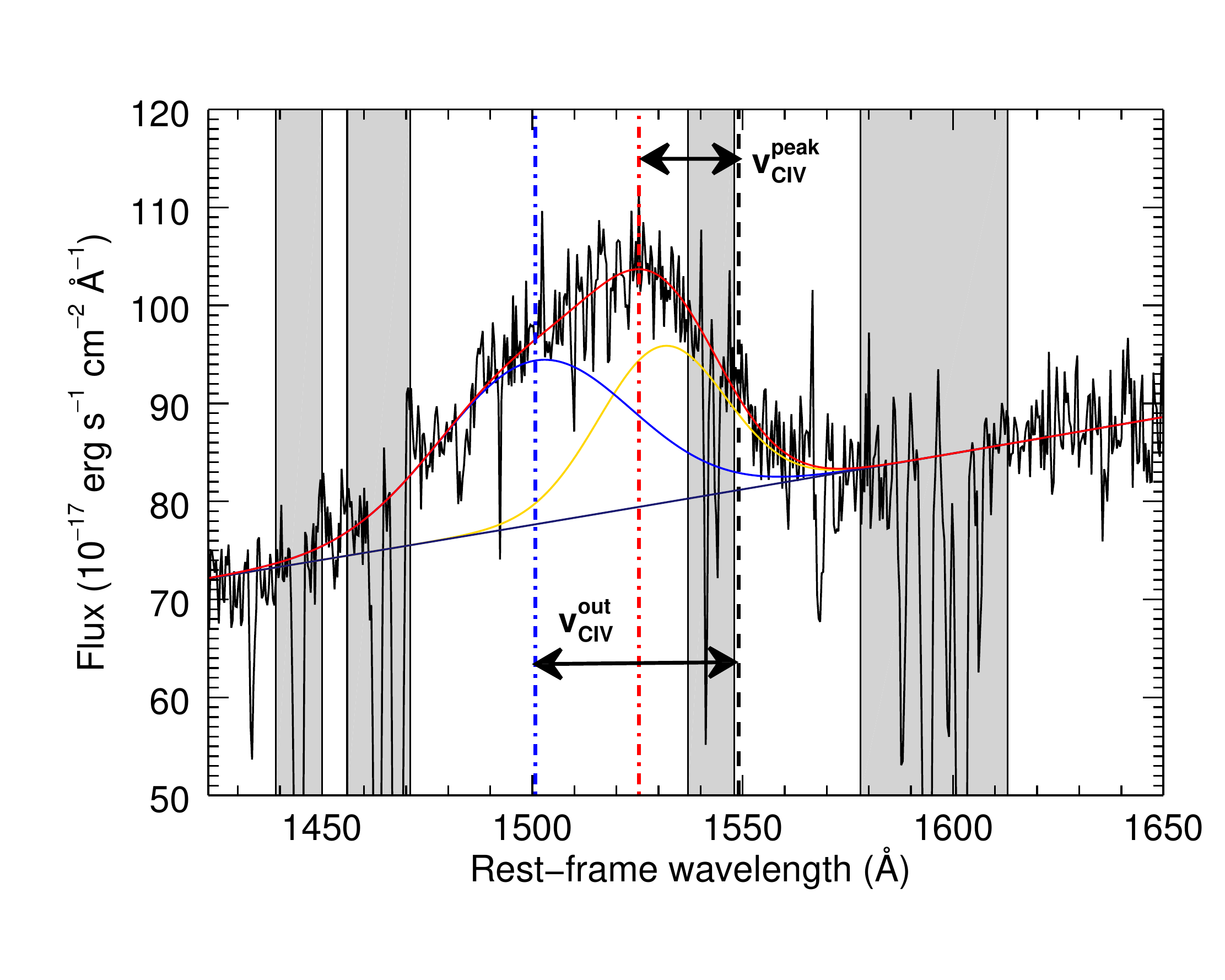}
    \includegraphics[width=1\columnwidth]{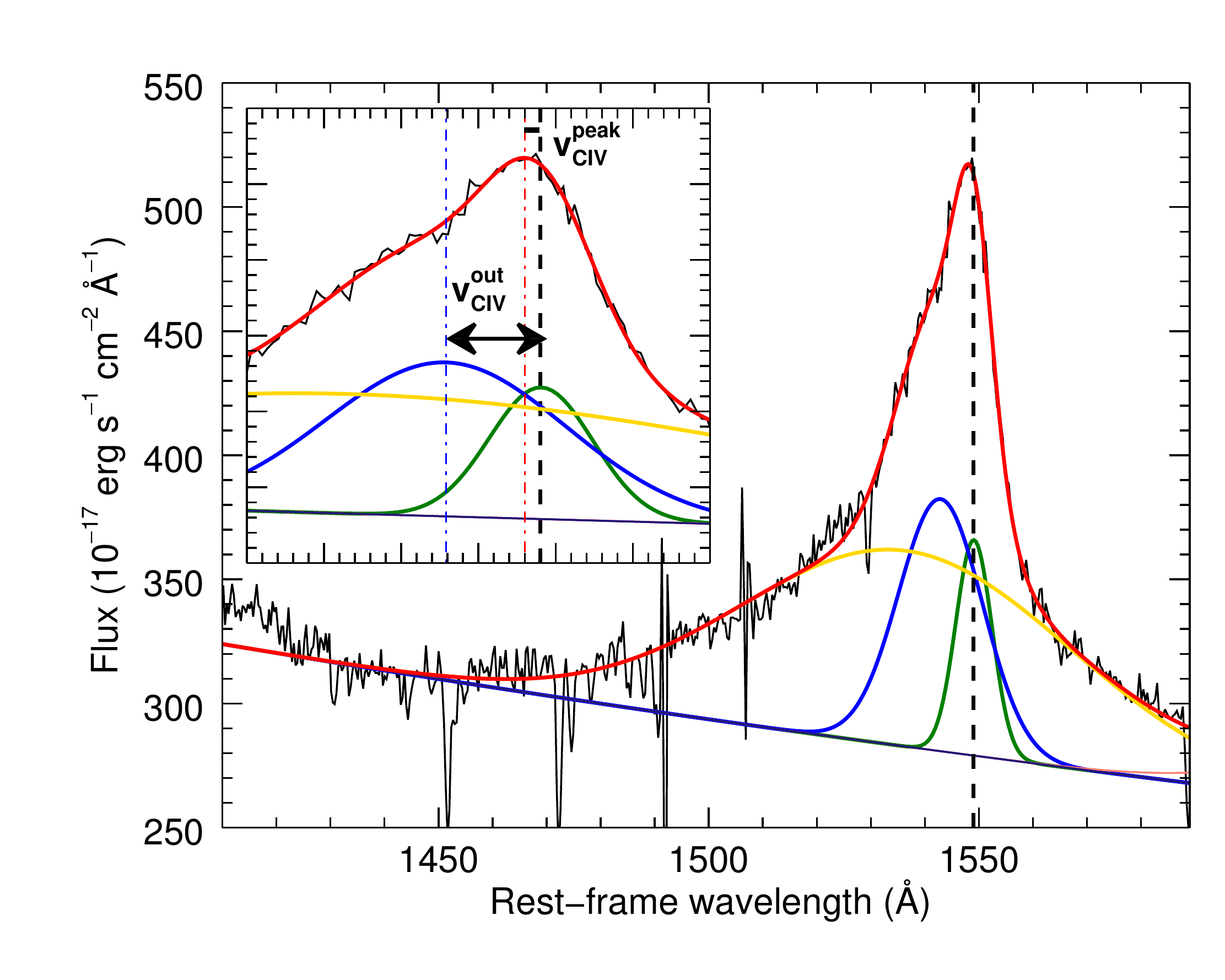}
     \caption{Parametrization of the CIV emission line profile of the \weak\ quasar J1422+4417 (Upper panel) and the \oiii\ quasar J0745+4734 (lower panel). In both panels the red, yellow and blue curves refer to the best-fit of the spectrum, virialized component and outflow component of the CIV emission line, respectively. The black dashed line indicates the $\rm\lambda_{CIV}$ at 1549 \AA\ . The red dotted-dashed line denotes the peak of the CIV emission line and \vciv\ respresent its velocity shift with respect to the systemic $\lambda$1549 \AA\ . The blue dotted-dashed line indicates the peak of the outflow component and v$\rm_{CIV}^{out}$ is its velocity shift with respect to the systemic redshift. In the lower panel the green curve refers to an additional Gaussian component required by the fit, in order to better described the CIV profile in this QSO.}\label{fig:BLUE}

    \end{figure}
We shifted the SDSS spectra to the systemic rest-frame using the redshift measured from H$\beta$ core component. 
The SDSS analysis was focused in the spectral region of CIV emission line at 1549 \AA , including spectral ranges free of emission/absorption features at both two sides of the CIV line centroid, in order to constrain the local underlying continuum. 

The CIV emission was fitted using a combination of up to three Gaussian profiles with spectral parameters (centroid, width and normalization) left free to vary. 
We found that for all but three QSOs the best-fit model consists on a combination of two Gaussian components. For the BAL QSOs J1157$+$2724 and J1549+1245, for which the blue side of CIV line profile is heavily affected by absorption, the CIV is modeled by one Gaussian component (see Fig. \ref{C6} for J1157+2724 and Fig. \ref{C15} for J1549+1245) and for the quasar J0745+4734, showing a very prominent peak at 1549 \AA , the best-fit model of the CIV profile is the combination of three Gaussian components (see Fig. \ref{C1}).

We considered in our fits the presence of FeII emission by using UV FeII templates from \cite{Vestergaard2001}, although FeII is not particularly strong in the CIV region. These templates were convolved with Gaussian profiles of different widths to account for the observed FeII-related emission features in the spectra.
If requested, we also included in the model one or two Gaussian components to account for HeII$\lambda$1640 \AA\ and/or OIII]$\lambda$1664 \AA\ emissions.

The resulting physical parameters of the CIV emission for the 18 WISSH QSOs are listed in Table \ref{tab:CIV}. 
We found a median value of REW$\rm_{CIV}$ = 21 \AA\ with the lowest value for the object J1521+5202 (REW$\rm_{CIV}$ = 8 \AA), confirming its WLQ nature (\citealt{Just2007}).
The CIV emission profile is broad with a median value of FWHM$\rm_{CIV}$ = 7,800 \kms\ and, in four cases, exceptional values of FWHM$\rm_{CIV}$ ($>$ 10,000 \kms) are detected.
Moreover we found that the peak of the CIV profile is blueshifted with respect to the systemic redshift with a velocity shift (\vciv\ hereafter) in the range of  \vciv = 200--7,500 \kms\ in all cases but J1326-0005, i.e. v$\rm_{CIV}^{peak}$ = -50 \kms\  (bearing in mind the presence of BAL features in this quasar).


The CIV emission line in the \weak\ QSOs shows broad, strongly blue and asymmetric profiles, with \vciv\ in the range of  \vciv\ = 2,500--7,500 \kms\ . This clearly indicates the presence of an outflowing component associated with this transition. According to the best-fit model, we refer to the Gaussian line with the most blueshifted centroid as the outflow component with \vout\ indicating its velocity shift with respect to the systemic (see Fig. \ref{fig:BLUE} upper panel).


The \oiii\ sample exhibit the FWHM$\rm_{CIV}$ in the range of 2,000-5,000 \kms , and lower blueshifts than those of the \weak\ QSOs, i.e. \vciv\ $\le$ 2,000 \kms . 
The profile of the CIV line in \oiii\ QSOs typically is more symmetric than in \weak\ objects, indicating a dominant contribution from emission of virialized gas. 
We therefore consider in this case the Gaussian component with the largest FWHM as the virialized one of the CIV emission line, while the second Gaussian component with a smaller FWHM is assumed to represent the emission associated with the wind (see Fig. \ref{fig:BLUE} lower panel).
It is worth noting that for the vast majority of our objects we measure a virialized component with a centroid not consistent ($>$ 2$\sigma$) with 1549 \AA\ (see Appendix \ref{sec:app_sdss}). This can be interpreted in terms of a low-velocity component likely associated with a virialized flow in a rotating accretion disk wind (\citealt{Young2007}; \citealt{Kashi2013}).
Finally, in case of J1538+0855, an \oiii\ object that also exhibits some distinctive properties of \weak\ sources (e.g. Appendix \ref{sec:app_1538}), the virialized and outflow components of the CIV emission are assumed as for \weak\ sources.

\begin{table}[]

	\makebox[1\columnwidth]{
	\small
\begin{threeparttable}
	\caption{Properties of CIV$\lambda1549$ emission line derived from parametric model fits (see Sect. \ref{sec:SDSS}). Objects in boldface belong to the \oiii\ sample. }\label{tab:CIV}
	\setlength\tabcolsep{1.7pt}
			\begin{tabular}{l ccccr}
				\hline
				\hline	
				SDSS Name & $\rm\lambda_{CIV}^{peak}$ &  \vciv \tnote{,a} & REW$\rm_{CIV}$&FWHM$\rm_{CIV}$&L$\rm_{CIV}$ \\
				& \footnotesize{(\AA)}&\footnotesize{(km s$^{-1})$}& \footnotesize{(\AA)} & \footnotesize{(km s$^{-1})$} &\footnotesize{(10$^{45}$ erg s$^{-1}$)} \\
				\hline
			{\bf{J0745+4734}}&1548$\pm1$&190$\pm120$&31.0\footnotesize{$_{-4.5}^{+0.2}$}&4320$\pm120$&8.3\footnotesize{$_{-0.1}^{+0.2}$}\\
			J0801+5210 &1535$\pm1$&2720$\pm180$&16.5$\pm0.6$&10930\footnotesize{$_{-160}^{+120}$}&2.9$\pm0.1$\\
			{\bf{J0900+4215}}&1546$\pm1$&560\footnotesize{$_{-110}^{+130}$}&31.6\footnotesize{$_{-0.7}^{+0.6}$}&4200$\pm150$&6.4\footnotesize{$_{-0.2}^{+0.3}$}\\
			J0958+2827 &1520$\pm1$&5640$\pm180$&19.0$\pm4.0$&9970\footnotesize{$_{-450}^{+510}$}&1.7$\pm0.2$\\
			J1106+6400&1534$\pm1$&2870\footnotesize{$_{-170}^{+150}$}&20.6$\pm0.3$&7830$\pm160$&2.8$\pm0.1$\\
			J1111+1336&1540$\pm1$&1920$\pm130$&20.0$\pm1.0$&8210\footnotesize{$_{-190}^{+200}$}&3.0$\pm0.2$\\
			J1157+2724\tnote{b}&1529\footnotesize{$_{-3}^{+4}$}&3840\footnotesize{$_{-570}^{+700}$}&71.0\footnotesize{$_{-14.0}^{+12.0}$}&7620\footnotesize{$_{-800}^{+650}$}&0.9$\pm0.1$\\
			J1201+0116&1531$\pm1$&3390$\pm130$&17.4$\pm0.9$&8460\footnotesize{$_{-250}^{+290}$}&1.8$\pm0.1$\\
			{\bf{J1201+1206}}&1544$\pm1$&980\footnotesize{$_{-110}^{+130}$}&37.0\footnotesize{$_{-2.0}^{+3.0}$}&5110$\pm110$&4.4\footnotesize{$_{-0.1}^{+0.2}$}\\
			J1236+6554&1532$\pm1$&3360$\pm110$&19.9\footnotesize{$_{-0.4}^{+0.3}$}&7380\footnotesize{$_{-130}^{+140}$}&2.9$\pm0.1$\\
			{\bf{J1326-0005}}\tnote{b}&1549$\pm1$&-50\footnotesize{$_{-130}^{+180}$}&82.0\footnotesize{$_{-7.0}^{+6.0}$}&2110\footnotesize{$_{-230}^{+240}$}&17.0$\pm2.0$\\
			J1421+4633&1523$\pm1$&5010\footnotesize{$_{-130}^{+290}$}&16.0$\pm2.0$&10240\footnotesize{$_{-280}^{+200}$}&2.1$\pm0.2$\\
			J1422+4417&1525$\pm1$&4670$\pm230$&20.0$\pm3.0$&12410\footnotesize{$_{-190}^{+160}$}&5.9$\pm0.1$\\
			J1521+5202&1511$\pm1$&7420$\pm200$&8.3$\pm0.6$&11500\footnotesize{$_{-350}^{+370}$}&2.1$\pm0.1$\\
			{\bf{J1538+0855}}&1538$\pm1$&2190$\pm110$&20.8$\pm0.2$&5160\footnotesize{$_{-30}^{+60}$}&4.0$\pm0.1$\\
			{\bf{J1549+1245}}\tnote{b}&1547$\pm1$&370\footnotesize{$_{-200}^{+250}$}&23.0$\pm1.0$&5720\footnotesize{$_{-210}^{+270}$}&3.8\footnotesize{$_{-0.3}^{+0.2}$}\\
			J2123-0050&1537$\pm1$&2330\footnotesize{$_{-150}^{+160}$}&12.4\footnotesize{$_{-0.7}^{+0.8}$}&6700\footnotesize{$_{-190}^{+170}$}&1.7$\pm0.1$\\
			J2346-0016&1532$\pm1$&3380\footnotesize{$_{-170}^{+130}$}&13.0$\pm2.0$&8600\footnotesize{$_{-230}^{+140}$}&1.2$\pm0.1$\\
				\hline
				
		\end{tabular}
			
				 \begin{tablenotes}[para,flushleft]
			\item[a] Velocity shifts of the CIV emission line are calculated from the peak of the multiple Gaussians modeling the CIV profile, with respect to the systemic redshift (see Fig. \ref{fig:BLUE}).
       			 \item[b] Sources J1157+2724 and J1549+1245 are classified as BAL QSOs according to the Balnicity Index (BI, \citealt{Weymann1991}) while J1326-0005 according to the modified absorption index (AI$_{1000} >$ 100, \citealt{Bruni2012}), a more conservative version of the \cite{Hall2002} AI definition. The detailed analysis of the BAL properties of the WISSH QSOs will be presented in a forthcoming paper (Bruni et al. in prep).
       			 For these objects the strong absorption in the CIV emission line affects the spectral parameters.
     				  \end{tablenotes}
       \end{threeparttable}
			}
			\end{table}

			\begin{figure}[]
             \centering
  \includegraphics[width=0.8\columnwidth]{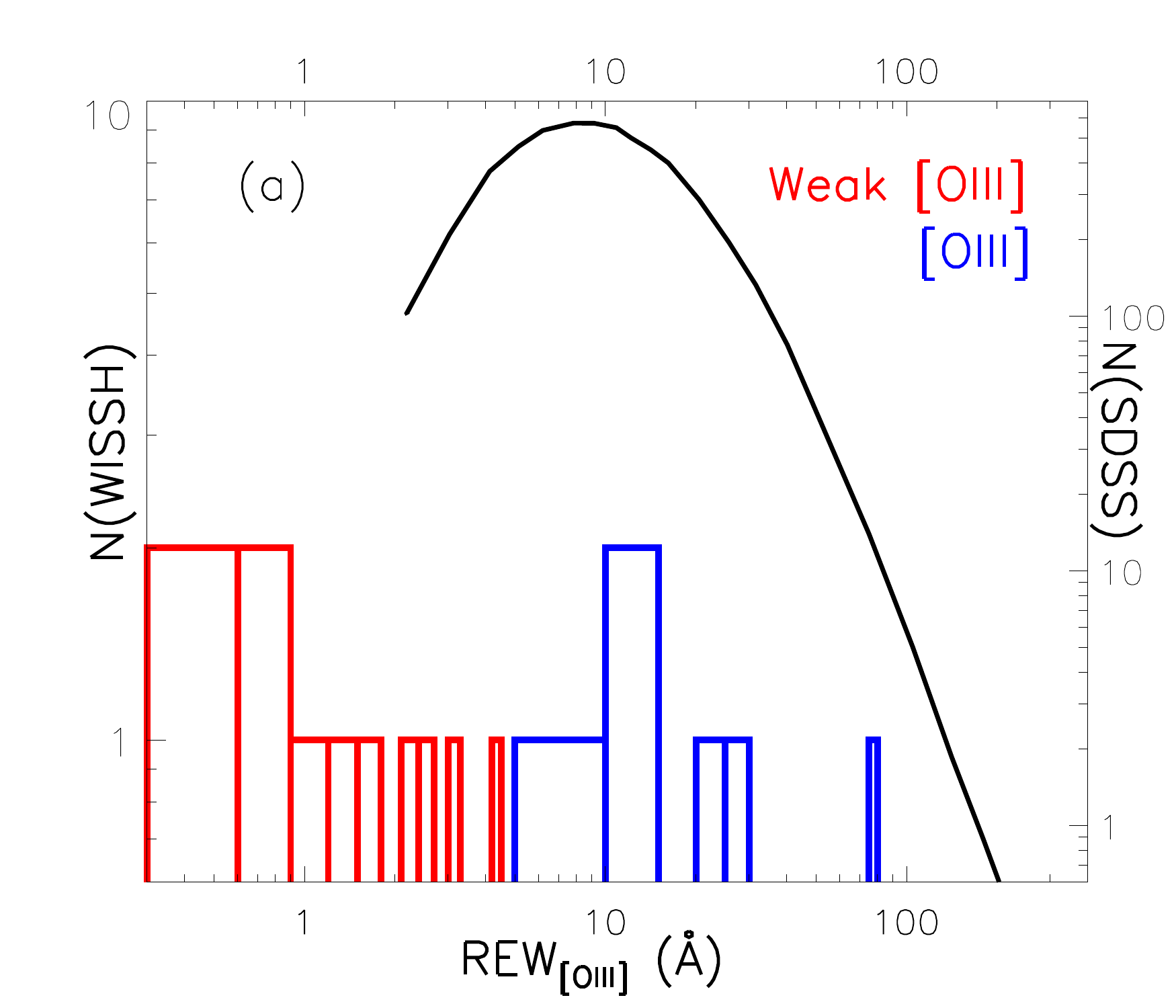}
   \includegraphics[width=0.8\columnwidth]{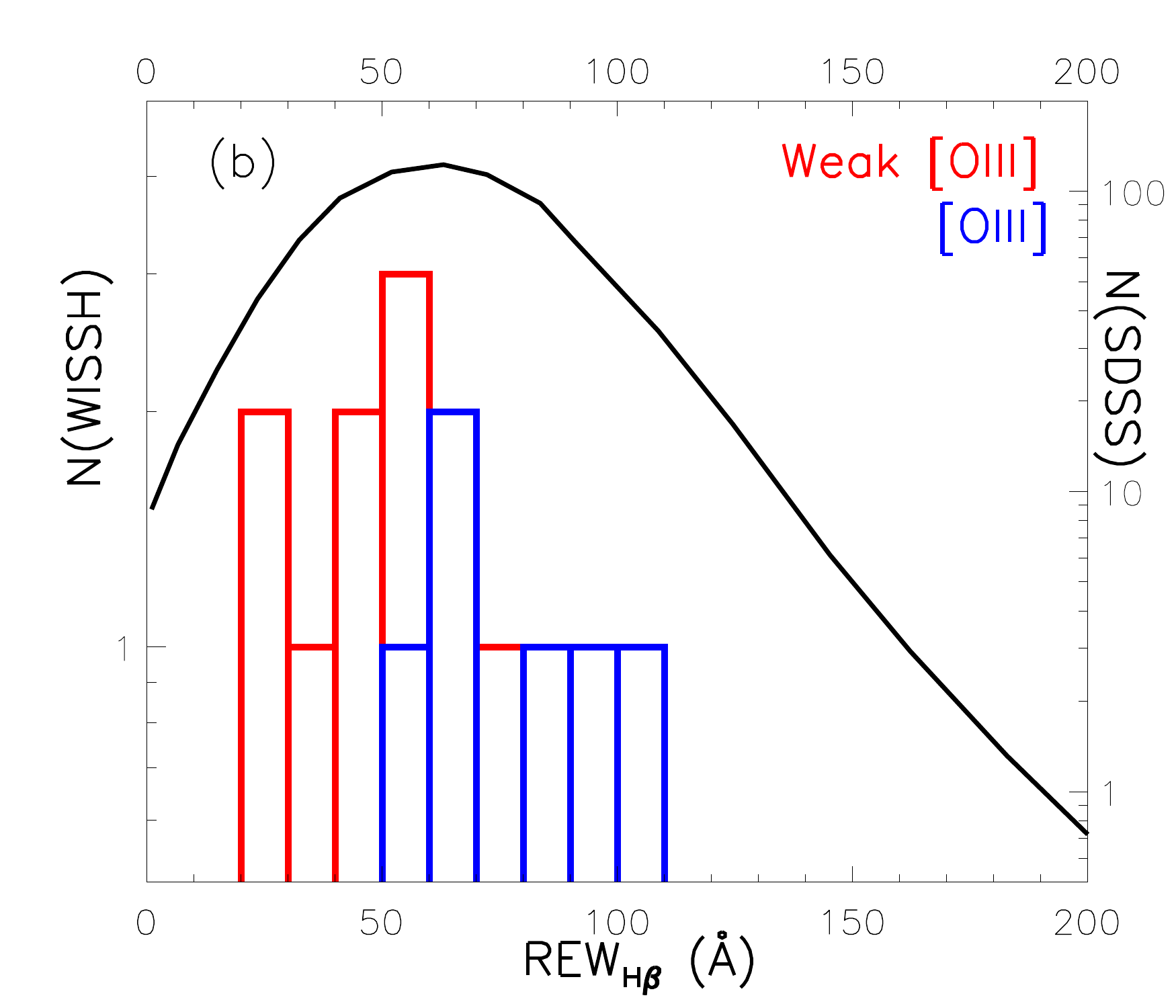}
      \caption{Observed distribution of the WISSH QSOs for the \oiii\ and \weak\ samples of {\it(a)} REW$\rm_{[OIII]}$ and {\it(b}) REW$\rm_{H\beta}$, compared to the best-fit of the REW$\rm_{[OIII]}$ and REW$\rm_{H\beta}$ observed distribution of SDSS DR7 AGNs of \cite{Bisogni2017} (black solid line).}
             \label{fig:distrib}%

    \end{figure}	
    
\begin{figure}[]
             \label{fig:SED}%
              \centering
  \includegraphics[width=1\columnwidth]{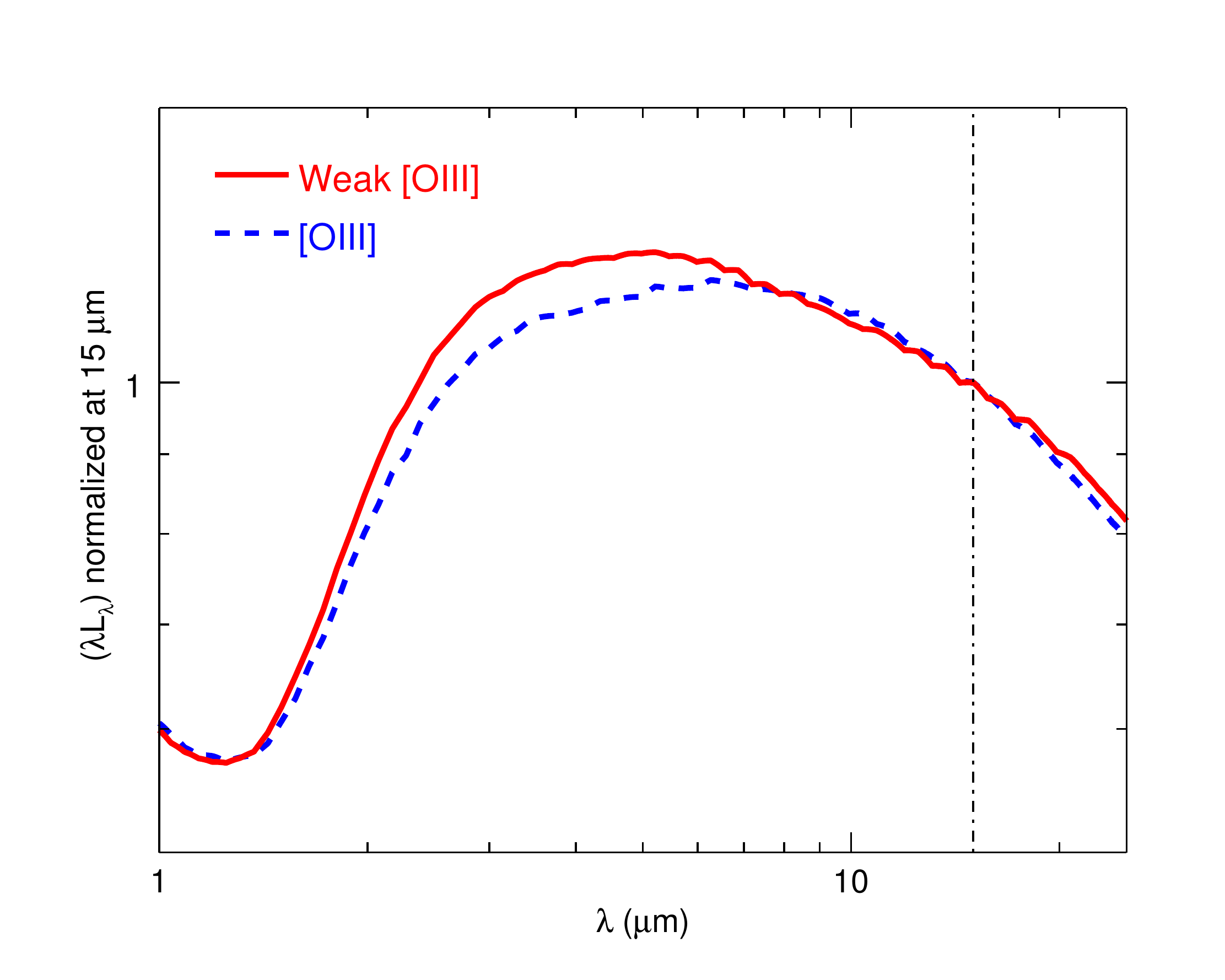}
   \caption{Average SEDs corresponding to WISE photometry of the WISSH \oiii\ (dashed blue line) and \weak\ samples (red line), normalized at 15{$\mu$m} (dot-dashed line).}

    \end{figure}

\section{Results}

 \subsection{Properties of the [OIII] emission: REW$\rm_{[OIII]}$ and orientation}\label{oiii_hbeta}

The observed equivalent width of the total profile of the [OIII] emission line (REW$\rm_{[OIII]}$)  can be used as indicator of the line of sight inclination as found by \cite{Risaliti2011} and \cite{Bisogni2017}, i.e. REW$\rm_{[OIII]}$ = REW$_{intrinsic}/cos(\theta$) (with $\theta$ as the angle between the accretion disc axis and the line of sight).
In these works they showed that the distribution of REW$\rm_{[OIII]}$ of a large sample of SDSS DR7 AGN  with 0.001 $< z <$ 0.8, hereafter SDSS distribution, shows a power-law tail with a slope of -3.5 at the largest REW$\rm_{[OIII]}$ values which is well reproduced by assuming an isotropic [OIII] emission (which is proportional to the intrinsic disc luminosity) and a random inclination of the accretion disk with respect to the line of sight.
This demonstrates that the inclination effect is the likely responsible of the large-REW$\rm_{[OIII]}$ power-law tail, whereby the higher REW$\rm_{[OIII]}$ the higher inclination. 

The \oiii\ sample of the WISSH QSOs exhibits REW$\rm_{[OIII]}$ $\approx$ 7-80 \AA , which is mostly due to the broad blueshifted component of the [OIII] emission. \cite{Risaliti2011} and \cite{Bisogni2017} reported that REW$\rm_{[OIII]}$ $\geq$ 25-30 \AA\ are associated with nearly edge-on AGN.  
Half of \oiii\ sources populate the high-tail of the SDSS REW$\rm_{[OIII]}$ distribution with REW$\rm_{[OIII]}$ \simgt\ 25 \AA\  (Fig. \ref{fig:distrib} {\it{top}}).
Therefore, a high inclination, i.e. $\theta$ $\approx$ 25º-73º (Bisogni et al. in prep.), can likely explain such large REW$\rm_{[OIII]}$.

The average value of REW$\rm_{[OIII]}$ for the \weak\ sample is $\approx$ 2 \AA .
Given their high bolometric luminosities, weak [OIII] emission is expected in WISSH QSOs due to over-ionization of the circumnuclear gas as found in \cite{Shen2014} by analyzing the rest-frame optical spectra of $\sim$ 20,000 Type-1 SDSS QSOs.
Another explanation for the reduced [OIII] emission could be linked to the ionization cone perpendicular to the galaxy disk, which would intercept a lower amount of ISM, resulting in a [OIII] line with a lower REW.

Furthermore, the \oiii\ WISSH QSOs typically have higher REW$\rm_{H\beta}$ values than the \weak\ sources (Fig. \ref{fig:distrib}, {\it{bottom}}), supporting a larger inclination scenario which leads to an observed lower underlying continuum.
We also note that the REW$\rm_{H\beta}$ distribution of the WISSH quasar is consistent with SDSS REW$\rm_{H\beta}$ distribution. As found in \cite{Bisogni2017}, we might expect a trend of the FWHM$\rm_{H\beta}^{BLR}$ also for the WISSH QSOs, i.e. the width of the H$\beta$ increases from low to high inclination (from small EW$_{[OIII]}$ to large EW$_{[OIII]}$). However, all the WISSH objects exhibit large values of FWHM$\rm_{H\beta}^{BLR}$ and we do not observe any significant trend. This can be likely explained with the presence of ultra-massive BHs  in these hyper-luminous AGN (see Sect. \ref{sec:mbh}), which heavily dilutes the effect of inclination.

 If the inclination plays a role to explain the large differences in REW$\rm_{[OIII]}$ between \oiii\ and \weak\ WISSH QSOs,  we also expect differences in the near- and mid-IR SED, as found by \cite{Bisogni2017b} in a large SDSS sample of QSOs. Face-on QSOs offer a direct view of the hottest dust component located in the innermost part of the torus, while in case of high-inclination sources the view of this region can be partially blocked. 
This is consistent with the result obtained by comparing the average SEDs corresponding to {\it WISE} photometry of the QSOs in the \oiii\ and \weak\ samples (Fig. \ref{fig:SED}) where sources with high REW$\rm_{[OIII]}$ show a flux deficit in the NIR part of the SED around 3 $\mu$m with respect to \weak\ sources.

\subsection{Single epoch H$\beta$-based SMBH masses and Eddington ratios} \label{sec:mbh}

	\begin{figure}[]
 \centering

 \includegraphics[width=1\columnwidth]{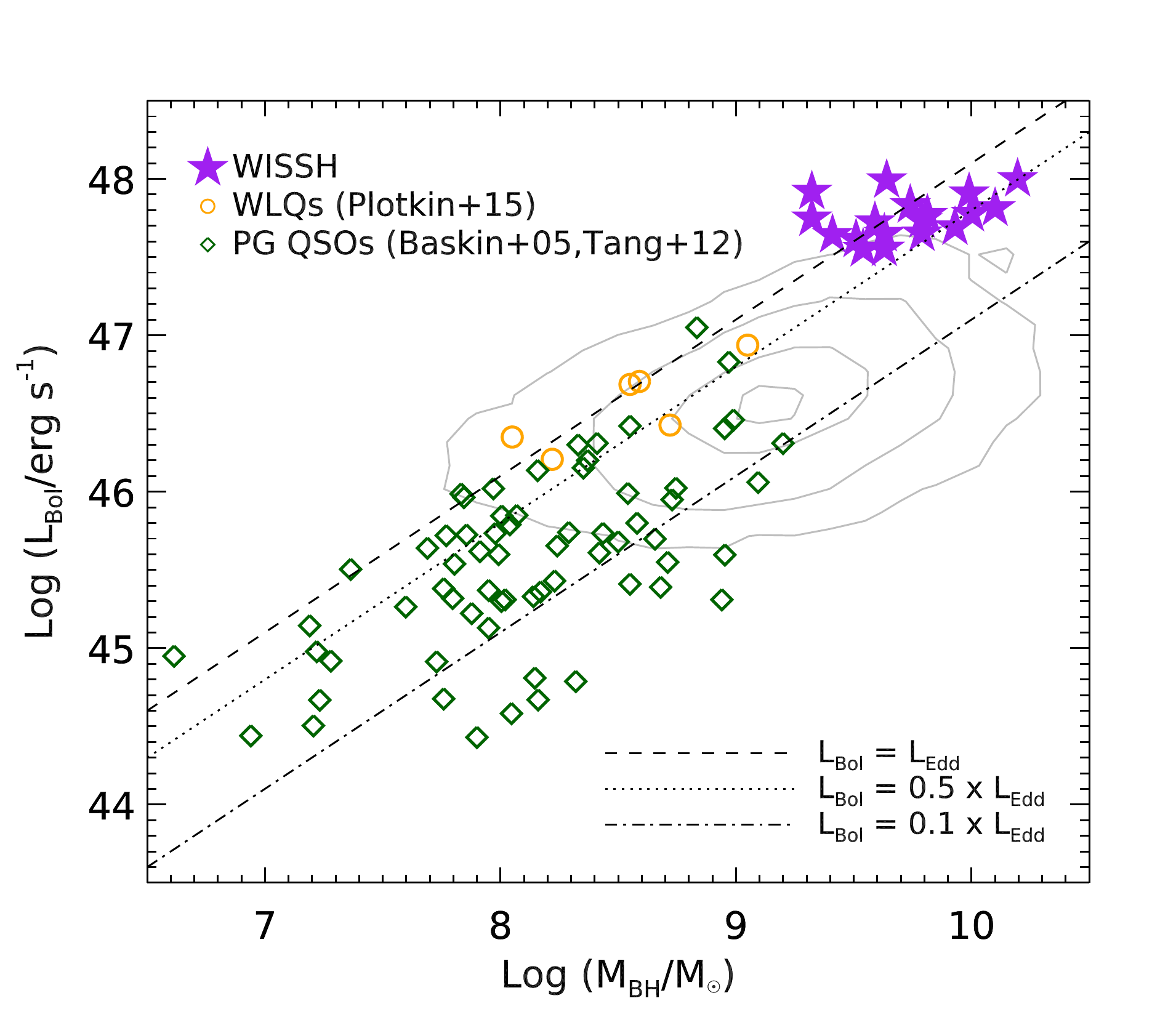}
\caption{Bolometric luminosity as a function of BH mass for the WISSH sample, compared to PG QSOs from \cite{Tang2012} and \cite{Baskin2005} and WLQs from \cite{Plotkin2015}. Luminosity in fractions of  0.1, 0.5 and 1 Eddington luminosity are respectively indicated with dot-dashed, dotted and dashed lines. Contours levels (0.01, 0.1, 0.5, and 0.9 relative to the peak) refer to SDSS DR7 QSOs from \cite{Shen2011}.}\label{fig:mbh}

 \end{figure}

	A common method of measuring the black hole mass (\mbh) is the so-called Single Epoch technique (SE), which depends on BLR size estimated from the continuum luminosity of the quasar and on the velocity dispersion of the BLR, derived from the FWHM of a specific broad emission line, i.e H$\alpha$, H$\beta$, MgII, CIV, Pa$\alpha$ and Pa$\beta$, depending on the redshift (\citealt{McGill2008}; \citealt{Trakhtenbrot2012}; \citealt{Matsuoka2013}; \citealt{Ricci2017_1}).
As mentioned in Sect. \ref{sec:SDSS}, the CIV emission line is affected by the presence of non virialized components (\citealt{Baskin2005}; \citealt{Richards2011}) which can bias the BH mass estimation. The H$\beta$ clouds are mostly dominated by virial motions, ascribing this emission line as the best estimator of the SMBH mass (\citealt{Denney2012}). 
In order to estimate M$\rm_{BH}$, we used SE relation for H$\beta$ reported in \cite{Bongiorno2014}, i.e. 	

\begin{equation}
\resizebox{1 \columnwidth}{!} 
{
$\rm Log(M_{BH}/M_\odot) = 6.7 + 2\times\rm Log\left(\frac{FWHM}{10^3 \rm\ km\ s^{-1}}\right) + 0.5\times Log\left(\frac{\lambda L_{\lambda}}{10^{44} \rm\ erg\ s^{-1}}\right)\hspace{0.2cm}$
}
\end{equation}


 For each quasar, we used the best-fit value of the FWHM of the broad component of H$\beta$ line derived in Sect. \ref{sec:analysis} (see Table \ref{tab:beta}) and the continuum luminosity at 5100 \AA .  Both the L$_{5100}$ and \lbol\ were obtained by fitting the UV to mid-IR photometric data (see Table \ref{tab:data1}), with  Type 1 AGN SED templates of \cite{Richards2006} using a SMC extinction law (\citealt{Prevot1984}) with colour excess E(B-V) as free parameter (Duras et al. in preparation). We list the results of H$\beta$-based M$\rm_{BH}$ for all the WISSH QSOs with rest-frame optical spectroscopy in Table \ref{tab:MBH}.
  
  We found that all the BHs have masses larger than 10$^{9}$ $M_\odot$, with 9 out of 18 QSOs hosting SMBHs with \mbh\ \simgt\ 5 $\times$ 10$^{9}$ $M_\odot$. 
 Based on these H$\beta$-based \mbh\ values and \lbol , we derived Eddington ratios \edd\ = \lbol /L$\rm_{Edd}$ = 0.4 -- 3.1 (with a median value of 1), where L$\rm_{Edd}$ = 1.26 $\times$ 10$^{38}$ (\mbh /M$_{\odot}$) \ergs , which are also reported in Table \ref{tab:MBH}. For the WISSH \oiii\ objects we also estimated the bolometric luminosity (L$\rm_{Bol}^{corr}$) corrected for the orientation effect using the mean inclination angles as determined in Bisogni et al. (2018, in preparation). Specifically, considering the expression for the observed REW$\rm_{[OIII]}$ distribution (eq. 4 in  \citealt{Risaliti2011})  as a function of the ratio between observed and intrinsic REW$\rm_{[OIII]}$, it is possible to retrieve the inclination angle probability distribution given the observed REW$\rm_{[OIII]}$ and, hence, the mean values of inclination angles.

 Fig.\ref{fig:mbh} shows the comparison of the  \mbh ,  \lbol\ and \edd\ measured for WISSH QSOs with those derived from (i) a sample of $\sim$23,000 SDSS QSOs with 1.5 $\leq$ z $\leq$ 2.2 with MgII-based \mbh\ (\citealt{Shen2011}, contour lines), hereafter "SDSS sample", (ii)  bright PG QSOs with z $<$ 0.5 with H$\beta$-based \mbh\ (\citealt{Tang2012}; \citealt{Baskin2005}, green diamonds) and (iii) the WLQs (see Sect. \ref{sec:intro} for the definition) with H$\beta$-based \mbh\ from \cite{Plotkin2015} (orange circles). The bolometric luminosity in fraction of 0.1, 0.5 and 1 Eddington luminosity is reported.
The WISSH QSOs are therefore powered by highly accreting SMBHs at the heaviest end of the \mbh\  function and allow to probe the extreme AGN accretion regime and the impact of this huge radiative output on the properties of the nuclear region and the surrounding host galaxy.

  \begin{table}[]
	\footnotesize
\makebox[1\columnwidth]{
	\small
	\begin{threeparttable}
	\caption{Logarithm of the bolometric luminosity, Logarithm of intrinsic luminosity at 5100 \AA , H$\beta$-based SMBH mass and Eddington ratio of the WISSH QSOs. Objects in boldface belong to the \oiii\ sample.}\label{tab:MBH}
	\setlength\tabcolsep{3pt}
	
			\begin{tabular}{lcccc}
				\hline
				\hline	
				SDSS Name &Log \lbol\ [Log L$\rm_{Bol}^{corr}$]\tnote{,a} &~~Log L$_{5100}$ & M$\rm_{BH}$\tnote{b}& \edd\ \\[0.5ex]
				(1) & (2) & (3) &(4) & (5) \\
				\hline
		
			{\bf{J0745+4734}} &48.0[48.2]&47.3 &15.7$\pm10.8$	  &0.5\footnotesize{$_{-0.3}^{+0.5}$}\\
			       J0801+5210 &47.8 	      &47.1 &6.2$\pm4.3$	  &0.7$\pm0.5$\\
			{\bf{J0900+4215}} &48.0[48.1]&47.2 &2.1$\pm1.5$	  &3.1\footnotesize{$_{-2.2}^{+0.6}$}\\
			      J0958+2827  &47.6 	      &46.9 &3.5$\pm2.4$	  &0.8$\pm0.6$\\
			       J1106+6400 &47.8 	      &47.0 &10.0$\pm6.9$	  &0.5$\pm0.3$\\
			       J1111+1336 &47.7 	      &47.0 &8.5$\pm5.9$	  &0.5$\pm0.3$\\
			       J1157+2724 &47.6 	      &47.0 &2.6$\pm1.8$	  &1.3$\pm0.9$\\
			       J1201+0116 &47.6	      &47.0 &3.2$\pm2.3$	  &1.0$\pm0.7$\\
			{\bf{J1201+1206}} &47.8[47.9]&47.1 &6.5$\pm4.5$	  &0.7\footnotesize{$_{-0.5}^{+0.3}$}\\
			       J1236+6554 &47.7 	      &47.0 &4.3$\pm3.0$	  &0.8$\pm0.6$\\
			{\bf{J1326-0005}} &47.8[48.3]&47.0 &2.1$\pm1.6$	  &2.1\footnotesize{$_{-1.6}^{+0.7}$}\\
		               J1421+4633 &47.7 	      &47.0 &6.2$\pm4.3$	  &0.6$\pm0.4$\\
			       J1422+4417 &48.0 	      &47.3 &4.4$\pm3.1$	  &1.8$\pm1.3$\\
			       J1521+5202 &47.9 	      &47.2 &9.98$\pm6.8$	  &0.7$\pm0.5$\\
			{\bf{J1538+0855}} &47.8[47.8]&47.1 &5.5$\pm3.8$	  &1.0\footnotesize{$_{-0.7}^{+0.8}$}\\
			{\bf{J1549+1245}} &47.8[47.9]&47.1 &12.6$\pm8.7$	  &0.4\footnotesize{$_{-0.3}^{+0.9}$}\\
			      J2123-0050  &47.7 	      &47.0 &3.9$\pm2.7$	  &1.1$\pm0.7$\\
			       J2346-0016 &47.5 	      &46.8 &4.3$\pm3.0$	  &0.7$\pm0.2$\\
				\hline
				\end{tabular}
			
				 \begin{tablenotes}[para,flushleft]
				 \item {\bf{Notes.}} The following information is listed: (1) SDSS name, (2) Logarithm of SED-based bolometric luminosity (in units of erg/s) (Duras et al. 2018, in prep), (3) Logarithm of SED-based intrinsic luminosity at 5100 \AA\ (in units of erg/s) (Duras et al. 2018, in prep), (4) H$\beta$-based SMBH mass (in units of 10$^9$ M$_\odot$), and (5) Eddington ratio.
				\item[a] Value of bolometric luminosity corrected for orientation effect, calculated as described in Sect. \ref{sec:mbh}.
				\item[b]  The error associated with the M$_{\rm BH}$  includes both the statistical uncertainties and the systematic uncertainty in the virial relation itself \citep[$\sim0.3$ dex; see][for a complete discussion]{Bongiorno2014}.
     				  \end{tablenotes}
       \end{threeparttable}
			}
			\end{table}

\subsection{Properties of the CIV emission line} \label{sec:shift}

\subsubsection{CIV velocity shift}

   \begin{figure*}[]
   \centering
    \includegraphics[width=1\linewidth]{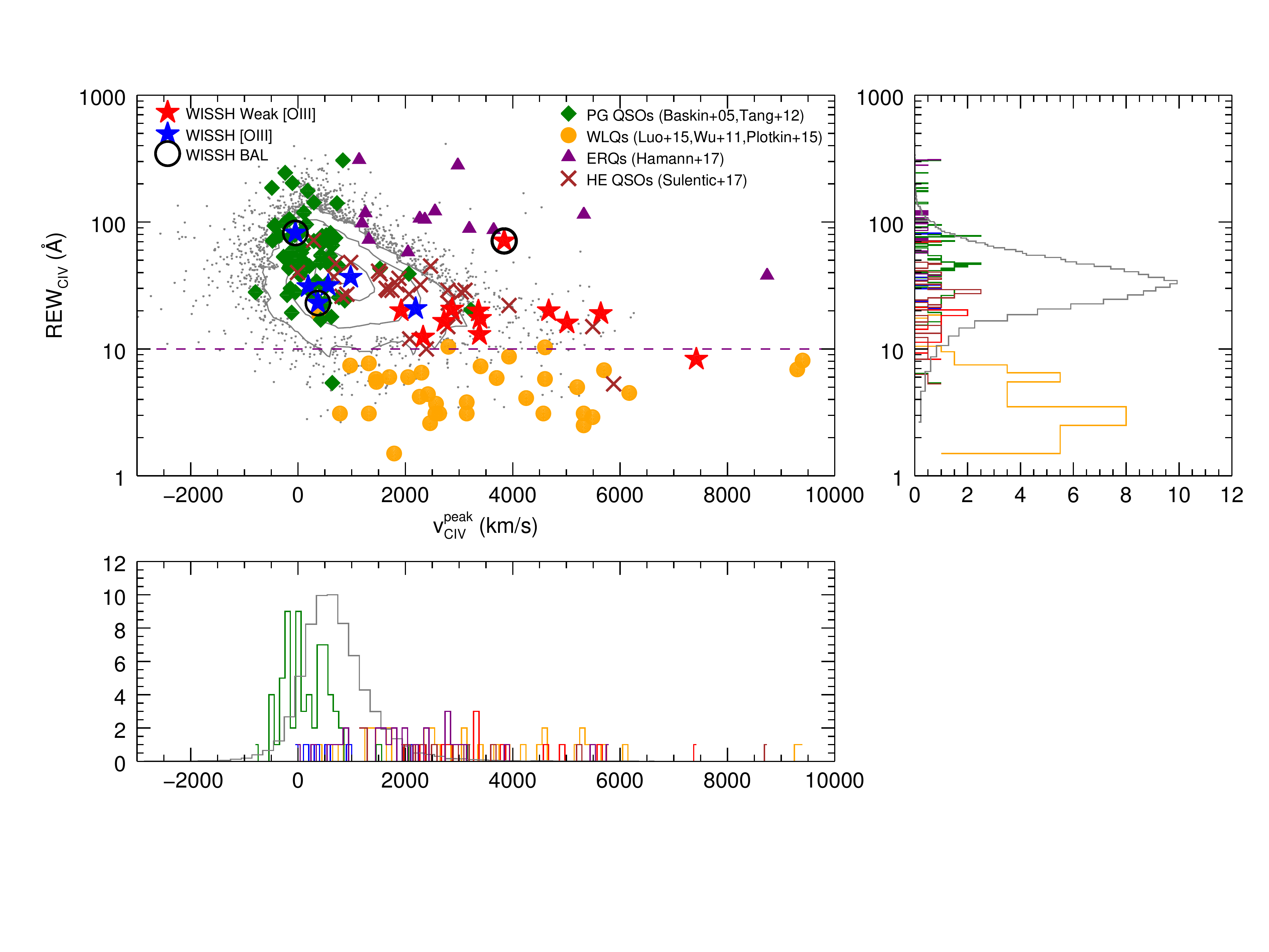}
       \caption{REW$\rm_{CIV}$ as a function of \vciv\ for different samples. The WISSH QSOs are indicated by blue and red stars, for \oiii\ and \weak\ sample respectively. The BAL WISSH QSOs are indicated with black circles. Contours levels (0.03, 0.1, 0.5, and 0.9 relative to the peak) refer to the SDSS DR7 sample from \cite{Shen2011}. The green diamonds indicate the PG QSOs from \cite{Baskin2005} and \cite{Tang2012}. The orange circles refer to the sample of WLQs from \cite{Plotkin2015}, \cite{Luo2015} and \cite{Wu2011}, the brown crosses indicate the HE QSOs from \cite{Sulentic2017}, the ERQs from \cite{Hamann2017} are indicated with purple triangles. The purple dashed line marks the REW$\rm_{CIV}$ = 10 \AA\ , which is the threshold for the WLQs. The distribution of the REW$\rm_{CIV}$ and \vciv\ are on the right and on the bottom, respectively.}
             \label{fig:ew_shift}%

    \end{figure*}

Previous works have found that the CIV emission line typically shows a velocity blueshift with respect to the systemic redshift (\citealt{Gaskell1982}), suggesting it is associated with outflowing gas in the BLR. As already mentioned in Sect. \ref{sec:SDSS}, for the WISSH objects we measured the CIV velocity shifts (\vciv) from the peak of the entire CIV emission line model fit with respect to the systemic redshift (similar approach is used in the work of \cite{Shen2011} for "SDSS sample").

As shown in Fig. \ref{fig:ew_shift}, we found an anti-correlation between the \vciv\ and the strenght of the CIV emission line for the WISSH QSOs. This result was already reported by previous studies, e.g. \cite{Corbin1996}; \cite{Richards2002}. 

We compare our findings with those derived for the SDSS sample\footnote{We used the improved redshift estimates of \cite{Hewett2010}, derived using emission lines not including the CIV.} (contours),  PG QSOs (\citealt{Tang2012}; \citealt{Baskin2005}), WLQs and PHL1811-analogs (\citealt{Plotkin2015}; \citealt{Luo2015} and  \citealt{Wu2011}), high luminosity QSOs from \cite{Sulentic2017} identified in the Hamburg ESO survey (hereafter HE QSOs) and Extremely Red QSOs (ERQs) from \cite{Hamann2017}, defined by a colour (i-WISE W3) $\ge$ 4.6 with a median bolometric luminosity of Log (L$\rm_{Bol}$/\ergs) $\sim$ 47.1$\pm0.3$. Remarkably,  \weak\ WISSH QSOs show extreme velocity shifts, comparable to the largest \vciv\ reported so far in WLQs. Furthermore, while WLQs show the distinctive property of a REW$\rm_{CIV}$ $\leq$ 10 \AA , the WISSH sample have a REW$\rm_{CIV}$ $\geq$ 10 \AA , demonstrating the existence of high-velocity outflows traced by high-ionization species also in sources with higher REW$\rm_{CIV}$ than those of the WLQs. Interestingly, the ERQs show both large blueshifts and very large (i.e. $\ge$ 100 \AA) REW$\rm_{CIV}$, which can be interpreted in terms of a CIV emitting region with a larger covering factor with respect to the ionizing continuum than normal QSOs (\citealt{Hamann2017}).

In Fig.~\ref{fig:shift_ewoiii} the REW$\rm_{[OIII]}$ as a function of the REW$\rm_{CIV}$ ({\it{left}}) and \vciv\ ({\it{right}}) is shown for the WISSH QSOs. We discovered an intriguing dichotomy between \oiii\ sample, which shows small values of \vciv\ ($\leq$ 2,000 \kms) with REW$_{CIV}$ $\geq$ 20 \AA\ and \weak\ sample, exhibiting \vciv\ $\geq$ 2,000 \kms\ with REW$_{CIV}$ $\leq$ 20 \AA . This is also supported by the same behavior shown by the QSOs in the \cite{Shen2016} and \cite{Tang2012} (grey points) and in the HE sample\footnote{For the HE QSOs sample we used the CIV and H$\beta$ broad emission line properties listed in \cite{Sulentic2017} and the [OIII] emission lines information in \cite{Sulentic2004,Sulentic2006} and \cite{Marziani2009}.} from \cite{Sulentic2017} (brown crosses), populating the same region of the plane REW$\rm_{[OIII]}$ -- \vciv\ . This dichotomy can be likely explained by assuming a polar geometry for the CIV winds, where the bulk of the emission is along the polar direction, against \oiii\ QSOs which are supposed to be viewed at high inclination.

     \begin{figure*}[]
\makebox[1\textwidth]{ 
  \includegraphics[width=0.5\linewidth]{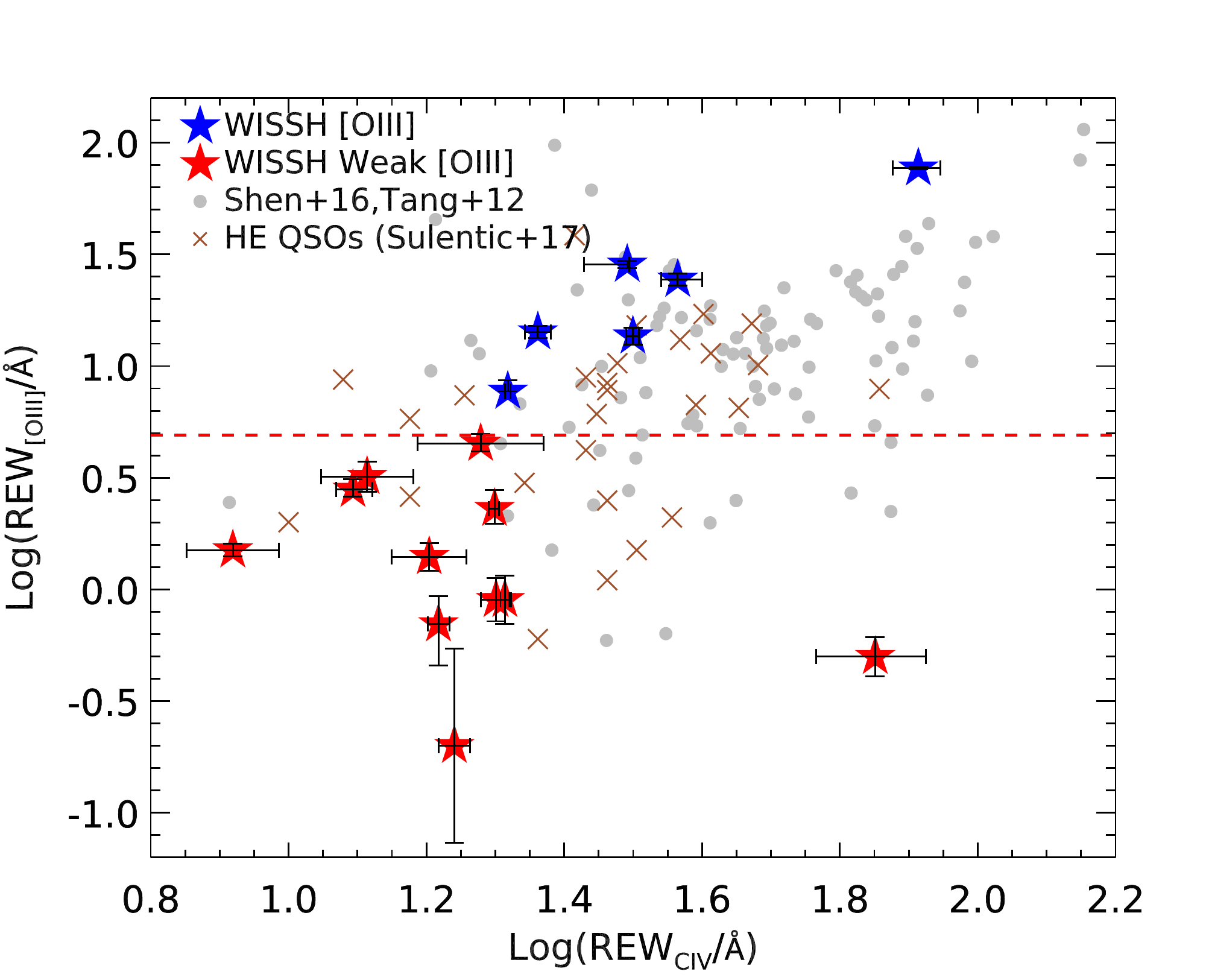}
  \includegraphics[width=0.5\linewidth]{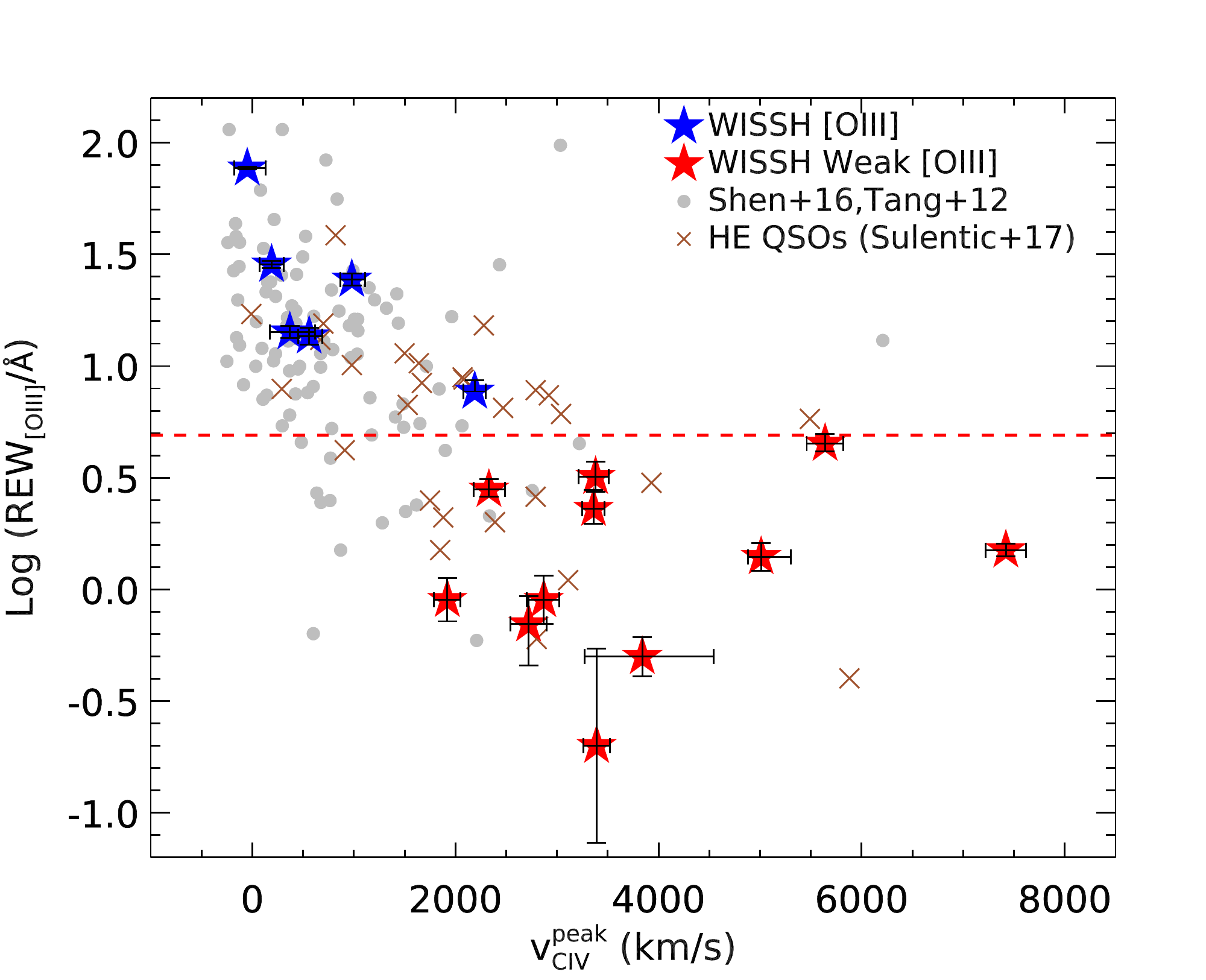}
  }   
     \caption{Rest-frame EW of [OIII]$\lambda$5007 as a function of  REW$\rm_{CIV}$ ({\it{left}}) and \vciv\ ({\it{right}}) for the WISSH targets with blue stars for the \oiii\ sample and red stars for the \weak\ sample, compared to the \cite{Shen2016}, \cite{Tang2012} and \citet[and reference therein]{Sulentic2017}. The dashed red line refers to the WISSH [OIII] dichotomy, with REW$\rm_{[OIII]}$ boundary value of 5 \AA .}
             \label{fig:shift_ewoiii}%

     \end{figure*}
  

 We also investigated the possible relation between the \vciv\ and the FWHM of the broad CIV emission line, by combining WISSH QSOs with other samples in Fig. \ref{fig:fwhmcivout_shift} ({\it{top}}). For this purpose we used 73 SDSS QSOs at 1.5 $<$ z $<$ 3.5 with available H$\beta$ information from \citealt{Shen2012} and \citealt{Shen2016}, 19 SDSS QSOs from \citealt{Coatman2016}, 66 radio-quiet non-BAL PG QSOs from \cite{Baskin2005} and \cite{Tang2012}, 6 WLQs from \cite{Plotkin2015} with z $\sim$ 1.4 -- 1.7, 28 HE QSOs from \citet[and reference therein]{Sulentic2017}. All these objects cover a luminosity range Log(L$\rm_{Bol}$/\ergs) = 44.5 -- 48.1. Hereafter, we refer to these samples as "H$\beta$ sample".
 As expected, sources with large \vciv\ (\simgt\ 2,000 \kms) show a CIV emission line with a broad profile (\simgt\ 6,000 \kms). This suggests that for these sources the line profile is the result of the combination of a virialized component plus a strongly outflowing one.
Fig. \ref{fig:fwhmcivout_shift} ({{\it{bottom panel}}}) shows the behavior of the FWHM of the outflow component of the CIV emission line resulting from our multi-component fit (see Sect. \ref{sec:SDSS}) as a function of the \vciv\ . Performing the Spearman rank correlation we found r = 0.6 and P-value = 5.2 $\times$ 10$^{-3}$, indicating that a large FWHM of CIV can be considered a proxy of the presence of a high velocity outflow.
 No clear trend is found between the FWHM$\rm_{H\beta}^{BLR}$ and \vciv , as shown in Fig. \ref{fig:civ_hbeta}. Interestingly, the WISSH QSOs show FWHM$\rm_{H\beta}^{BLR}$ $\geq$ 4,000 \kms , even in those sources with large \vciv , while previous works claimed the presence of large \vciv\ in QSOs with FWHM$\rm_{H\beta}^{BLR}$ $<$ 4,000 \kms , such as low luminosity Population A QSOs (\citealt{Sulentic2007}; \citealt{Marziani2010}) and WLQs (\citealt{Plotkin2015}). This indicates that the inclusion of QSOs with extreme luminosities as the WISSH ones allows to extend the detection of large CIV blueshifts to sources with FWHM$\rm_{H\beta}^{BLR}$ $>$ 4,000 \kms\ (a similar result has been also reported for the HE sample in \cite{Sulentic2017}).

\begin{figure}[]
\includegraphics[width=1\columnwidth]{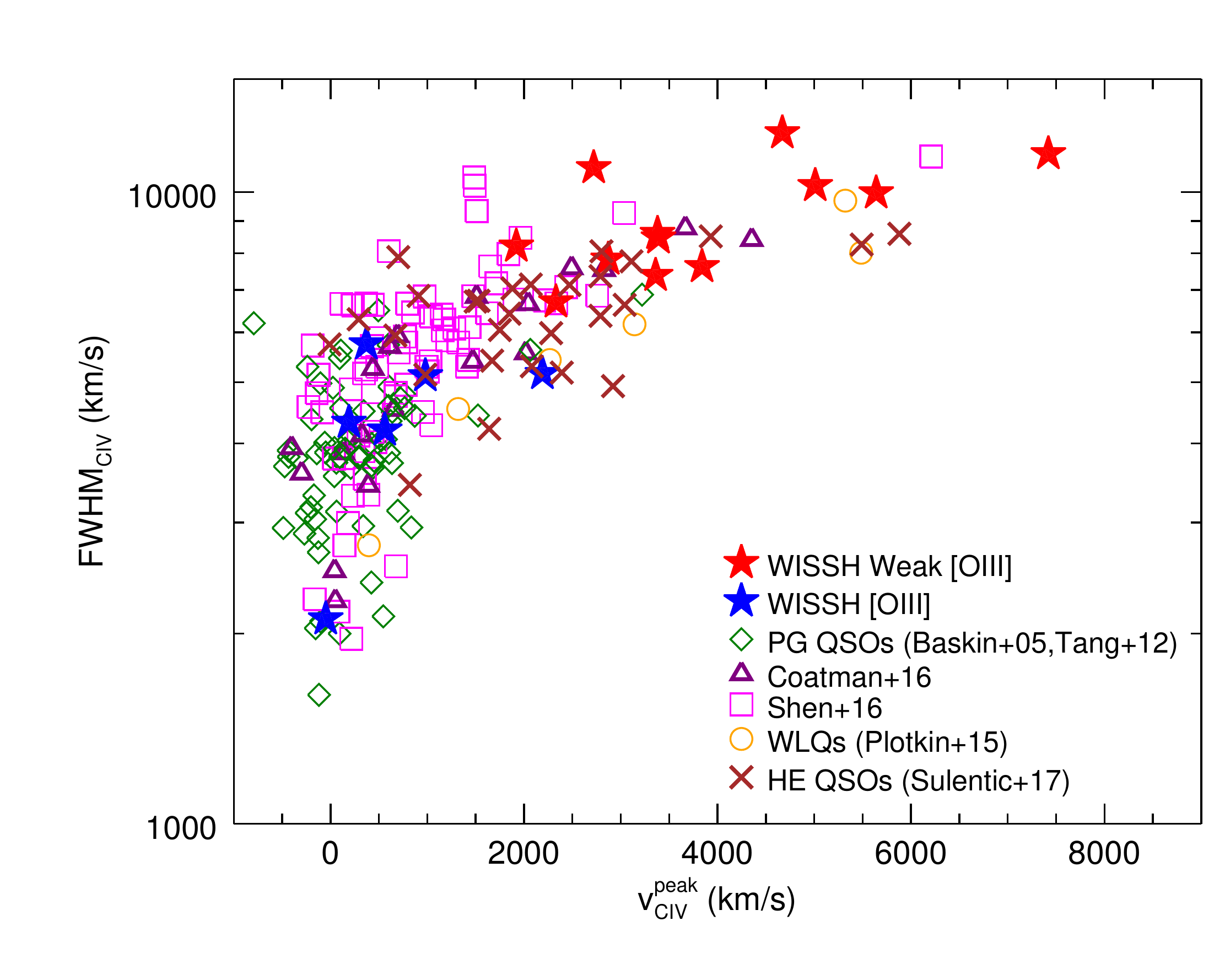}
\includegraphics[width=1\columnwidth]{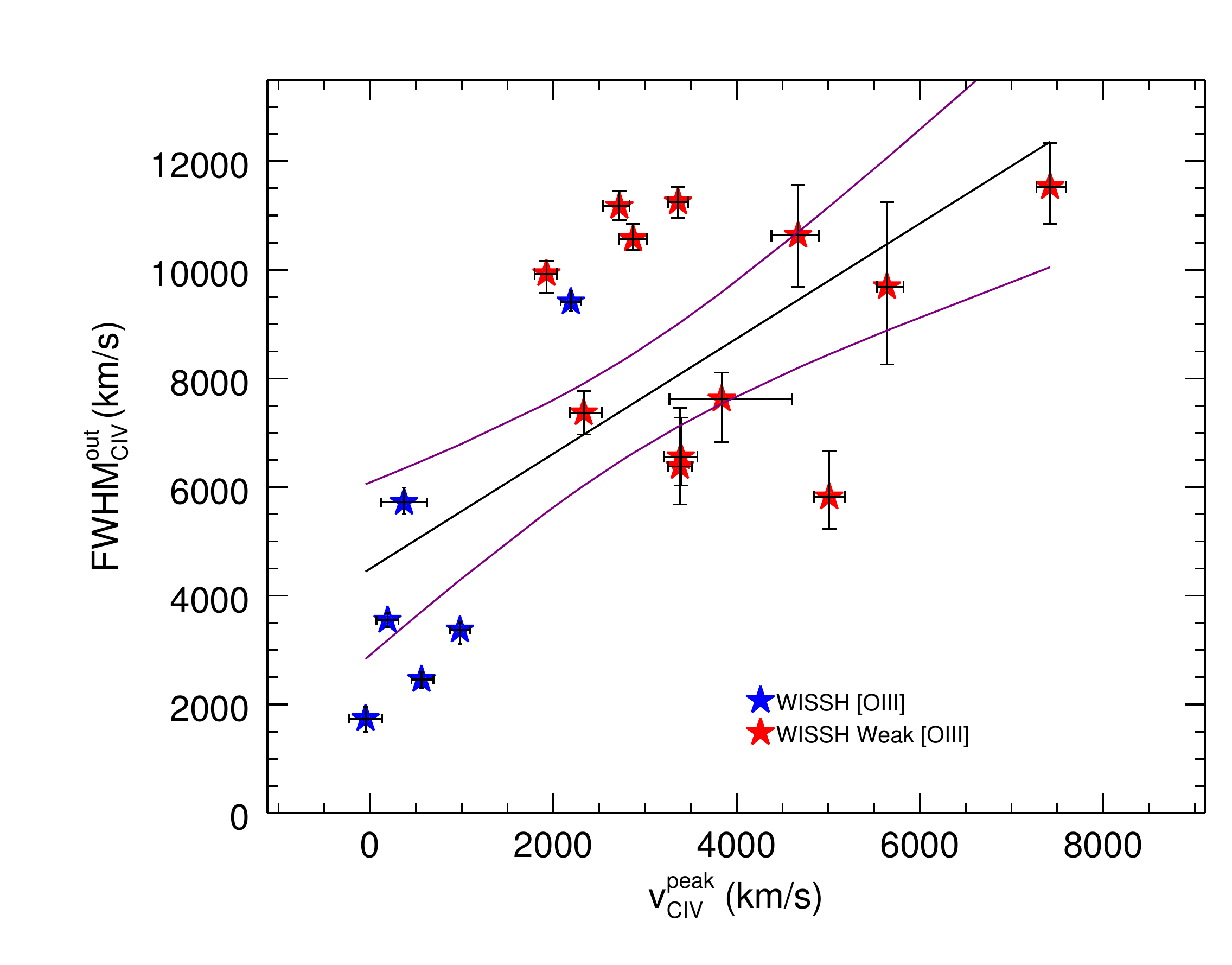}
\caption{FWHM$\rm_{CIV}$ of the entire emission line profile as a function of \vciv\ for the WISSH sample, compared to PG QSOs from \cite{Baskin2005} and \cite{Tang2012} (green diamonds), WLQs from Plotkin et al 2015 (orange circles), SDSS QSOs from \cite{Shen2016} and \cite{Coatman2016} (magenta squares and purple triangles respectively) and HE QSOs from \cite{Sulentic2017} ({\it{top}}) and FWHM(CIV) of the outflow component ({{\it{bottom}}}), as a function of \vciv\ for the WISSH sample. The black solid line indicates the best linear fits to the data and the purple dashed lines correspond to the 68\% confidence interval.}
\label{fig:fwhmcivout_shift}%

    \end{figure}

 \begin{figure}[]
\includegraphics[width=1\columnwidth]{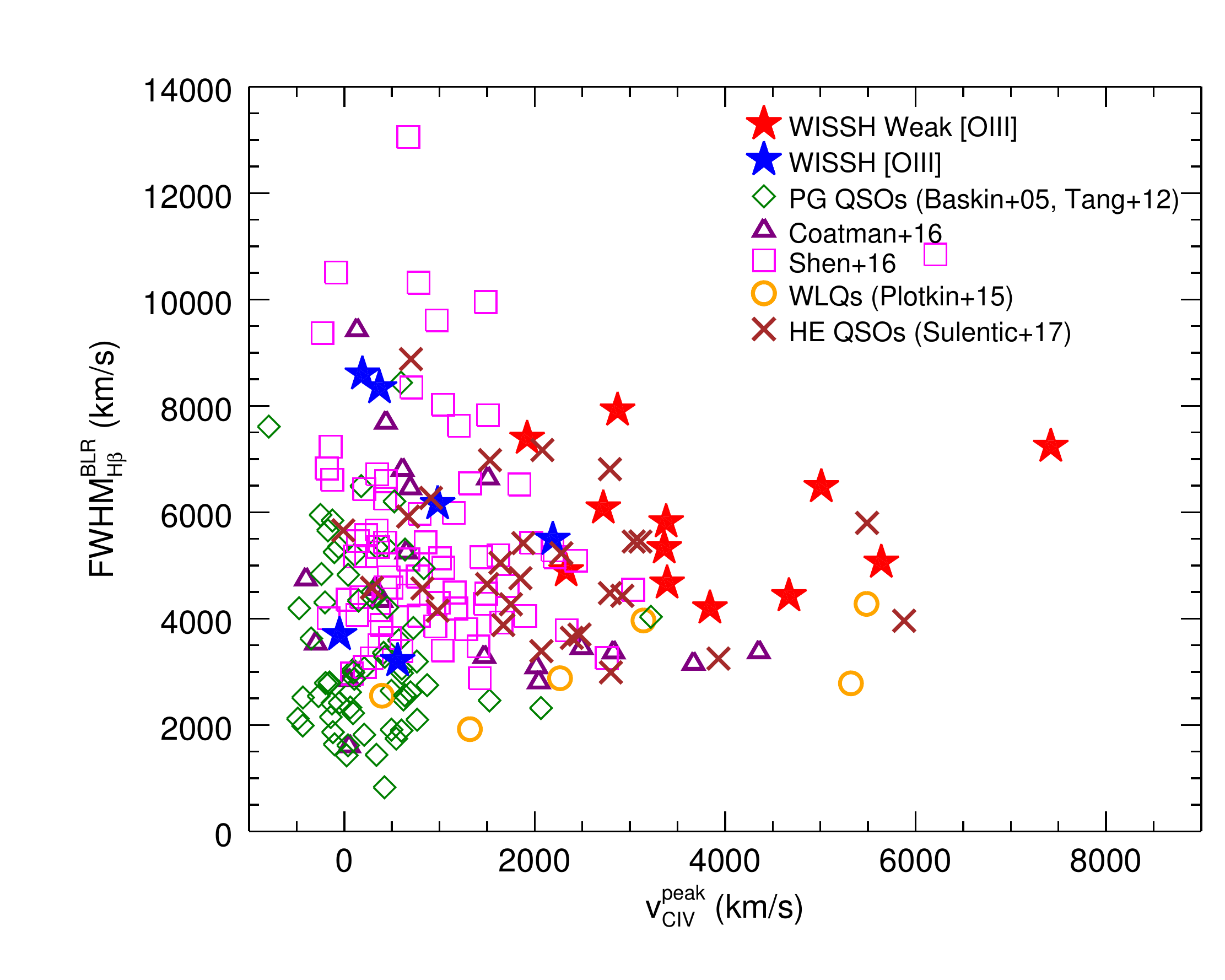}
  \caption{FWHM$\rm_{H\beta}^{BLR}$ as a function of \vciv\  for the WISSH sample, compared to PG QSOs from \cite{Baskin2005} and \cite{Tang2012} (green diamonds), WLQs from \cite{Plotkin2015} (orange circles), SDSS QSOs from \cite{Shen2016} and \cite{Coatman2016} (magenta squares and purple triangles, respectively) and HE QSOs from \cite{Sulentic2017}.}
\label{fig:civ_hbeta}%

  \end{figure}



\subsubsection{Mass and kinetic power of CIV winds}\label{sec:ekin}


\begin{figure}[h]
  \includegraphics[width=1\columnwidth]{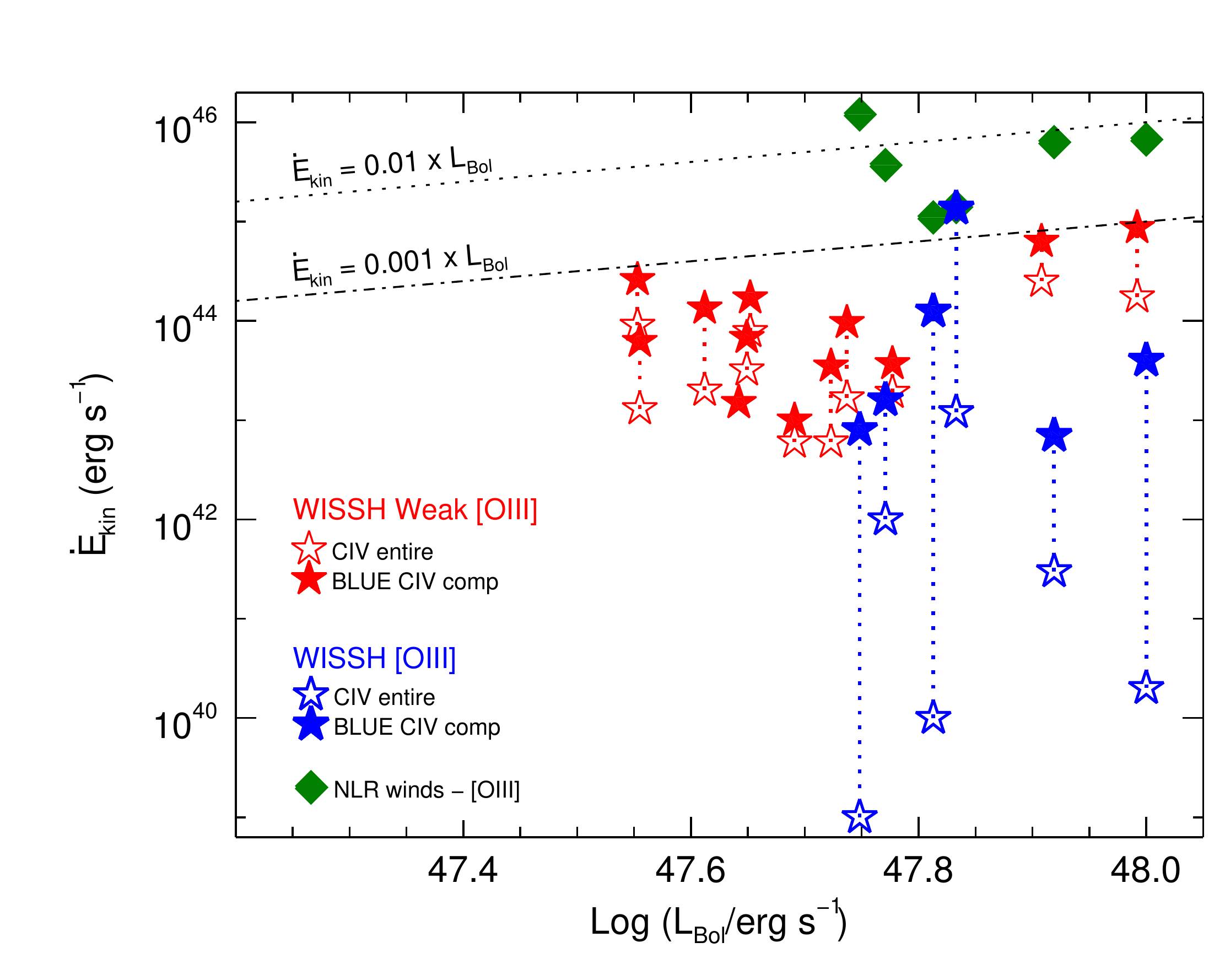}
  \caption{Kinetic power of the BLR winds as a function of \lbol\ . Red filled stars refer to values of \ekin\ of the \weak\ QSOs derived from the CIV outflow  component using v$\rm_{CIV}^{out}$. Blue filled stars refer to values of \ekin\ of the \oiii\ QSOs derived from the entire  CIV profile using \vmax\ . Blue and red empty stars refer to values of \ekin\ derived from the entire CIV profile using \vciv\ , for the \oiii\ and \weak\ QSOs, respectively. BLR winds are compared with WISSH NLR ionized outflows from Paper I (green diamonds).}
\label{fig:ekin}%

    \end{figure}

We estimated the ionized gas mass (\mion) of the outflow associated with the outflow component of the CIV emission line for the \weak\ sources according to the following formula from \cite{Marziani2016}. This relation assumes a constant density scenario for the CIV emitting gas (i.e. n(C$^{3+}$)/n(C) = 1) and takes into account the carbon abundance:

\begin{equation}\label{civ3}
\rm{M_{ion}} = 9.5 \times 10^{2} \times  L\rm_{45} (CIV) \times Z\rm_{5}^{-1} \times n\rm_{9}^{-1}\ (\mathrm{M_{\odot}})
\end{equation}

where  L$_{45}$(CIV) is the luminosity of the outflow component in units of 10$^{45}$ \ergs , Z$_{5}$ is the metallicity in units of 5 Z$_{\odot}$ (i.e. the typical value measured for the BLR in high-$z$ luminous QSOs, \citealt{Nagao2006}) and n$_{9}$ is the gas density in units of 10$^{9}$ cm$^{-3}$.
For the \weak\ sample we estimated M$\rm_{ion}$ = 150 -- 1,100 M$_{\odot}$, assuming Z = 5 Z$_{\sun}$ and n = 10$^{9.5}$ cm$^{-3}$ (see Table \ref{tab:ekin1}).
From equation (\ref{civ3}) it is possible to derive the mass outflow rate (\mdot), i.e. the mass of ionized gas passing through a sphere of radius R:

\begin{equation}\label{civ4}
\rm\dot{M}_{ion} = 3  \times (M\rm_{ion} \times v)/R_{1}\   \mathrm{(M_\odot\ yr^{-1})}
\end{equation}
 
where  v = \vout\ and R$_{1}$ is the outflow radius in units of 1 pc. The  BLR radius-luminosity relation (R$\rm_{BLR}$ $\propto$ L$\rm_{5100}^{0.5}$, e.g. \citealt{Bentz2009}), indicates a typical radius of  $\sim$ 1 pc for the WISSH QSOs with luminosity L$\rm_{5100}$ $>$ 6 $\times$ 10$^{46}$ \ergs.
We inferred \mdot\ in the range $\approx$ 3 -- 30  M$_{\odot}$ yr$^{-1}$, which is similar to the accretion rate for a quasar with \lbol\ corresponding to the  median \lbol\ of the WISSH range, i.e. $\sim$ 70 $\rm M_\odot yr^{-1}$   for \lbol\ = 4 $\times$ 10$^{47}$ \ergs , assuming a radiative efficiency of 10$\%$.
Accordingly, the kinetic power of the outflow can be expressed as: 

\begin{equation}\label{civ5}
\rm\dot{E}_{kin} = \frac{1}{2}\  $\mdot$\  \times\ \rm{v^{2}}\ ($\ergs$)
\end{equation}
 
 The resulting values of $\rm\dot{E}_{kin}$ associated with the outflow component as a function of \lbol\  are plotted in Fig. \ref{fig:ekin} as red filled symbols for objects in the \weak\ sample.
 
 Since the \oiii\ QSOs also show an outflow component of the CIV emission (see Fig. \ref{fig:BLUE}), we were able to provide an estimate of \mion , \mdot\ and \ekin\ also for them. Moreover, as discussed in Sect. \ref{oiii_hbeta}, \oiii\ QSOs are supposed to be viewed at high inclination.
 Accordingly, we estimated \mdot\ and \ekin\ by assuming the so-called maximum velocity v$\rm_{CIV}^{max}$ = \vout\ $+$ 2$\times\sigma\rm_{CIV}\rm^{out}$, where $\rm\sigma_{CIV}^{out}$ is the velocity dispersion of the outflow component, taking into account projection effects and can be indeed considered representative of the bulk velocity of the outflow in case of biconically symmetric outflowing gas (see Paper I and references therein).  In Fig. \ref{fig:ekin} the \ekin\ values associated with the outflow component  in \oiii\ sources are plotted as blue filled stars.

As mentioned in Sect. \ref{sec:shift}, the bulk of virialized component of the CIV emission line is located bluewards 1549 \AA\ in our fits for both the \weak\ and (even slightly) \oiii\ QSOs. This means that it can be also associated with an outflow and, therefore, we also derived \mion , \mdot\ and \ekin\ using the formula (\ref{civ3}), (\ref{civ4}), (\ref{civ5}), considering its luminosity in addition to that of the  outflow component and adopting  \vciv\ in the calculations of the outflow rates (see Table \ref{tab:ekin2}). 
The kinetic power derived by considering the entire profile of the CIV emission line  are shown in Fig. \ref{fig:ekin} as red and blue empty symbols for the \weak\ and \oiii\ samples, respectively. 
In the case of \weak\ sources they are a factor of $\approx$ 2-10 smaller than the values based only on the luminosity and velocity of the outflow component (\vciv\ $<$ \vout), due to the \ekin\ $\propto$ v$^3$ dependence. In the case of \oiii\ sources, the \ekin\ calculated in this way are a factor of $\geq$ 100-1,000 smaller than those derived by using \vmax .
Accordingly, they can be considered as very conservative estimates of \ekin .

			  \begin{table}[h!]
\small
\begin{threeparttable}

	\caption{Properties of the CIV outflows derived from the outflow component of the CIV emission line. Objects in boldface belong to the \oiii\ sample.}\label{tab:ekin1}
	\setlength\tabcolsep{1pt}
			\begin{tabular}{lccccc}
				\hline
				\hline
				
				SDSS Name & L$\rm_{CIV}$  &\mion	& v$\rm_{CIV}^{out/max}$\tnote{,a}& \mdot\ &\ekin \\[0.5ex]
									&(10$^{45}$\ergs) &(M$_{\odot}$)&(\kms )& (M$_{\odot}$ yr$^{-1}$) & ($10^{42}$\ergs)\\[0.5ex]
									(1)&(2)&(3)&(4)&(5)&(6)\\[0.5ex]
				\hline
			{\bf{J0745+4734}}&1.90\footnotesize{$_{-0.03}^{+0.05}$}&570& 4200\footnotesize{$_{-130}^{+120}$}&7.3&40\\
			J0801+5210&2.6$\pm0.1$&770& 5050\footnotesize{$_{-130}^{+190}$}&11.9&100\\
			
		 	{\bf{J0900+4215}}&1.4$\pm0.1$&420& 2580\footnotesize{$_{-160}^{+130}$}&3 &7\\			
			
			J0958+2827&0.7$\pm0.2$&220& 10700\footnotesize{$_{-1040}^{+1400}$}&7&260\\
			
			J1106+6400&2.3$\pm0.1$&690& 3820\footnotesize{$_{-170}^{+160}$}&8&40\\
				
			J1111+1336&2.6$\pm0.2$&780 & 2360\footnotesize{$_{-170}^{+160}$} & 6& 10\\

	        J1157+2724&0.9$\pm0.1$&270& 3840\footnotesize{$_{-570}^{+700}$} &3&15 \\	
												
			J1201+0116&0.5$\pm0.1$&150& 9830\footnotesize{$_{-250}^{+450}$} &4 &130 \\
				
			{\bf{J1201+1206}}&1.1$\pm0.1$&340&3640\footnotesize{$_{-220}^{+140}$}&4& 20\\
												
			J1236+6554&2.4$\pm0.1$&710& 4620\footnotesize{$_{-130}^{+150}$} &10 &68\\
								
			{\bf{J1326-0005}}&6.9$\pm0.1$&2090 & 1570\footnotesize{$_{-230}^{+200}$} &10&8 \\
			
			J1421+4633&0.5$\pm0.1$ &160& 10380\footnotesize{$_{-270}^{+460}$} &5 &170\\
									
			J1422+4417&3.7$\pm0.1$&1100& 9350\footnotesize{$_{-740}^{+440}$}&31&870\\
							
			J1521+5202&1.1$\pm0.1$&330& 12570\footnotesize{$_{-680}^{+1450}$} &13 &630\\
			
				{\bf{J1538+0855}}&1.9$\pm0.1$&590& 13400\footnotesize{$_{-330}^{+250}$}&24&1360\\
									 
			{\bf{J1549+1245}}&3.8\footnotesize{$_{-0.3}^{+0.2}$}&1070 &5230\footnotesize{$_{-220}^{+300}$}&16&125\\
			
			J2123-0050&1.2$\pm0.1$&350& 4720\footnotesize{$_{-240}^{+320}$}&5&35 \\
		
			J2346-0016&0.5$\pm0.1$&150 & 7580\footnotesize{$_{-310}^{+480}$} &3 &62 \\
				\hline
				\end{tabular}
				 \begin{tablenotes}[para,flushleft]
					{\bf{Notes.}} The following information are listed: (1) SDSS name, (2) CIV luminosity (in units of 10$^{45}$ \ergs), (3) ionized gas mass (in units of M$_{\odot}$), (4) velocity shift of the outflow component (in units of \kms), (5) ionized gas mass rate (in units of M$_{\odot}$/yr$^{-1}$) and (6) kinetic power (units of 10$^{42}$ \ergs).
					
		        \item[a] We used v$\rm_{CIV}^{out}$ and v$\rm_{CIV}^{max}$ in the calculation of the outflow properties of the \weak\ and \oiii\ QSOs respectively (see Sect. \ref{sec:ekin}).
       				
     				  \end{tablenotes}
       \end{threeparttable}
 
			\end{table}

  \begin{table}[h!]
  \small
\centering
	\caption{Properties of the CIV outflows derived from the entire CIV emission line. Objects in boldface belong to the \oiii\ sample.}\label{tab:ekin2}
	\setlength\tabcolsep{1pt}
\begin{threeparttable}
			\begin{tabular}{lccccc}
				\hline
				\hline
				SDSS Name & L$\rm_{CIV}$  &\mion	& \vciv & \mdot\ &\ekin \\[0.5ex]
									&(10$^{45}$\ergs) &(M$_{\odot}$)&(\kms )& (M$_{\odot}$ yr$^{-1}$) & ($10^{42}$\ergs)\\[0.5ex]
									(1)&(2)&(3)&(4)&(5)&(6)\\[0.5ex]
				\hline
			{\bf{J0745+4734}}&8.3\footnotesize{$_{-0.1}^{+0.2}$}&2500 & 190$\pm120$ &2 &0.02 \\	
			J0801+5210&2.9$\pm0.1$&860& 2720$\pm180$ &7&17\\
			
		 {\bf{J0900+4215}}&6.4\footnotesize{$_{-0.2}^{+0.3}$}&1940& 560\footnotesize{$_{-110}^{+130}$}&3&0.3\\
			
			J0958+2827&1.7$\pm0.2$&520& 5640$\pm180$ &9&90 \\
			
			J1106+6400&2.8$\pm0.1$&840 & 2870\footnotesize{$_{-170}^{+150}$} &7&19 \\
				
			J1111+1336&3.0$\pm0.2$ & 900& 1920$\pm130$&5 & 6\\
								
			J1157+2724&0.9$\pm0.1$&270& 3840\footnotesize{$_{-570}^{+700}$} &3&15 \\
			
			J1201+0116&1.8$\pm0.1$&540& 3390$\pm130$&6&20 \\
				
			{\bf{J1201+1206}}&4.4\footnotesize{$_{-0.1}^{+0.2}$}&1350& 980\footnotesize{$_{-110}^{+130}$}&4 &1\\
												
			J1236+6554&2.9$\pm0.1$&870& 3360$\pm110$&9&32 \\
								
			{\bf{J1326-0005}}&17$\pm2$&4990 &50\footnotesize{$_{-130}^{+180}$}&1&0.001 \\
			
			J1421+4633&2.1$\pm0.2$&640 &5010\footnotesize{$_{-130}^{+290}$}&10 &78 \\
									
			J1422+4417&5.9$\pm0.1$&1780 &4670$\pm230$ & 25 &174\\
							
			J1521+5202&2.1$\pm0.1$&620& 7420$\pm200$&14 &250 \\	
			
				{\bf{J1538+0855}}&4.0$\pm0.1$&1210& 2190$\pm110$&8&12\\
									 
			{\bf{J1549+1245}}&3.8\footnotesize{$_{-0.3}^{+0.2}$}&1130 & 370\footnotesize{$_{-200}^{+250}$}&1 &0.01 \\
			
			J2123-0050&1.7$\pm0.1$&520& 2330\footnotesize{$_{-150}^{+160}$}&4 &6\\
		
			J2346-0016&1.2$\pm0.1$&350&3380\footnotesize{$_{-170}^{+130}$} &4&13\\
				\hline
				\end{tabular}
				\begin{tablenotes}[para,flushleft]
				
					{\bf{Notes.}} The following information are listed: (1) SDSS name, (2) CIV luminosity (in units of 10$^{45}$ \ergs), (3) ionized gas mass (in units of M$_{\odot}$), (4) velocity shift (in units of \kms), (5) ionized gas mass rate (in units of M$_{\odot}$/yr$^{-1}$) and (6) kinetic power (units of 10$^{42}$ \ergs).\\	
 \end{tablenotes}
  \end{threeparttable}
			\end{table}

 As shown in Fig. \ref{fig:ekin} the bulk of kinetic power of the BLR winds discovered in WISSH QSOs is \ekin\ $\sim$ 10$^{-5}$ $\times$ \lbol\ for both \weak\ and \oiii\ QSOs. 
It is instructive to compare these \ekin\ with those inferred for ionized NLR winds traced by [OIII]. About 20$\%$ of the BLR winds detected in WISSH QSOs shows kinetic powers comparable (i.e. 10$^{-3}$ $<$ \ekin/\lbol\ $<$ 10$^{-2}$) to those estimated for NLR winds in \oiii\ sources (green diamonds). Remarkably, in one case, i.e. J1538+0855, the outflows associated with CIV and [OIII] have consistent \ekin\ values (see Appendix \ref{sec:app_1538}). This would suggest that we are possibly revealing the same outflow in two different gas phases at increasing distance from the AGN (in an energy conserving scenario).
This further supports the idea that this object represents a hybrid showing a mixture of the distinctive properties of the two populations of {\it{Weak}} - and \oiii\ QSOs.\\
Hereafter, we consider the outflow parameters derived by using \vout\ and \vmax\  for the \weak\ and the \oiii\ QSOs, respectively, as the most representative ones.
   Fig \ref{fig:mass_load} ({\it{left panel}}), shows a comparison of \ekin\ as a function of \lbol\ between the NLR and BLR winds revealed in WISSH QSOs and a large collection of [OIII]-based NLR winds from a heterogeneous AGN sample reported in \cite{Fiore2017}. 
A sizable fraction of BLR winds in \weak\ and \oiii\ QSOs are as powerful as NLR winds in less luminous QSOs, with \ekin\ $<$ 10$^{43-44}$ \ergs. This suggests that BLR winds should be taken into account to obtain a complete census of strong AGN-driven winds  and comprehensively evaluate their effects of depositing energy and momentum into the ISM.\\
 In order to give an idea of the possible uncertainties affecting the calculations of \ekin , in Fig. \ref{fig:mass_load} ({\it{left panel}}) we show the maximum and minimum values obtained by considering a very large range of variation for the two fundamental parameters  {\it{n$\rm_{e}$}} and {\it{Z}}. More specifically,  the lower bound corresponds to the assumption of n$_{e}$ = 10$^{10}$ cm$^{-3}$ based on the presence of the semiforbidden line CIII]$\lambda$1909 \AA\ (\citealt{Ferland1984}), and Z = 8 Z$_{\odot}$  (\citealt{Nagao2006}), while the upper bound corresponds to the assumption of n$_{e}$ = 10$^{9}$ cm$^{-3}$, based on the absence of forbidden lines such as [OIII]$\lambda$4363 \AA\ (\citealt{Ferland1984}) and Z = 3 Z$_{\odot}$  (\citealt{Nagao2006}).\\
 Fig. \ref{fig:mass_load} ({\it{right panel}}) displays the outflow momentum load (i.e. the ouflow momentum rate $\rm\dot{P}_{out}$ ($\equiv$ $\rm\dot{M}_{out}$ $\times$ v$\rm_{out}$) normalized to  the AGN radiation momentum rate $\rm\dot{P}_{AGN}$ $\equiv$ \lbol /$c$) as a function of outflows velocity for different classes of outflow derived by \cite{Fiore2017}, compared to those measured for the winds traced by the blueshifted CIV emission line in WISSH. 
 The BLR winds in WISSH show velocities between those measured for X-ray Ultra-Fast (v $> 10^4$ \kms) outflows (UFOs, crosses) and [OIII]-based outflows (v $<$ 2,000 \kms, triangles). This matches well with interpreting the outflow velocity distribution as a proxy of the distribution in radial distance from the AGN, i.e. from the innermost region of the accretion disk (tens of gravitational radii for the UFOs, e.g. \citealt{Nardini2015}; \citealt{Tombesi2012,Tombesi2013}; \citealt{Gofford2015}) up to kpc-scale in case of the [OIII] winds (e.g.  \citealt{Harrison2012}; \citealt{Carniani2015}; \citealt{Cresci2015}; Paper I).\\
The BLR winds typically exhibit a low momentum load, i.e $\rm\dot{P}_{out}$/$\rm\dot{P}_{AGN}$ \simlt\ 0.1, which is a range poorly sampled by other ionized winds.
 Furthermore, the BLR winds seem to represent the low-power, low-velocity analogs of UFOs. The latter show  $\rm\dot{P}_{out}$/$\rm\dot{P}_{AGN}$ $\sim$ 1, as expected in case of quasi-spherical winds with electron scattering optical depth $\tau \sim$  1 produced by systems accreting at \edd\ $\sim$ 1 (e.g. \citealt{King2010}). UFOs are found to have a large covering factor (\citealt{Nardini2015}; \citealt{King2015}). BLR clouds in luminous AGN are expected to have a covering factor of $\sim$ 0.1 (\citealt{Netzer1990}). We can thus speculate that the low  $\rm\dot{P}_{out}/\rm\dot{P}_{AGN}$ of BLR winds with respect to that of UFOs may be due to both a lower N$\rm_H$ and a lower covering factor of the CIV outflowing gas with respect to the fast highly ionized gas responsible for the UFOs.

\begin{figure*}[]
\makebox[1\textwidth]{ 
  \includegraphics[width=0.515\linewidth]{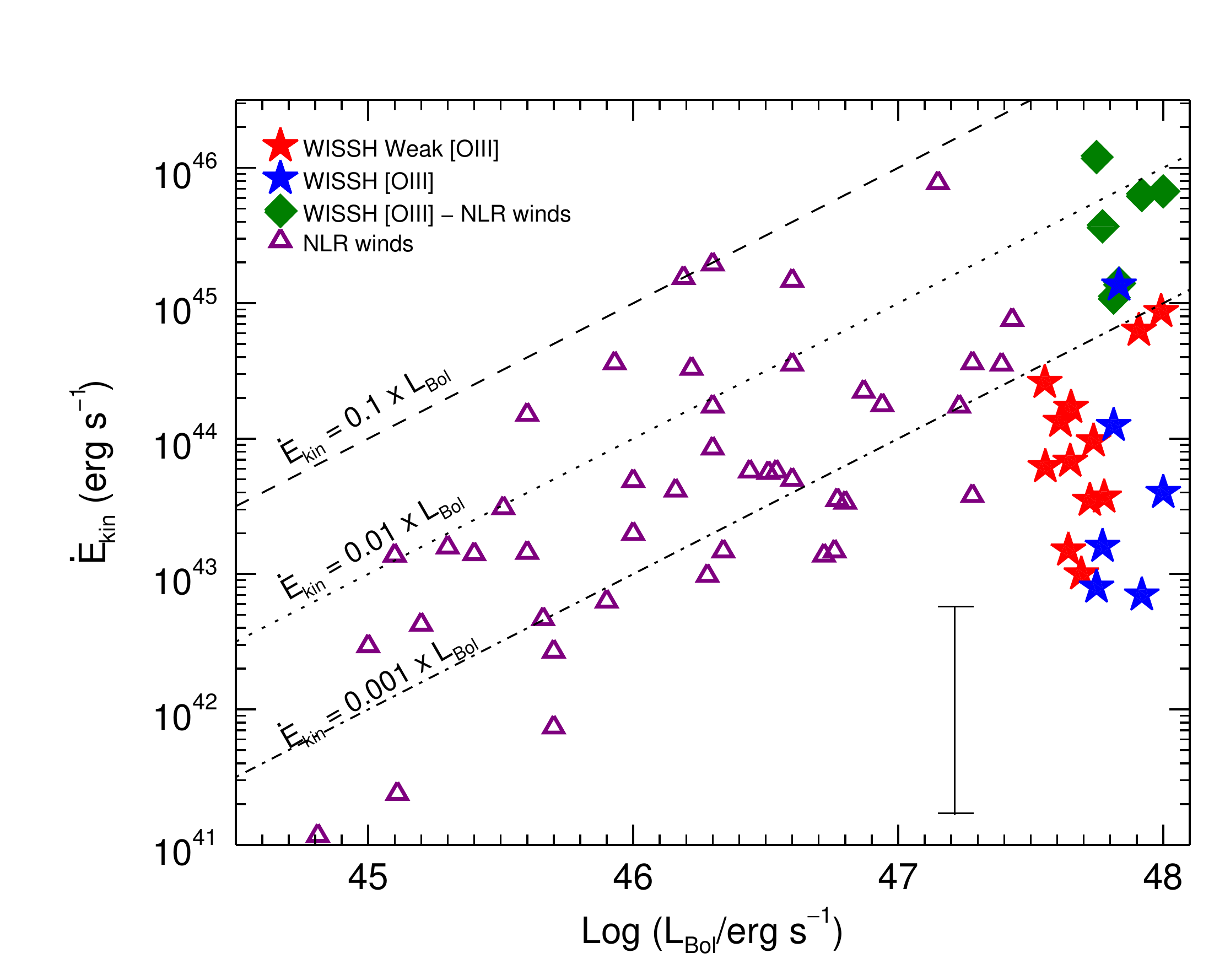}
  \includegraphics[width=0.52\linewidth]{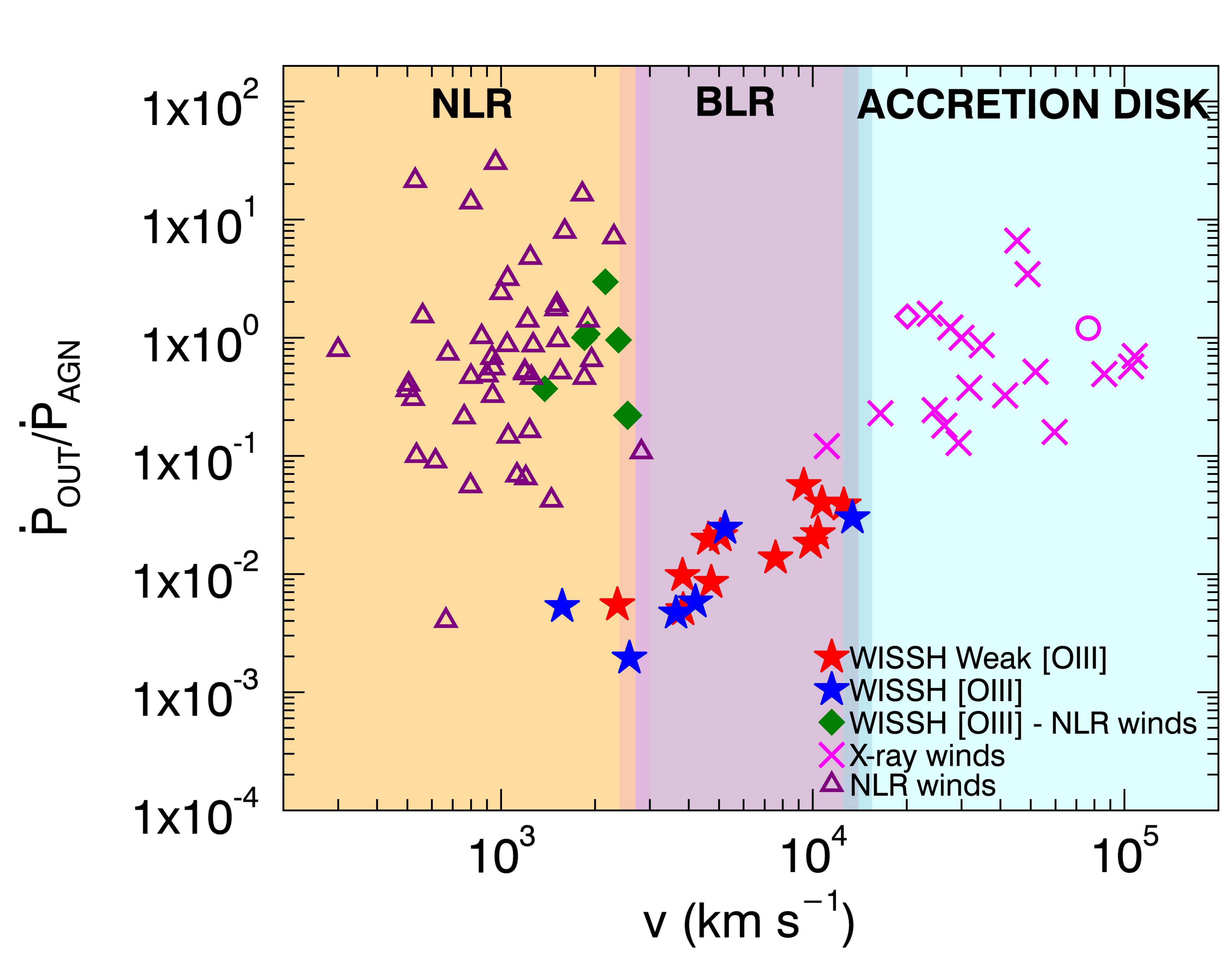}
  }
  \caption{{\it{(left)}} Kinetic power of the CIV outflow component as a function of \lbol\ for the WISSH QSOs (red and blue stars) compared with WISSH NLR ionized outflows from Paper I (green diamonds) and other samples from literature, (e.g. \citealt{Fiore2017} for details) (purple triangles). The error bar (bottom right corner) is calculated as described in Sect. \ref{sec:ekin}. {\it{(right)}} Wind momentum load as a function of the outflow velocity. The BLR winds traced by the CIV outflow components from the WISSH QSOs (red and blue stars) are compared with X-ray winds (magenta crosses) and Ionized winds (green triangle). X-ray winds for Markarian 231 and IRASF11119+13257 are represented with magenta diamond and circle, respectively.} 
\label{fig:mass_load}%

    \end{figure*}


\section{What is the physical driver of the CIV velocity shift?}\label{driver}

In order to shed light on the main physical driver of the large blueshifts of the CIV emission line observed in QSOs,  we investigated the dependence of CIV velocity shift on fundamental AGN parameters such as \lbol , \edd\ and the UV-to-X-ray continuum slope (\aox). Previous studies indeed found that  \vciv\ is correlated with all these three quantities (\citealt{Marziani2016}, \citealt{Richards2011}), but it still unclear  which is the fundamental dependency. 
We proceeded as follows: we initially identified the main driver  of \vciv\  between \lbol\ and \edd\  and  then we studied the dependency of \vciv\ on this parameter with that on \aox .

\subsection{Velocity shifts vs. L$\rm_{Bol}$ and \edd}\label{Lbol}

We compared the WISSH sample to QSOs for which the BH mass and Eddington ratio are derived from the H$\beta$ emission line, the "H$\beta$ sample", because of the large uncertainties of BH mass estimation from CIV emission line, affected by non-virialized component. All these sources therefore have reliable SMBH masses and \edd .

Fig. \ref{fig:shift_corr} shows  \vciv\ as a function of \lbol\ (a) and \edd\ (b) for the WISSH  and  the "H$\beta$ sample" QSOs.
The blueshifts are clearly correlated with both \lbol\ and \edd . Specifically, we found a stronger correlation with  \lbol\ (Spearman rank r = 0.43 and P-value = 1.8 $\times$ 10$^{-9}$) than with \edd\ (Spearman rank r = 0.33 and P-value = 7.1 $\times$ 10$^{-6}$), bearing in mind the large scatter affecting both relations.

Moreover, performing a least squares regression, we found that \vciv\ $\propto L\rm_{Bol}^{0.28\pm0.04}$. This is consistent with a radiation-driven wind scenario (\citealt{Laor2002}), for which we indeed expect a terminal outflow velocity v$_{t}$ $\propto \sqrt{L/R_{\ge BLR}} \propto L^{0.25}$ (valid for R $\ge$ R$\rm_{BLR}$, e.g. \citealt{Netzer1993}; \citealt{Kaspi2000};  \citealt{Bentz2009}).
A similar dependence is also found for \vciv\ as a function of \mbh . This lends further support to the radiative wind scenario and suggests that the radiation pressure is dominant over the Keplerian velocity field (for which a dependence of $\propto$ M$\rm_{BH}^{0.5}$ is expected).


In order to determine which is the fundamental variable between \lbol\ and \edd , we studied correlations between the residuals from (i) \edd\--\lbol\ and \vciv\--\lbol\ relations and (ii)  \lbol\--\edd\ and \vciv\--\edd\ relations (see Appendix B in \citealt{Bernardi2005} for further details about residuals analysis). 
More specifically, we tested the hypothesis that the bolometric luminosity is the fundamental variable. In this case we expect that: (i) no significant correlation between the residuals obtained from the \edd\--\lbol\ ($\rm\Delta_{\lambda, L}$) and \vciv\--\lbol\ ($\rm\Delta_{v, L}$) relations; (ii) a correlation between residuals obtained from \lbol\--\edd\  ($\rm\Delta_{L, \lambda}$) and \vciv\--\edd\  ($\rm\Delta_{v, \lambda}$) relations; (iii) the slope of this correlation should be the same of the \vciv\--\lbol\ relation.
The $\rm\Delta_{\lambda, L}$--$\rm\Delta_{v, L}$ residuals are plotted in Fig. \ref{fig:shift_corr}c. In this case we derived a Spearman rank r = 0.11 and P-value = 0.16, which indicates no correlation between parameters as expected from (i). 
In Fig. \ref{fig:shift_corr}d the $\rm\Delta_{L, \lambda}$ and $\rm\Delta_{v, \lambda}$ residuals are plotted. In this case we found a strong correlation with a Spearman rank r = 0.32 and P-value = 1.1 $\times$ 10$^{-5}$, with a slope consistent (2$\sigma$) with that we measured for \vciv\--\lbol\ relation.

We also created 1000 {\it{bootstrap}} samples from the residuals shown in Fig. \ref{fig:shift_corr}d and calculated the corresponding Spearman rank, r. From the original residuals we derived r = 0.33 with a 95\% confidence interval of 0.18-0.46, which is defined as the interval spanning from the 2.5th to the 97.5th percentile of the resampled values.

By combining these results, we conclude that \lbol\ is the fundamental variable with respect to the \edd\ and it can be considered as the main driver of the observed CIV blueshifts with respect to \edd .


\begin{figure*}[]
 \centering
\centering
\makebox[1\textwidth]{

\includegraphics[width=0.5\linewidth]{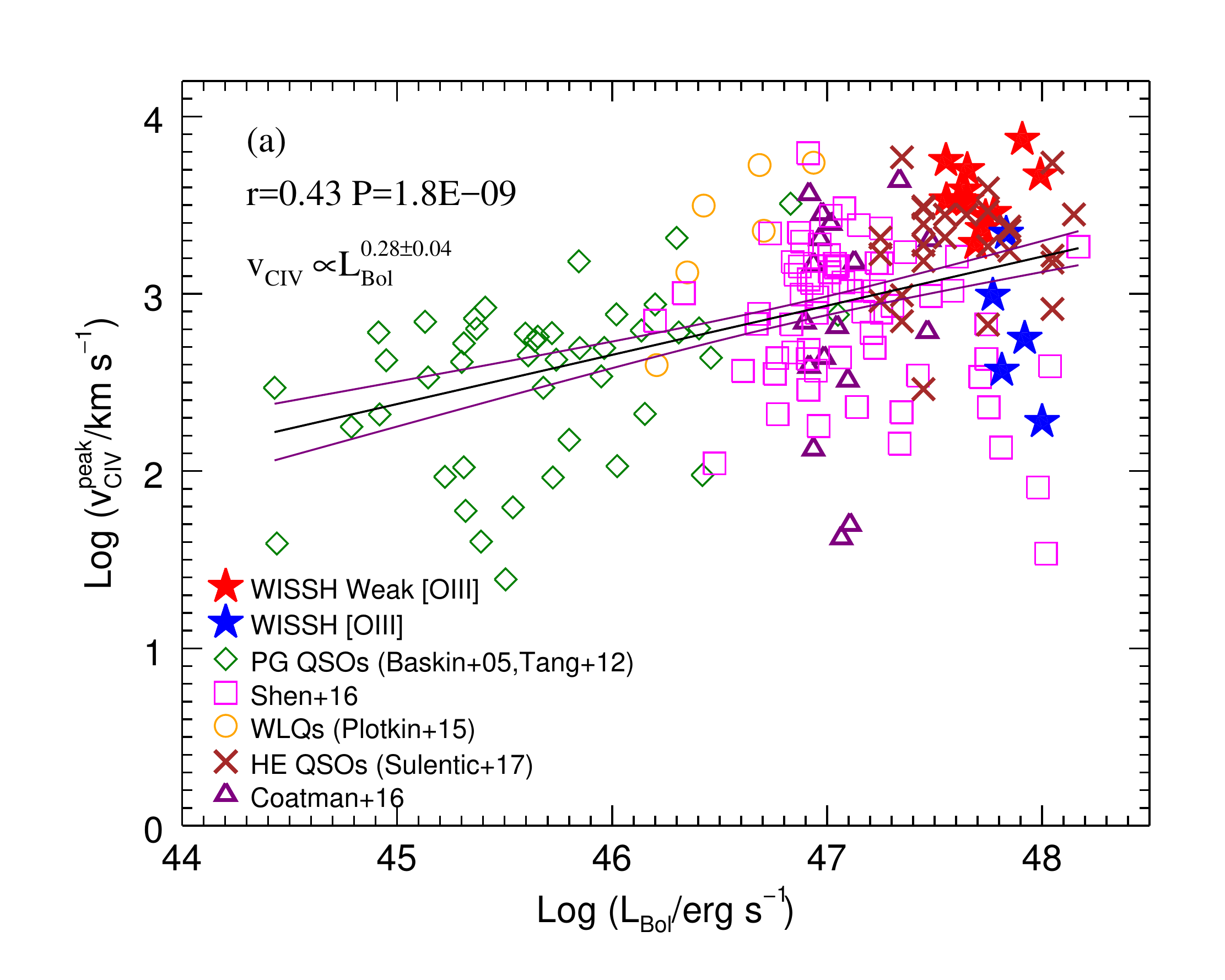}
 \includegraphics[width=0.5\linewidth]{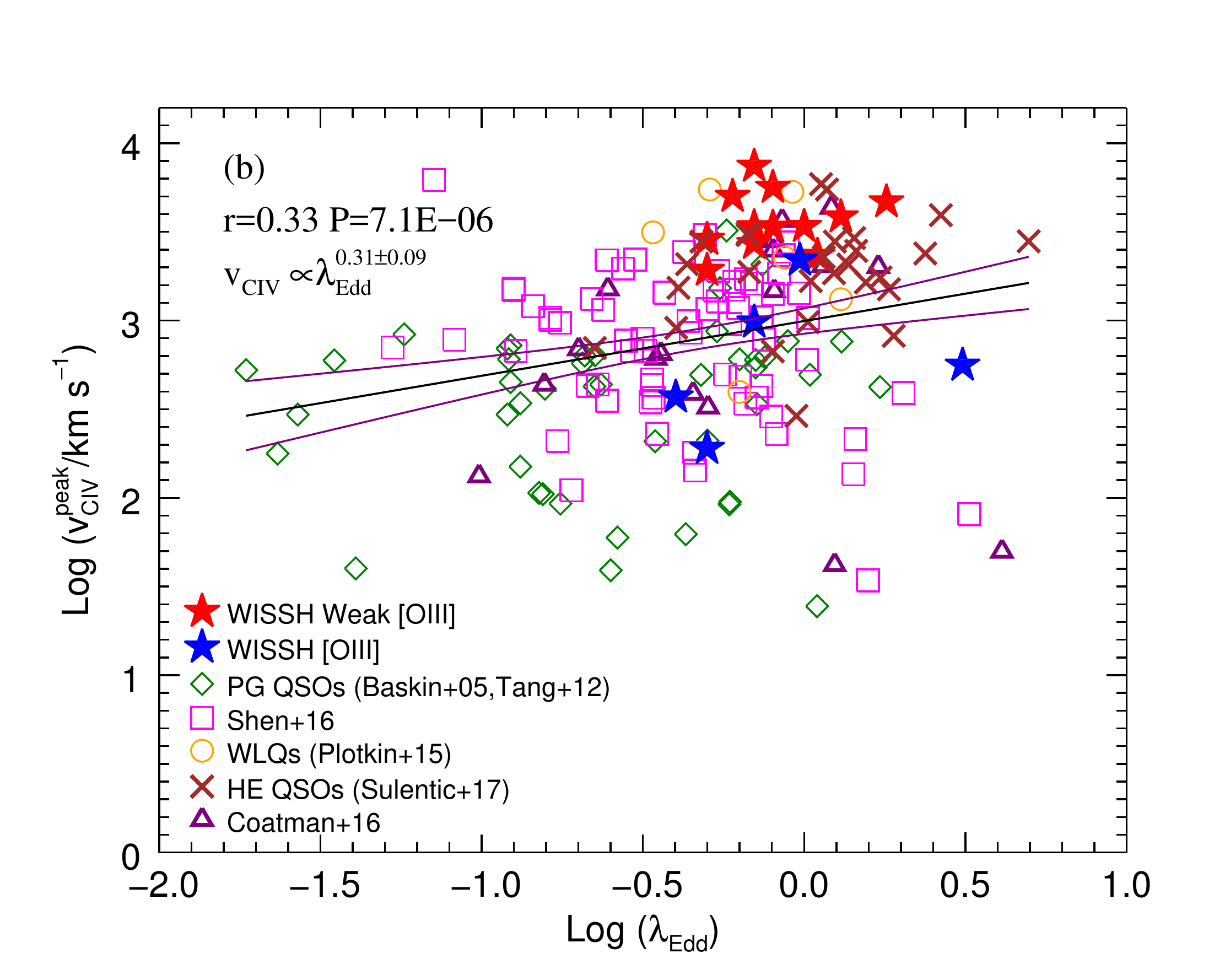}
  }

\makebox[1\textwidth]{
\includegraphics[width=0.5\linewidth]{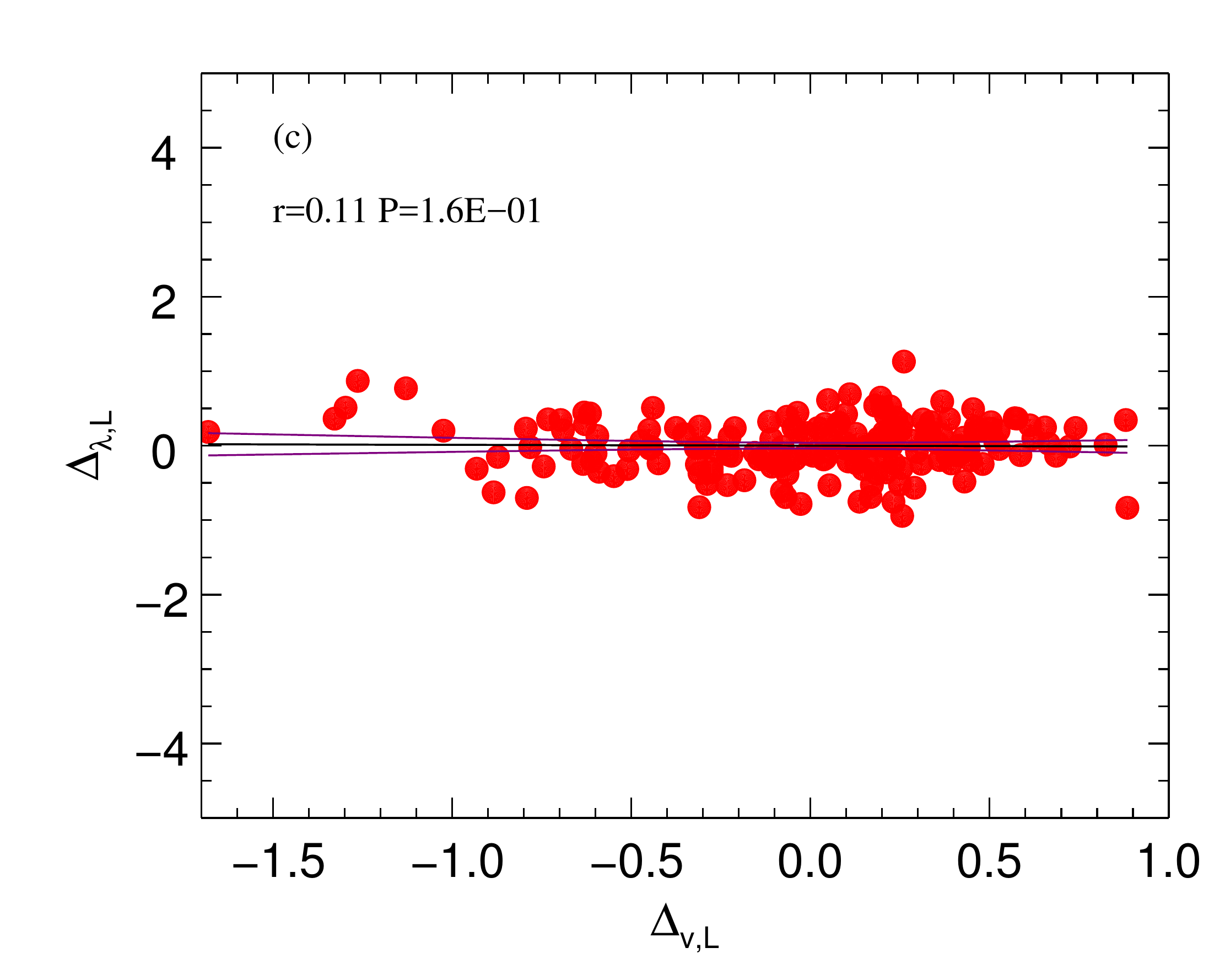}
 \includegraphics[width=0.5\linewidth]{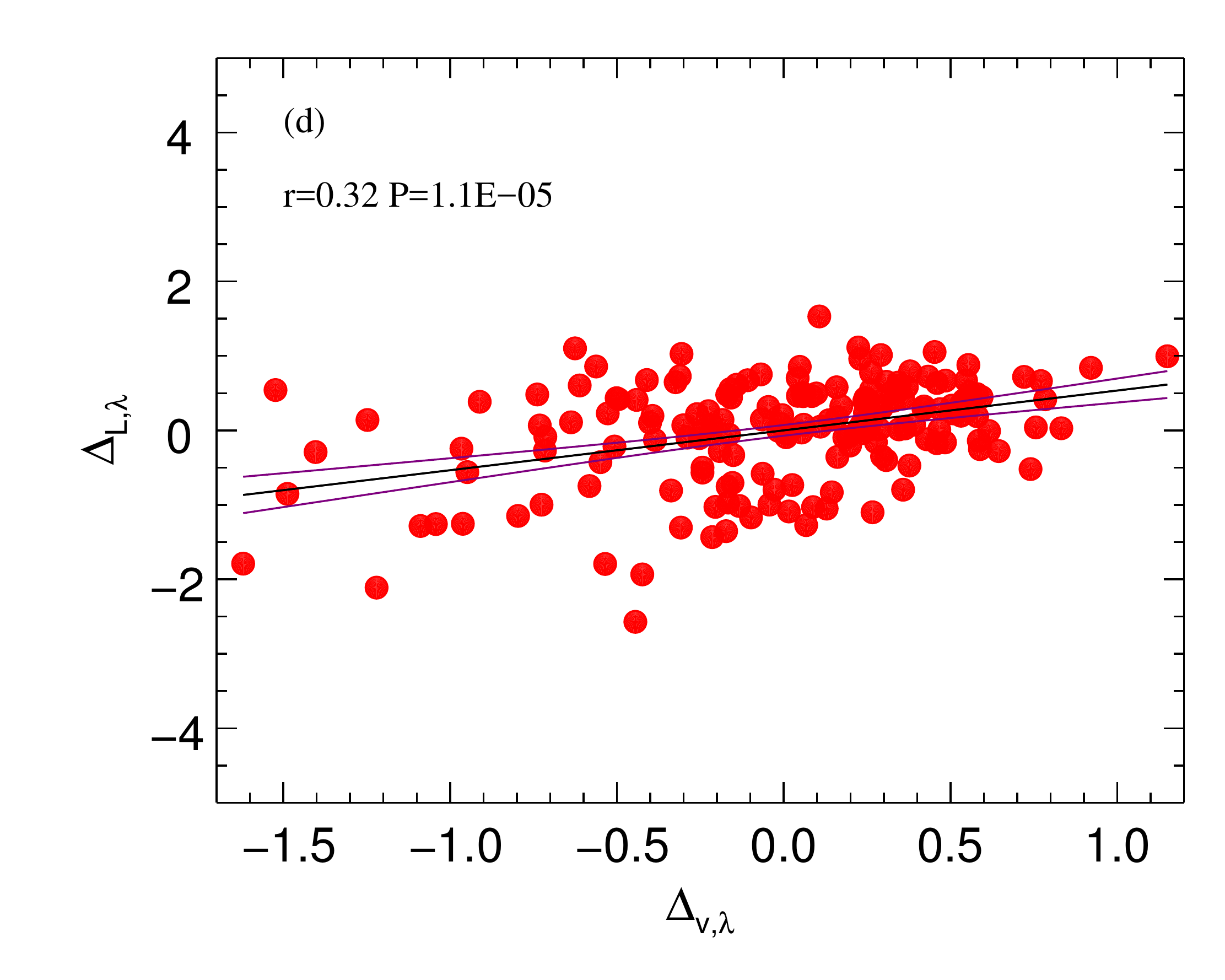}
}
  \caption{\vciv\  as a function of \lbol\ \textit{(a)} and as a function of \edd\ \textit{(b)}, for the WISSH sample (\textit{blue/red stars}) compared to the PG QSOs from \cite{Baskin2005} and \cite{Tang2012} (green diamonds), WLQs from \cite{Plotkin2015} (orange circles) and SDSS QSOs from \cite{Shen2016} and \cite{Coatman2016} (magenta squares and purple triangles) and HE QSOs from \citet[and reference therein]{Sulentic2017}. The L$\rm_{Bol}$ for the HE sample has been derived from L$_{5100}$ applying a bolometric correction of 5.6, as suggested by \cite{Runnoe2012} for the luminosity range of the HE sample. \textit{(c)} Residuals plot of \edd\ - \lbol\ and \vciv\ - \lbol\ relations. \textit{(d)} Residuals plot of \lbol\ - \edd\ and \vciv\ - \edd\ relations. The black lines indicate the best linear fits to the data and the purple lines correspond to the 68\% confidence interval.}
\label{fig:shift_corr}%



  \end{figure*}

  \subsection{Velocity shifts vs. L$\rm_{Bol}$ and $\alpha\rm_{ox}$}\label{aox}

We have performed the same analysis described in Sect. \ref{Lbol} considering \lbol\ and \aox , to investigate the primary driver of the CIV blueshifts.
  Fig. \ref{fig:shift_corr_aox} shows the \vciv\  as a function of  \lbol\ and \aox\  for 14 WISSH QSOs with both H$\beta$ and X-ray measurements. For the WISSH QSOs the \aox\ was derived using the monochromatic luminosities at 2500 \AA\ obtained from broad-band SED fitting and the absorption-corrected luminosities at 2 keV (\citealt{Martocchia2017}). We compare our findings with those derived for 170 radio-quiet, broad-line QSOs from the \cite{Wu2009} sample and 29 WLQs (\citealt{Luo2015}; \citealt{Wu2011}) with available UV and X-ray information. 
  
Performing a least-square fit also for this large sample, we confirm the results reported in Sect. \ref{Lbol} about the presence of correlation between \vciv\ and \lbol , deriving a dependence \vciv\ $\propto$ \lbol $^{{0.25}\pm 0.06}$ (Spearman rank r = 0.30  and P-value = 1.1 $\times$ 10$^{-5}$; see Fig. \ref{fig:shift_corr_aox}a).
   We found a very strong anti-correlation between the \vciv\ and \aox\ (Fig. \ref{fig:shift_corr_aox}b), i.e. \vciv\ $\propto$ \aox $^{-1.07\pm0.16}$ (Spearman rank r = -0.46 and P-value = 8.7 $\times$ 10$^{-13}$), confirming the results reported in \cite{Richards2011}.

We also performed the correlations analysis between the residuals from (i) \vciv\--\lbol\ ($\rm\Delta_{v,L}$) and \aox\--\lbol\ ($\rm\Delta_{\alpha ox,L}$) relations and (ii)  \vciv\--\aox\ ($\rm\Delta_{v, \alpha ox}$) and \lbol\--\aox\ ($\rm\Delta_{L,\alpha ox}$) relations,  based on the hypothesis that \lbol\ is the fundamental parameter. However, the results reported in Fig. \ref{fig:shift_corr_aox}c,d  are at odds with this hypothesis, indicating a clear anti-correlation between $\rm\Delta_{v, L}$ and $\rm\Delta_{\alpha ox, L}$, and no correlation between $\rm\Delta_{v,\alpha ox}$ and $\rm\Delta_{L,\alpha ox}$. Furthermore, the slope of the residuals $\rm\Delta_{v, L}$-$\rm\Delta_{\alpha ox, L}$ is consistent with that found for \aox\ - \vciv\ relation.
From a statistical point of view, this points to \aox\ as the primary driver of the blueshifts of the CIV emission line observed in these QSOs.
 
We note that there is a well-known strong anti-correlation between \aox\ and the UV luminosity ($\sim$ \lbol\ for Type I QSOs), (e.g. \citealt{Vignali2003}; \citealt{Steffen2006};  \citealt{Lusso2010}; \citealt{Martocchia2017}) according to which the steeper the \aox\ the higher the luminosity.  Therefore, both selecting steep \aox\ or high \lbol\ allows to pick up fast outflows. We can conclude that the strength and the slope of the ionizing continuum is the main driver of the BLR winds. 
 In Table \ref{tab:corr} a summary of correlations of the CIV velocity shifts with physical quantities as \lbol , \edd , \mbh\ and \aox\ is reported. Results from residuals correlations are also listed.

  The dependence of the velocity shift on both \lbol\ and \aox\ is in agreement with our scenario of radiation driven wind, according to which a strong UV continuum is necessary to launch the wind but the level of Extreme UV (EUV) and X-ray emission (i.e. up to 2 keV) is crucial to determine its existence, since strong X-ray radiation can easily overionize the gas and hamper an efficient line-driving mechanism. On the contrary, an X-ray weaker emission with respect to the optical/UV accretion disk emission can allow UV line opacity (\citealt{Leighly2004}; \citealt{Richards2011}).
Furthermore, as suggested by \cite{Wu2009}, the increasing steepness of the \aox\ (i.e. the soften of EUV--X-ray emission) at progressively higher UV luminosities may also explain the weakness of the CIV emission line (ionization potential 64.45 eV) in luminous QSOs: the strength of the line indeed depends on the number of the photons available to produce it, whereby a deficit of such ionizing photons leads to CIV weaker emission lines, as typically observed in \weak\ QSOs and WLQs.

   \begin{figure*}[]
 \centering
\makebox[1\textwidth]{

\includegraphics[width=0.5\linewidth]{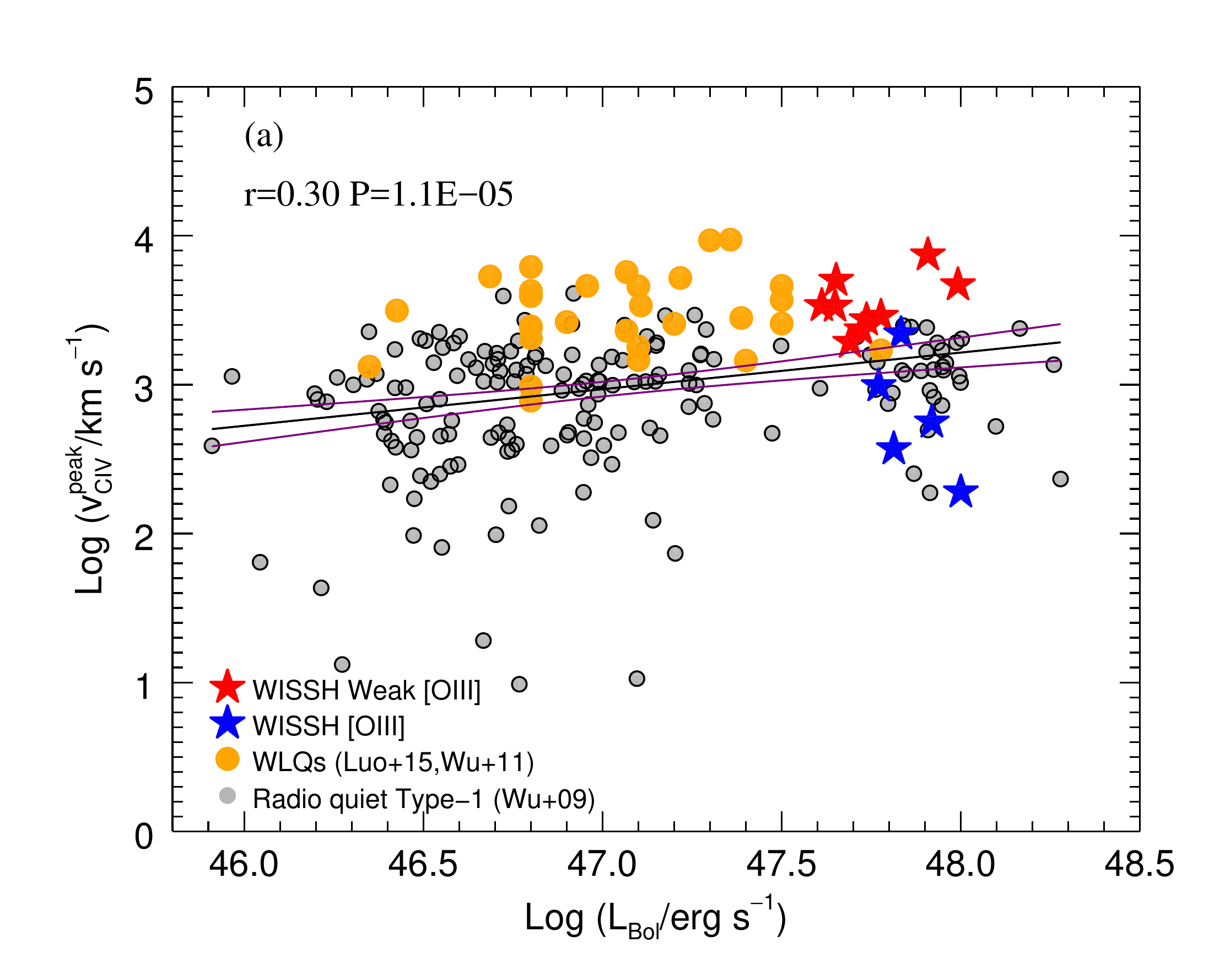}
 \includegraphics[width=0.5\linewidth]{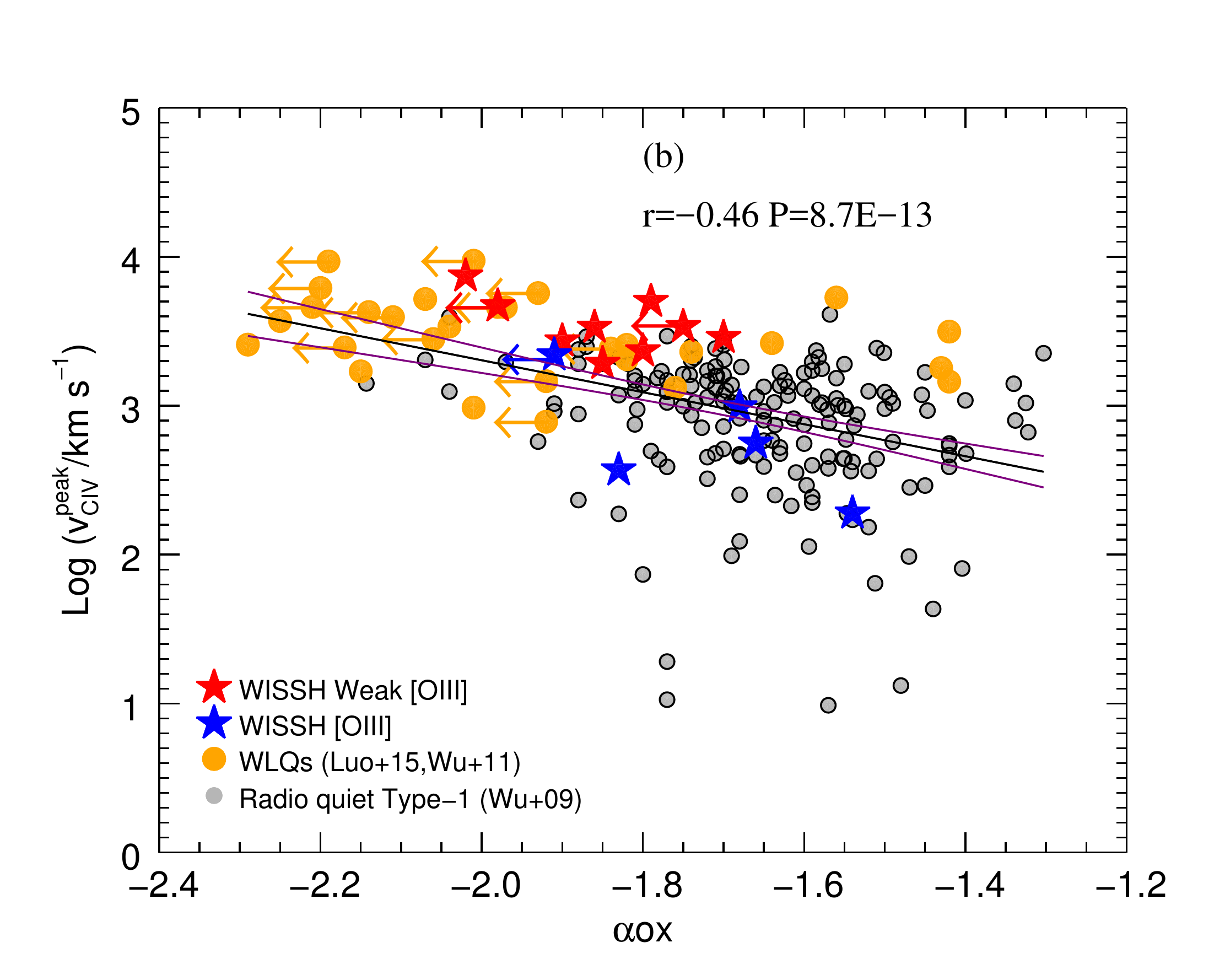}
}
\makebox[1\textwidth]{
\includegraphics[width=0.5\linewidth]{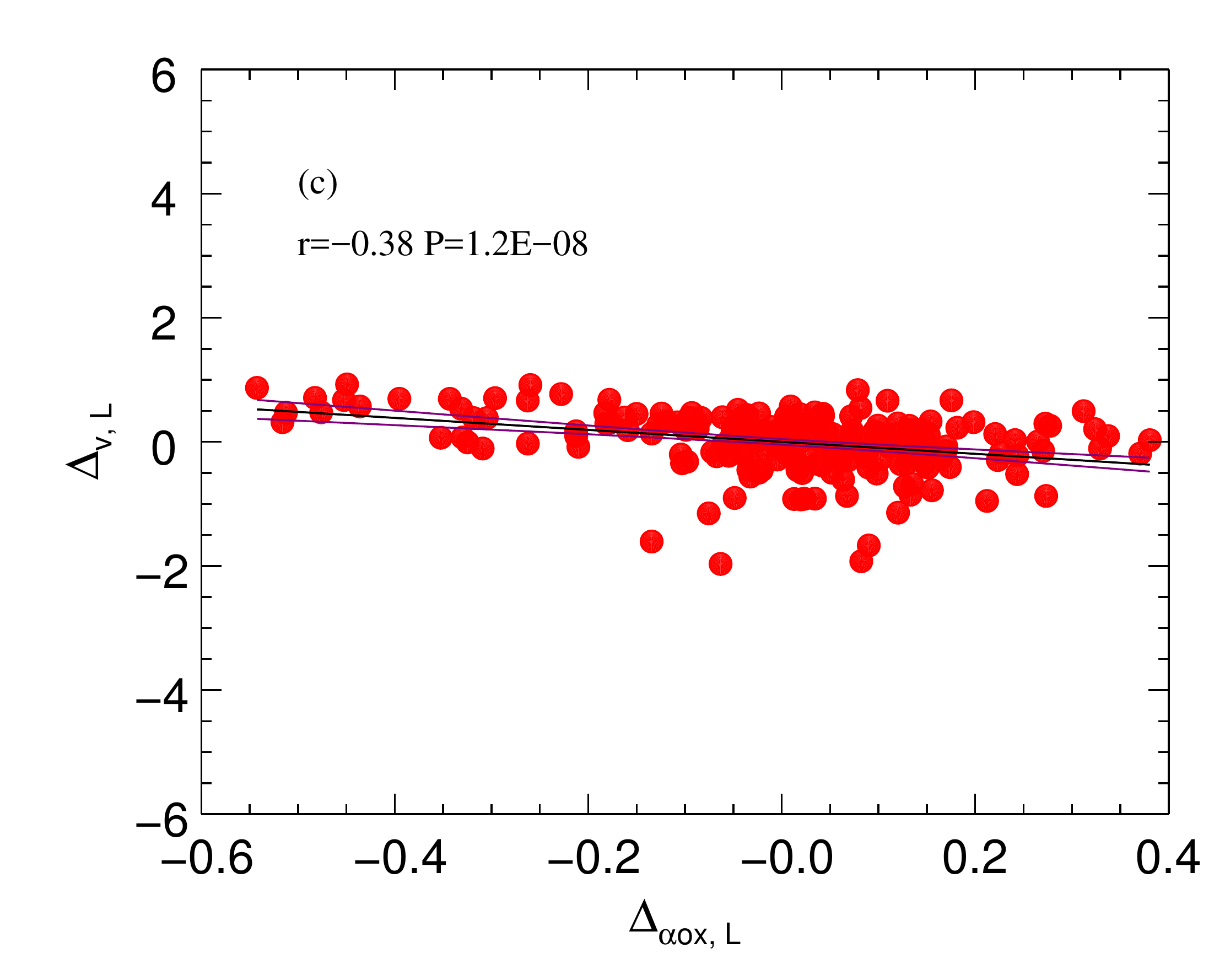}
 \includegraphics[width=0.5\linewidth]{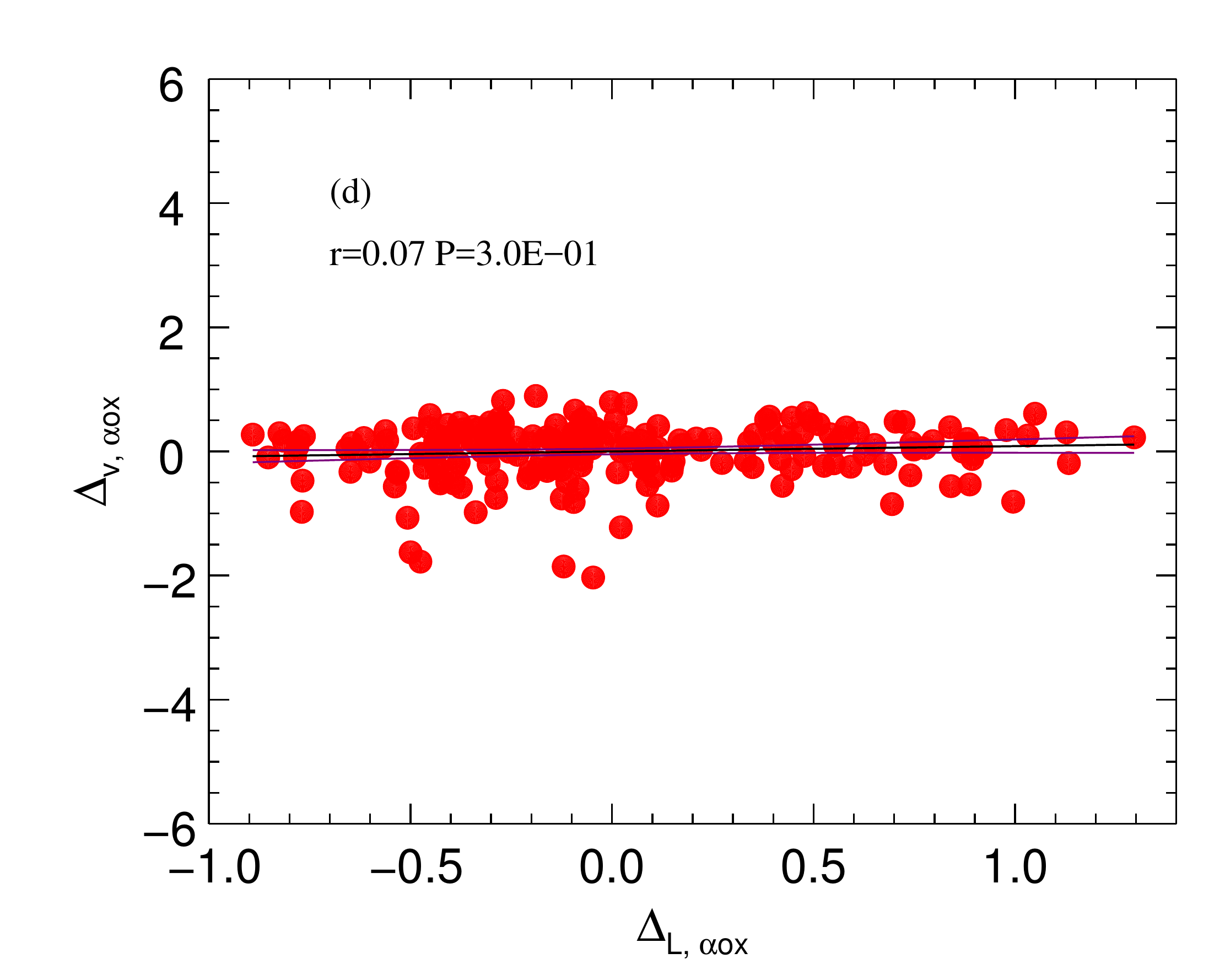}
}

  \caption{\vciv\ as a function of \lbol\ \textit{(a)} and  \aox\ \textit{(b)}, for the WISSH sample (\textit{blue/red stars}) compared to Radio quiet Type-1 QSOs from \cite{Wu2009} (black circles) and WLQs from \cite{Wu2011} and \cite{Luo2015} (orange circles). \textit{(c)} Residuals plot of \vciv\ - \lbol\ and  \aox\ - \lbol\  relations. \textit{(d)} Residuals plot of \vciv\ - \aox\ and  \lbol\ - \aox\  relations. The black lines represent the best linear fits to the data and the purple lines correspond to the 68\% confidence interval.}
\label{fig:shift_corr_aox}%

  \end{figure*}

 \begin{table}[]
 \center
\caption{Correlations of \vciv\ with AGN fundamental parameters.}\label{tab:corr}
	\setlength\tabcolsep{10pt}
	\begin{threeparttable}
			\begin{tabular}{lcc c}
\hline		
				\hline	
			 Correlation & ~Slope	& r &~ P-value \\
	(1)&(2)&(3)&(4)\\

\hline
	           \vciv\ vs. \lbol\ \tnote{a}&	0.28$\pm0.04$&	0.43& 1.8$\times$ 10$^{-9}$ \\
				\vciv\ vs. \edd\ \tnote{a}& 0.31$\pm0.09$& 0.33& 7.1$\times$ 10$^{-6}$\\
				\vciv\ vs. \mbh\ \tnote{a}& 0.30$\pm0.05$& 0.34& 3.5 $\times$ 10$^{-6}$\\
				$\rm\Delta_{\lambda, L}$ vs. $\rm\Delta_{v, L}$ \tnote{a}& -0.01$\pm0.06$ & 0.11 & 1.6 $\times$ 10$^{-1}$ \\
				$\rm\Delta_{L, \lambda}$ vs. $\rm\Delta_{v, \lambda}$ \tnote{a}& 0.54$\pm0.10$ & 0.32 & 1.1 $\times$ 10$^{-5}$ \\ 
\hline			
				\vciv\ vs. \lbol\ \tnote{b}&	0.25$\pm0.06$&0.30&1.1 $\times$ 10$^{-5}$ \\
				\vciv\ vs. \aox\ \tnote{b} &   -1.07$\pm0.16$&-0.46& 8.7 $\times$ 10$^{-13}$\\
				$\rm\Delta_{v,L}$ vs. $\rm\Delta_{\alpha ox,L}$ \tnote{b}& -0.97$\pm0.18$ & -0.38 & 1.2 $\times$ 10$^{-8}$ \\
				$\rm\Delta_{v, \alpha ox}$ vs. $\rm\Delta_{L,\alpha ox}$ \tnote{b}& 0.09$\pm0.06$ & 0.07 & 3.0 $\times$ 10$^{-1}$ \\
				
				\hline
				\end{tabular}
				 \begin{tablenotes}
			{\footnotesize{\item {\bf{Notes.}} The following information is listed. (1) Correlations of \vciv\ with \lbol , \edd , \mbh\ and \aox , and between residuals (see Sect. \ref{Lbol} and \ref{aox}), (2) least square regression slope, (3) Spearman rank and (4) null hypothesis probability.
	   		   	    \item[a] Correlations derived for the WISSH QSOs and other QSO samples from literature with H$\beta$ measurements, as detailed in Sect. \ref{Lbol}.
				    \item[b] Correlations derived for the WISSH QSOs with both H$\beta$ and X-ray measurements, as detailed in Sect. \ref{aox}}.} 
					  \end{tablenotes}
		 \end{threeparttable}
			\end{table}

\section{Summary and Conclusions}\label{summary}

We have presented the results from the analysis of thirteen sources in the WISSH quasar sample with NIR spectroscopy from LBT/LUCI or VLT/SINFONI observations, in addition to five sources discussed in Paper I. We stress that these objects have been selected on the basis of a redshift for which the LBT/LUCI spectrum covers the [OIII] wavelength range and, therefore, can be considered as randomly selected from the entire WISSH QSOs sample. 
Our analysis has been performed with the goal of deriving the properties of (i) H$\beta$ and [OIII] emission lines from the optical rest-frame spectra and (ii) CIV emission line by using rest-frame UV SDSS spectra.
Our main findings can be summarized as follows:

\begin{itemize}

\item All the 18 WISSH QSOs considered here exhibit SMBHs with mass larger than 10$^{9}$ $M_\odot$, with 50$\%$ of them hosting very massive SMBHs with \mbh\ \simgt\ 5 $\times$ 10$^{9}$ $M_\odot$. Based on these H$\beta$-based \mbh\ values, we derived  Eddington ratios  0.4 $<$ \edd\ $<$ 3.1. This supports the evidence that the WISSH QSOs are powered by highly accreting SMBHs at the massive end of the SMBH mass function.

\item According to the REW$\rm_{[OIII]}$ $\geq$ or $<$ 5 \AA , the WISSH QSOs can be divided into two samples: \oiii\ and \weak\ samples, respectively. In particular 6 sources exhibit REW$\rm_{[OIII]}$ $\approx$ 7-70 \AA , showing [OIII] profile dominated by a broad blueshifted component and 11 sources have REW$\rm_{[OIII]}$ $\approx$ 0.3-3 \AA .

As reported in Sect. \ref{oiii_hbeta}, there is more than one explanation to ascribe to the WISSH REW$\rm_{[OIII]}$ distribution. One is in terms of orientation effect, with the \oiii\ sample likely seen at high ($\approx$25-73 deg) inclination.  This leads to a lower continuum luminosity and, hence, high REW$\rm_{[OIII]}$ values. On the contrary, the \weak\ sources (included J1422+4417, see Sect. \ref{sec:OIII_analysis}) are likely associated with nearly face-on AGNs. 
But the weak [OIII] emission for the WISSH QSOs is expected to be due to the high bolometric luminosities of these sources. Indeed the [OIII] emission tends to decrease because of the over-ionization from the central engine, as found by \cite{Shen2014}. A further explanation for the difference in REW$\rm_{[OIII]}$ between the \oiii\ and \weak\ may be ascribed by the presence of a kpc-scale ionization cone oriented along the galaxy disk for the \oiii\ objects, leading to a larger amount of gas to be ionized, while a ionization cone oriented perpendicular to the galaxy disk in the case of \weak\ sample leads to a lower content of [OIII] gas to be ionized.

 \item  Most of the WISSH QSOs exhibits 10 $\simlt$ REW$\rm_{CIV}$ $\simlt$ 20 \AA\ with the peak of the CIV emission line profile blueshifted with respect to the systemic redshift (\vciv\ $\sim$ 2,000--8,000 \kms), indicating that the emitting gas is outflowing. This suggests that the luminosity-based selection criterion of WISSH is very effective in collecting strong CIV winds. 
Historically, such large CIV blueshifted have been associated with fainter CIV emitting QSOs, i.e. the WLQs (\citealt{Plotkin2015}), although very recently \cite{Hamann2017} have reported the existence of ERQs showing CIV winds and extremely large REW$\rm_{CIV}$ (i.e. $>$ 100 \AA). We found that QSOs belonging to the \weak\ sample, representing the 70$\%$ of the WISSH sample analyzed here, show a broad asymmetric blueward and relatively weak (i.e. REW$\rm_{CIV}$ $<$ 20 \AA\ ) CIV line profile (see Fig. \ref{fig:shift_ewoiii}), with the peak of the entire CIV profile extremely blueshifted (\vciv\ $\ge$ 2,000 \kms\ up to 8,000 \kms). On the contrary, the 30$\%$ of the WISSH QSOs (i.e. \oiii\ sample) has peaky CIV profile with REW$\rm_{CIV}$ $>$ 20 \AA\ and \vciv\ $\le$ 2,000 \kms . 
High-luminosity, optically-selected HE QSOs from \cite{Sulentic2017} follow a similar trend by populating the same region of the WISSH QSOs in the plane REW$\rm_{CIV}$-\vciv\ as shown in Fig. \ref{fig:ew_shift} and REW$\rm_{[OIII]}$-\vciv\ in Fig. \ref{fig:shift_ewoiii}.

 This highlights a dichotomy in the detection of NLR and BLR winds in WISSH QSOs, which could be likely due to inclination effects in a polar geometry scenario for the CIV winds, as suggested in Sect. \ref{sec:shift}. 

\item We found that a sizable fraction of WISSH QSOs exhibit BLR winds traced by CIV blueshifted emission with \mdot\ $\approx$ 10-100 M$_{\odot}$ yr$^{-1}$, comparable with the median accretion rate of WISSH QSOs. The 20$\%$ of the BLR winds detected in WISSH QSOs shows kinetic powers  10$^{-3}$ $<$ \ekin/\lbol\ $<$ 10$^{-2}$), i.e. comparable to those measured for NLR winds associated with [OIII] broad/blueshifted emission lines detected in the [OIII] sample. The remaining 80\%  of the BLR winds shows kinetic powers comparable to those traced by [OIII] in less luminous AGNs. This suggests that the BLR winds should be taken into account to obtain a complete census of strong AGN-driven outflows and comprehensively evaluate their effects of depositing energy and momentum into the ISM.

\item The BLR winds traced by the CIV emission line exhibit a velocity range (v$\rm_{CIV}^{out/max}$ $\approx$ 2,000-13,000 \kms) between those measured for UFO (at sub-pc scale) and NLR winds (at kpc-scale), interpreting the outflow velocity distribution as a proxy of the distribution in radial distance for different classes of outflows (see Fig. \ref{fig:mass_load}). Moreover, the BLR winds exhibit a $\rm\dot{P}_{out}$/$\rm\dot{P}_{AGN}$ $\simlt$ 0.1, which is lower than that found for UFOs (i.e. $\rm\dot{P}_{out}$/$\rm\dot{P}_{AGN}$ $\sim$ 1). In this case the momentum rate of the outflow is comparable to that of the AGN, suggesting a covering factor for the UFOs of the order of unity. Since the covering factor for the BLR is expected to be $\sim$ 0.1, we can thus speculate that the momentum load of the nuclear winds may reflect the covering factor of the outflowing gas.

\item We investigated the dependence of \vciv\ on fundamental AGN physical parameters such as \lbol , \edd\ and \aox . We found a stronger correlation with \lbol\ than  \edd\  (see Fig. \ref{fig:shift_corr}), with a dependence consistent with a radiatively driven winds scenario (\vciv\  $\propto$ \lbol $^{0.28\pm0.04}$). Moreover, we further studied the dependence of \vciv\ on \aox , finding a stronger correlation with \aox\ than \lbol . This relation is expected due to the dependence of  \aox\  on \lbol\ (\citealt{Vignali2003}), i.e. the steeper \aox\ the larger \lbol .  This indicates that the shape of the ionizing continuum could be considered as the primary driver of the blueshifts of the CIV emission line.


 \end{itemize}

\begin{acknowledgements}
We thank the anonymous referee for helpful comments that improved the paper.
We  acknowledge very useful discussions with P. Marziani, B. Husemann and D. Kakkad. We thank M. Vestergaard for kindly providing the UV FeII templates.
The scientific results reported in this article are based on observations made by the LBT, the
European Southern Observatory (ESO program 093.A-0175(A,B)) and the SDSS.
The LBT is an international collaboration among institutions in the United
States, Italy and Germany. LBT Corporation partners are: The University of Arizona
on behalf of the Arizona Board of Regents; Istituto Nazionale di Astrofisica,
Italy; LBT Beteiligungsgesellschaft, Germany, representing the Max-Planck Society,
the Astrophysical Institute Potsdam, and Heidelberg University; The Ohio
State University, and The Research Corporation, on behalf of The University of
Notre Dame, University of Minnesota and University of Virginia.
Funding for the Sloan Digital Sky Survey IV has been provided by the Alfred P. Sloan Foundation, the U.S. Department of Energy Office of Science, and the Participating Institutions. SDSS acknowledges support and resources from the Center for High-Performance Computing at the University of Utah. The SDSS web site is www.sdss.org.
The SDSS is managed by the Astrophysical Research Consortium for the Participating Institutions. The Participating Institutions are the American Museum of Natural History, Astrophysical Institute Potsdam, University of Basel, University of Cambridge, Case Western Reserve University, University of Chicago, Drexel University, Fermilab, the Institute for Advanced Study, the Japan Participation Group, Johns Hopkins University, the Joint Institute for Nuclear Astrophysics, the Kavli Institute for Particle Astrophysics and Cosmology, the Korean Scientist Group, the Chinese Academy of Sciences (LAMOST), Los Alamos National Laboratory, the Max-Planck-Institute for Astronomy (MPIA), the Max-Planck-Institute for Astrophysics (MPA), New Mexico State University, Ohio State University, University of Pittsburgh, University of Portsmouth, Princeton University, the United States Naval Observatory, and the University of Washington.
This research has made use of the NASA/IPAC Extragalactic Database (NED), which is operated by the Jet Propulsion Laboratory, California Institute of Technology, under contract with the National Aeronautics and Space Administration. This research project was supported by the DFG Cluster of Excellence ‘Origin and Structure of the Universe’ (www.universe-cluster.de). We acknowledge financial support from PRIN-INAF 2014 (Windy Black Holes combing galaxy evolution). A. Bongiorno and E. Piconcelli
acknowledge financial support from INAF under the contract PRIN-INAF-2012. L. Zappacosta acknowledges financial support under ASI/INAF contract
 I/037/12/0. G. Bruni acknowledges financial support under the INTEGRAL ASI-INAF agreement 2013-025.R01. F. Tombesi acknowledges support from the 2014 MIUR grant “Rita Levi Montalcini”. S. Bisogni is supported by the National Aeronautics and Space Administration through Chandra Award Number AR7-18013 X issued by the Chandra X-ray Observatory Center, which is operated by the Smithsonian Astrophysical Observatory for and on behalf of the National Aeronautics Space Administration under contract NAS8-03060
We are grateful to M. Fumana for his assistance in LBT data
reduction.

\end{acknowledgements}

\bibliographystyle{aa} 
\bibliography{bib} 

\begin{thebibliography}{110}
\expandafter\ifx\csname natexlab\endcsname\relax\def\natexlab#1{#1}\fi

\bibitem[{{Ahn} {et~al.}(2014){Ahn}, {Alexandroff}, {Allende Prieto}, {Anders},
  {Anderson}, {Anderton}, {Andrews}, {Aubourg}, {Bailey}, {Bastien}, \&
  et~al.}]{Ahn2014}
{Ahn}, C.~P., {Alexandroff}, R., {Allende Prieto}, C., {et~al.} 2014, \apjs,
  211, 17

\bibitem[{{Baskin} \& {Laor}(2005)}]{Baskin2005}
{Baskin}, A. \& {Laor}, A. 2005, \mnras, 356, 1029

\bibitem[{{Bentz} {et~al.}(2009){Bentz}, {Peterson}, {Netzer}, {Pogge}, \&
  {Vestergaard}}]{Bentz2009}
{Bentz}, M.~C., {Peterson}, B.~M., {Netzer}, H., {Pogge}, R.~W., \&
  {Vestergaard}, M. 2009, \apj, 697, 160

\bibitem[{{Bernardi} {et~al.}(2005){Bernardi}, {Sheth}, {Nichol}, {Schneider},
  \& {Brinkmann}}]{Bernardi2005}
{Bernardi}, M., {Sheth}, R.~K., {Nichol}, R.~C., {Schneider}, D.~P., \&
  {Brinkmann}, J. 2005, \aj, 129, 61

\bibitem[{{Bischetti} {et~al.}(2017){Bischetti}, {Piconcelli}, {Vietri},
  {Bongiorno}, {Fiore}, {Sani}, {Marconi}, {Duras}, {Zappacosta}, {Brusa},
  {Comastri}, {Cresci}, {Feruglio}, {Giallongo}, {La Franca}, {Mainieri},
  {Mannucci}, {Martocchia}, {Ricci}, {Schneider}, {Testa}, \&
  {Vignali}}]{Bischetti2017}
{Bischetti}, M., {Piconcelli}, E., {Vietri}, G., {et~al.} 2017, \aap, 598, A122

\bibitem[{{Bisogni} {et~al.}(2017{\natexlab{a}}){Bisogni}, {Marconi}, \&
  {Risaliti}}]{Bisogni2017}
{Bisogni}, S., {Marconi}, A., \& {Risaliti}, G. 2017{\natexlab{a}}, \mnras,
  464, 385

\bibitem[{{Bisogni} {et~al.}(2017{\natexlab{b}}){Bisogni}, {Marconi},
  {Risaliti}, \& {Lusso}}]{Bisogni2017b}
{Bisogni}, S., {Marconi}, A., {Risaliti}, G., \& {Lusso}, E.
  2017{\natexlab{b}}, ArXiv e-prints

\bibitem[{{Bongiorno} {et~al.}(2014){Bongiorno}, {Maiolino}, {Brusa},
  {Marconi}, {Piconcelli}, {Lamastra}, {Cano-D{\'{\i}}az}, {Schulze},
  {Magnelli}, {Vignali}, {Fiore}, {Menci}, {Cresci}, {La Franca}, \&
  {Merloni}}]{Bongiorno2014}
{Bongiorno}, A., {Maiolino}, R., {Brusa}, M., {et~al.} 2014, \mnras, 443, 2077

\bibitem[{{Boroson}(2002)}]{Boroson2002}
{Boroson}, T.~A. 2002, \apj, 565, 78

\bibitem[{{Boroson} \& {Green}(1992)}]{Boroson1992}
{Boroson}, T.~A. \& {Green}, R.~F. 1992, \apjs, 80, 109

\bibitem[{{Bruni} {et~al.}(2012){Bruni}, {Mack}, {Salerno},
  {Montenegro-Montes}, {Carballo}, {Benn}, {Gonz{\'a}lez-Serrano}, {Holt}, \&
  {Jim{\'e}nez-Luj{\'a}n}}]{Bruni2012}
{Bruni}, G., {Mack}, K.-H., {Salerno}, E., {et~al.} 2012, \aap, 542, A13

\bibitem[{{Brusa} {et~al.}(2015){Brusa}, {Bongiorno}, {Cresci}, {Perna},
  {Marconi}, {Mainieri}, {Maiolino}, {Salvato}, {Lusso}, {Santini}, {Comastri},
  {Fiore}, {Gilli}, {La Franca}, {Lanzuisi}, {Lutz}, {Merloni}, {Mignoli},
  {Onori}, {Piconcelli}, {Rosario}, {Vignali}, \& {Zamorani}}]{Brusa2015}
{Brusa}, M., {Bongiorno}, A., {Cresci}, G., {et~al.} 2015, \mnras, 446, 2394

\bibitem[{{Cano-D{\'{\i}}az} {et~al.}(2012){Cano-D{\'{\i}}az}, {Maiolino},
  {Marconi}, {Netzer}, {Shemmer}, \& {Cresci}}]{Cano-Diaz2012}
{Cano-D{\'{\i}}az}, M., {Maiolino}, R., {Marconi}, A., {et~al.} 2012, \aap,
  537, L8

\bibitem[{{Carniani} {et~al.}(2015){Carniani}, {Marconi}, {Maiolino},
  {Balmaverde}, {Brusa}, {Cano-D{\'{\i}}az}, {Cicone}, {Comastri}, {Cresci},
  {Fiore}, {Feruglio}, {La Franca}, {Mainieri}, {Mannucci}, {Nagao}, {Netzer},
  {Piconcelli}, {Risaliti}, {Schneider}, \& {Shemmer}}]{Carniani2015}
{Carniani}, S., {Marconi}, A., {Maiolino}, R., {et~al.} 2015, \aap, 580, A102

\bibitem[{{Cicone} {et~al.}(2014){Cicone}, {Maiolino}, {Sturm},
  {Graci{\'a}-Carpio}, {Feruglio}, {Neri}, {Aalto}, {Davies}, {Fiore},
  {Fischer}, {Garc{\'{\i}}a-Burillo}, {Gonz{\'a}lez-Alfonso},
  {Hailey-Dunsheath}, {Piconcelli}, \& {Veilleux}}]{Cicone2014}
{Cicone}, C., {Maiolino}, R., {Sturm}, E., {et~al.} 2014, \aap, 562, A21

\bibitem[{{Coatman} {et~al.}(2016){Coatman}, {Hewett}, {Banerji}, \&
  {Richards}}]{Coatman2016}
{Coatman}, L., {Hewett}, P.~C., {Banerji}, M., \& {Richards}, G.~T. 2016,
  \mnras, 461, 647

\bibitem[{{Coatman} {et~al.}(2017){Coatman}, {Hewett}, {Banerji}, {Richards},
  {Hennawi}, \& {Prochaska}}]{Coatman2017}
{Coatman}, L., {Hewett}, P.~C., {Banerji}, M., {et~al.} 2017, \mnras, 465, 2120

\bibitem[{{Corbin} \& {Boroson}(1996)}]{Corbin1996}
{Corbin}, M.~R. \& {Boroson}, T.~A. 1996, \apjs, 107, 69

\bibitem[{{Cresci} {et~al.}(2015){Cresci}, {Mainieri}, {Brusa}, {Marconi},
  {Perna}, {Mannucci}, {Piconcelli}, {Maiolino}, {Feruglio}, {Fiore},
  {Bongiorno}, {Lanzuisi}, {Merloni}, {Schramm}, {Silverman}, \&
  {Civano}}]{Cresci2015}
{Cresci}, G., {Mainieri}, V., {Brusa}, M., {et~al.} 2015, \apj, 799, 82

\bibitem[{{Davies}(2007)}]{Davies2007}
{Davies}, R.~I. 2007, \mnras, 375, 1099

\bibitem[{{Denney}(2012)}]{Denney2012}
{Denney}, K.~D. 2012, \apj, 759, 44

\bibitem[{{Di Matteo} {et~al.}(2005){Di Matteo}, {Springel}, \&
  {Hernquist}}]{DiMatteo2005}
{Di Matteo}, T., {Springel}, V., \& {Hernquist}, L. 2005, \nat, 433, 604

\bibitem[{{Diamond-Stanic} {et~al.}(2009){Diamond-Stanic}, {Fan}, {Brandt},
  {Shemmer}, {Strauss}, {Anderson}, {Carilli}, {Gibson}, {Jiang}, {Kim},
  {Richards}, {Schmidt}, {Schneider}, {Shen}, {Smith}, {Vestergaard}, \&
  {Young}}]{Diamond2009}
{Diamond-Stanic}, A.~M., {Fan}, X., {Brandt}, W.~N., {et~al.} 2009, \apj, 699,
  782

\bibitem[{{Duras} {et~al.}(2017){Duras}, {Bongiorno}, {Piconcelli}, {Bianchi},
  {Pappalardo}, {Valiante}, {Bischetti}, {Feruglio}, {Martocchia}, {Schneider},
  {Vietri}, {Vignali}, {Zappacosta}, {La Franca}, \& {Fiore}}]{Duras2017}
{Duras}, F., {Bongiorno}, A., {Piconcelli}, E., {et~al.} 2017, \aap, 604, A67

\bibitem[{{Fabian}(2012)}]{Fabian2012}
{Fabian}, A.~C. 2012, \araa, 50, 455

\bibitem[{{Fan} {et~al.}(1999){Fan}, {Strauss}, {Gunn}, {Lupton}, {Carilli},
  {Rupen}, {Schmidt}, {Moustakas}, {Davis}, {Annis}, {Bahcall}, {Brinkmann},
  {Brunner}, {Csabai}, {Doi}, {Fukugita}, {Heckman}, {Hennessy}, {Hindsley},
  {Ivezi{\'c} }, {Knapp}, {Lamb}, {Munn}, {Pauls}, {Pier}, {Rockosi},
  {Schneider}, {Szalay}, {Tucker}, \& {York}}]{Fan1999}
{Fan}, X., {Strauss}, M.~A., {Gunn}, J.~E., {et~al.} 1999, \apjl, 526, L57

\bibitem[{{Faucher-Gigu{\`e}re} \& {Quataert}(2012)}]{Faucher2012}
{Faucher-Gigu{\`e}re}, C.-A. \& {Quataert}, E. 2012, \mnras, 425, 605

\bibitem[{{Ferland} \& {Elitzur}(1984)}]{Ferland1984}
{Ferland}, G.~J. \& {Elitzur}, M. 1984, \apjl, 285, L11

\bibitem[{{Ferland} {et~al.}(2013){Ferland}, {Porter}, {van Hoof}, {Williams},
  {Abel}, {Lykins}, {Shaw}, {Henney}, \& {Stancil}}]{Ferland2013}
{Ferland}, G.~J., {Porter}, R.~L., {van Hoof}, P.~A.~M., {et~al.} 2013, \rmxaa,
  49, 137

\bibitem[{{Feruglio} {et~al.}(2015){Feruglio}, {Fiore}, {Carniani},
  {Piconcelli}, {Zappacosta}, {Bongiorno}, {Cicone}, {Maiolino}, {Marconi},
  {Menci}, {Puccetti}, \& {Veilleux}}]{Feruglio2015}
{Feruglio}, C., {Fiore}, F., {Carniani}, S., {et~al.} 2015, \aap, 583, A99

\bibitem[{{Feruglio} {et~al.}(2010){Feruglio}, {Maiolino}, {Piconcelli},
  {Menci}, {Aussel}, {Lamastra}, \& {Fiore}}]{Feruglio2010}
{Feruglio}, C., {Maiolino}, R., {Piconcelli}, E., {et~al.} 2010, \aap, 518,
  L155

\bibitem[{{Fiore} {et~al.}(2017){Fiore}, {Feruglio}, {Shankar}, {Bischetti},
  {Bongiorno}, {Brusa}, {Carniani}, {Cicone}, {Duras}, {Lamastra}, {Mainieri},
  {Marconi}, {Menci}, {Maiolino}, {Piconcelli}, {Vietri}, \&
  {Zappacosta}}]{Fiore2017}
{Fiore}, F., {Feruglio}, C., {Shankar}, F., {et~al.} 2017, \aap, 601, A143

\bibitem[{{Freudling} {et~al.}(2013){Freudling}, {Romaniello}, {Bramich},
  {Ballester}, {Forchi}, {Garc{\'{\i}}a-Dabl{\'o}}, {Moehler}, \&
  {Neeser}}]{reflex}
{Freudling}, W., {Romaniello}, M., {Bramich}, D.~M., {et~al.} 2013, \aap, 559,
  A96

\bibitem[{{Gaskell}(1982)}]{Gaskell1982}
{Gaskell}, C.~M. 1982, \apj, 263, 79

\bibitem[{{Genzel} {et~al.}(2014){Genzel}, {F{\"o}rster Schreiber}, {Rosario},
  {Lang}, {Lutz}, {Wisnioski}, {Wuyts}, {Wuyts}, {Bandara}, {Bender}, {Berta},
  {Kurk}, {Mendel}, {Tacconi}, {Wilman}, {Beifiori}, {Brammer}, {Burkert},
  {Buschkamp}, {Chan}, {Carollo}, {Davies}, {Eisenhauer}, {Fabricius},
  {Fossati}, {Kriek}, {Kulkarni}, {Lilly}, {Mancini}, {Momcheva}, {Naab},
  {Nelson}, {Renzini}, {Saglia}, {Sharples}, {Sternberg}, {Tacchella}, \& {van
  Dokkum}}]{Genzel2014}
{Genzel}, R., {F{\"o}rster Schreiber}, N.~M., {Rosario}, D., {et~al.} 2014,
  \apj, 796, 7

\bibitem[{{Gofford} {et~al.}(2015){Gofford}, {Reeves}, {McLaughlin}, {Braito},
  {Turner}, {Tombesi}, \& {Cappi}}]{Gofford2015}
{Gofford}, J., {Reeves}, J.~N., {McLaughlin}, D.~E., {et~al.} 2015, \mnras,
  451, 4169

\bibitem[{{Gofford} {et~al.}(2013){Gofford}, {Reeves}, {Tombesi}, {Braito},
  {Turner}, {Miller}, \& {Cappi}}]{Gofford2013}
{Gofford}, J., {Reeves}, J.~N., {Tombesi}, F., {et~al.} 2013, \mnras, 430, 60

\bibitem[{{Granato} {et~al.}(2004){Granato}, {De Zotti}, {Silva}, {Bressan}, \&
  {Danese}}]{Granato2004}
{Granato}, G.~L., {De Zotti}, G., {Silva}, L., {Bressan}, A., \& {Danese}, L.
  2004, \apj, 600, 580

\bibitem[{{Gunn} {et~al.}(2006){Gunn}, {Siegmund}, {Mannery}, {Owen}, {Hull},
  {Leger}, {Carey}, {Knapp}, {York}, {Boroski}, {Kent}, {Lupton}, {Rockosi},
  {Evans}, {Waddell}, {Anderson}, {Annis}, {Barentine}, {Bartoszek}, {Bastian},
  {Bracker}, {Brewington}, {Briegel}, {Brinkmann}, {Brown}, {Carr},
  {Czarapata}, {Drennan}, {Dombeck}, {Federwitz}, {Gillespie}, {Gonzales},
  {Hansen}, {Harvanek}, {Hayes}, {Jordan}, {Kinney}, {Klaene}, {Kleinman},
  {Kron}, {Kresinski}, {Lee}, {Limmongkol}, {Lindenmeyer}, {Long}, {Loomis},
  {McGehee}, {Mantsch}, {Neilsen}, {Neswold}, {Newman}, {Nitta}, {Peoples},
  {Pier}, {Prieto}, {Prosapio}, {Rivetta}, {Schneider}, {Snedden}, \&
  {Wang}}]{Gunn2006}
{Gunn}, J.~E., {Siegmund}, W.~A., {Mannery}, E.~J., {et~al.} 2006, \aj, 131,
  2332

\bibitem[{{Hall} {et~al.}(2002){Hall}, {Anderson}, {Strauss}, {York},
  {Richards}, {Fan}, {Knapp}, {Schneider}, {Vanden Berk}, {Geballe}, {Bauer},
  {Becker}, {Davis}, {Rix}, {Nichol}, {Bahcall}, {Brinkmann}, {Brunner},
  {Connolly}, {Csabai}, {Doi}, {Fukugita}, {Gunn}, {Haiman}, {Harvanek},
  {Heckman}, {Hennessy}, {Inada}, {Ivezi{\'c}}, {Johnston}, {Kleinman},
  {Krolik}, {Krzesinski}, {Kunszt}, {Lamb}, {Long}, {Lupton}, {Miknaitis},
  {Munn}, {Narayanan}, {Neilsen}, {Newman}, {Nitta}, {Okamura}, {Pentericci},
  {Pier}, {Schlegel}, {Snedden}, {Szalay}, {Thakar}, {Tsvetanov}, {White}, \&
  {Zheng}}]{Hall2002}
{Hall}, P.~B., {Anderson}, S.~F., {Strauss}, M.~A., {et~al.} 2002, \apjs, 141,
  267

\bibitem[{{Hamann} {et~al.}(2017){Hamann}, {Zakamska}, {Ross}, {Paris},
  {Alexandroff}, {Villforth}, {Richards}, {Herbst}, {Brandt}, {Cook}, {Denney},
  {Greene}, {Schneider}, \& {Strauss}}]{Hamann2017}
{Hamann}, F., {Zakamska}, N.~L., {Ross}, N., {et~al.} 2017, \mnras, 464, 3431

\bibitem[{{Harrison} {et~al.}(2014){Harrison}, {Alexander}, {Mullaney}, \&
  {Swinbank}}]{Harrison2014}
{Harrison}, C.~M., {Alexander}, D.~M., {Mullaney}, J.~R., \& {Swinbank}, A.~M.
  2014, \mnras, 441, 3306

\bibitem[{{Harrison} {et~al.}(2012){Harrison}, {Alexander}, {Swinbank},
  {Smail}, {Alaghband-Zadeh}, {Bauer}, {Chapman}, {Del Moro}, {Hickox},
  {Ivison}, {Men{\'e}ndez-Delmestre}, {Mullaney}, \& {Nesvadba}}]{Harrison2012}
{Harrison}, C.~M., {Alexander}, D.~M., {Swinbank}, A.~M., {et~al.} 2012,
  \mnras, 426, 1073

\bibitem[{{Hewett} \& {Wild}(2010)}]{Hewett2010}
{Hewett}, P.~C. \& {Wild}, V. 2010, \mnras, 405, 2302

\bibitem[{{Just} {et~al.}(2007){Just}, {Brandt}, {Shemmer}, {Steffen},
  {Schneider}, {Chartas}, \& {Garmire}}]{Just2007}
{Just}, D.~W., {Brandt}, W.~N., {Shemmer}, O., {et~al.} 2007, \apj, 665, 1004

\bibitem[{{Kashi} {et~al.}(2013){Kashi}, {Proga}, {Nagamine}, {Greene}, \&
  {Barth}}]{Kashi2013}
{Kashi}, A., {Proga}, D., {Nagamine}, K., {Greene}, J., \& {Barth}, A.~J. 2013,
  \apj, 778, 50

\bibitem[{{Kaspi} {et~al.}(2000){Kaspi}, {Smith}, {Netzer}, {Maoz}, {Jannuzi},
  \& {Giveon}}]{Kaspi2000}
{Kaspi}, S., {Smith}, P.~S., {Netzer}, H., {et~al.} 2000, \apj, 533, 631

\bibitem[{{King} \& {Pounds}(2015)}]{King2015}
{King}, A. \& {Pounds}, K. 2015, \araa, 53, 115

\bibitem[{{King}(2010)}]{King2010}
{King}, A.~R. 2010, \mnras, 402, 1516

\bibitem[{{Kormendy} \& {Richstone}(1995)}]{Kormendy1995}
{Kormendy}, J. \& {Richstone}, D. 1995, \araa, 33, 581

\bibitem[{{Laor} \& {Brandt}(2002)}]{Laor2002}
{Laor}, A. \& {Brandt}, W.~N. 2002, \apj, 569, 641

\bibitem[{{Leighly}(2004)}]{Leighly2004}
{Leighly}, K.~M. 2004, \apj, 611, 125

\bibitem[{{Luo} {et~al.}(2015){Luo}, {Brandt}, {Hall}, {Wu}, {Anderson},
  {Garmire}, {Gibson}, {Plotkin}, {Richards}, {Schneider}, {Shemmer}, \&
  {Shen}}]{Luo2015}
{Luo}, B., {Brandt}, W.~N., {Hall}, P.~B., {et~al.} 2015, \apj, 805, 122

\bibitem[{{Lusso} {et~al.}(2010){Lusso}, {Comastri}, {Vignali}, {Zamorani},
  {Brusa}, {Gilli}, {Iwasawa}, {Salvato}, {Civano}, {Elvis}, {Merloni},
  {Bongiorno}, {Trump}, {Koekemoer}, {Schinnerer}, {Le Floc'h}, {Cappelluti},
  {Jahnke}, {Sargent}, {Silverman}, {Mainieri}, {Fiore}, {Bolzonella}, {Le
  F{\`e}vre}, {Garilli}, {Iovino}, {Kneib}, {Lamareille}, {Lilly}, {Mignoli},
  {Scodeggio}, \& {Vergani}}]{Lusso2010}
{Lusso}, E., {Comastri}, A., {Vignali}, C., {et~al.} 2010, \aap, 512, A34

\bibitem[{{Markwardt}(2009)}]{Markwardt2009}
{Markwardt}, C.~B. 2009, in Astronomical Society of the Pacific Conference
  Series, Vol. 411, Astronomical Data Analysis Software and Systems XVIII, ed.
  D.~A. {Bohlender}, D.~{Durand}, \& P.~{Dowler}, 251

\bibitem[{{Martocchia} {et~al.}(2017){Martocchia}, {Piconcelli}, {Zappacosta},
  {Duras}, {Vietri}, {Vignali}, {Bianchi}, {Bischetti}, {Bongiorno}, {Brusa},
  {Lanzuisi}, {Marconi}, {Mathur}, {Miniutti}, {Nicastro}, {Bruni}, \&
  {Fiore}}]{Martocchia2017}
{Martocchia}, S., {Piconcelli}, E., {Zappacosta}, L., {et~al.} 2017, ArXiv
  e-prints

\bibitem[{{Marziani} {et~al.}(2016){Marziani}, {Mart{\'{\i}}nez Carballo},
  {Sulentic}, {Del Olmo}, {Stirpe}, \& {Dultzin}}]{Marziani2016}
{Marziani}, P., {Mart{\'{\i}}nez Carballo}, M.~A., {Sulentic}, J.~W., {et~al.}
  2016, \apss, 361, 29

\bibitem[{{Marziani} {et~al.}(2010){Marziani}, {Sulentic}, {Negrete},
  {Dultzin}, {Zamfir}, \& {Bachev}}]{Marziani2010}
{Marziani}, P., {Sulentic}, J.~W., {Negrete}, C.~A., {et~al.} 2010, \mnras,
  409, 1033

\bibitem[{{Marziani} {et~al.}(2009){Marziani}, {Sulentic}, {Stirpe}, {Zamfir},
  \& {Calvani}}]{Marziani2009}
{Marziani}, P., {Sulentic}, J.~W., {Stirpe}, G.~M., {Zamfir}, S., \& {Calvani},
  M. 2009, \aap, 495, 83

\bibitem[{{Marziani} {et~al.}(2001){Marziani}, {Sulentic}, {Zwitter},
  {Dultzin-Hacyan}, \& {Calvani}}]{Marziani2001}
{Marziani}, P., {Sulentic}, J.~W., {Zwitter}, T., {Dultzin-Hacyan}, D., \&
  {Calvani}, M. 2001, \apj, 558, 553

\bibitem[{{Matsuoka} {et~al.}(2013){Matsuoka}, {Silverman}, {Schramm},
  {Steinhardt}, {Nagao}, {Kartaltepe}, {Sanders}, {Treister}, {Hasinger},
  {Akiyama}, {Ohta}, {Ueda}, {Bongiorno}, {Brandt}, {Brusa}, {Capak}, {Civano},
  {Comastri}, {Elvis}, {Lilly}, {Mainieri}, {Masters}, {Mignoli}, {Salvato},
  {Trump}, {Taniguchi}, {Zamorani}, {Alexander}, \&
  {Schawinski}}]{Matsuoka2013}
{Matsuoka}, K., {Silverman}, J.~D., {Schramm}, M., {et~al.} 2013, \apj, 771, 64

\bibitem[{{McGill} {et~al.}(2008){McGill}, {Woo}, {Treu}, \&
  {Malkan}}]{McGill2008}
{McGill}, K.~L., {Woo}, J.-H., {Treu}, T., \& {Malkan}, M.~A. 2008, \apj, 673,
  703

\bibitem[{{Menci} {et~al.}(2008){Menci}, {Fiore}, {Puccetti}, \&
  {Cavaliere}}]{Menci2008}
{Menci}, N., {Fiore}, F., {Puccetti}, S., \& {Cavaliere}, A. 2008, \apj, 686,
  219

\bibitem[{{Nagao} {et~al.}(2006){Nagao}, {Marconi}, \& {Maiolino}}]{Nagao2006}
{Nagao}, T., {Marconi}, A., \& {Maiolino}, R. 2006, \aap, 447, 157

\bibitem[{{Nardini} {et~al.}(2015){Nardini}, {Reeves}, {Gofford}, {Harrison},
  {Risaliti}, {Braito}, {Costa}, {Matzeu}, {Walton}, {Behar}, {Boggs},
  {Christensen}, {Craig}, {Hailey}, {Matt}, {Miller}, {O'Brien}, {Stern},
  {Turner}, \& {Ward}}]{Nardini2015}
{Nardini}, E., {Reeves}, J.~N., {Gofford}, J., {et~al.} 2015, Science, 347, 860

\bibitem[{{Netzer}(1990)}]{Netzer1990}
{Netzer}, H. 1990, in Active Galactic Nuclei, ed. R.~D. {Blandford},
  H.~{Netzer}, L.~{Woltjer}, T.~J.-L. {Courvoisier}, \& M.~{Mayor}, 57--160

\bibitem[{{Netzer} \& {Laor}(1993)}]{Netzer1993}
{Netzer}, H. \& {Laor}, A. 1993, \apjl, 404, L51

\bibitem[{{Peterson}(1997)}]{Peterson1997}
{Peterson}, B.~M. 1997, {An Introduction to Active Galactic Nuclei}

\bibitem[{{Plotkin} {et~al.}(2010){Plotkin}, {Anderson}, {Brandt},
  {Diamond-Stanic}, {Fan}, {MacLeod}, {Schneider}, \& {Shemmer}}]{Plotkin2010}
{Plotkin}, R.~M., {Anderson}, S.~F., {Brandt}, W.~N., {et~al.} 2010, \apj, 721,
  562

\bibitem[{{Plotkin} {et~al.}(2015){Plotkin}, {Shemmer}, {Trakhtenbrot},
  {Anderson}, {Brandt}, {Fan}, {Gallo}, {Lira}, {Luo}, {Richards}, {Schneider},
  {Strauss}, \& {Wu}}]{Plotkin2015}
{Plotkin}, R.~M., {Shemmer}, O., {Trakhtenbrot}, B., {et~al.} 2015, \apj, 805,
  123

\bibitem[{{Prevot} {et~al.}(1984){Prevot}, {Lequeux}, {Prevot}, {Maurice}, \&
  {Rocca-Volmerange}}]{Prevot1984}
{Prevot}, M.~L., {Lequeux}, J., {Prevot}, L., {Maurice}, E., \&
  {Rocca-Volmerange}, B. 1984, \aap, 132, 389

\bibitem[{{Ricci} {et~al.}(2017){Ricci}, {La Franca}, {Onori}, \&
  {Bianchi}}]{Ricci2017_1}
{Ricci}, F., {La Franca}, F., {Onori}, F., \& {Bianchi}, S. 2017, \aap, 598,
  A51

\bibitem[{{Richards} {et~al.}(2011){Richards}, {Kruczek}, {Gallagher}, {Hall},
  {Hewett}, {Leighly}, {Deo}, {Kratzer}, \& {Shen}}]{Richards2011}
{Richards}, G.~T., {Kruczek}, N.~E., {Gallagher}, S.~C., {et~al.} 2011, \aj,
  141, 167

\bibitem[{{Richards} {et~al.}(2006){Richards}, {Lacy}, {Storrie-Lombardi},
  {Hall}, {Gallagher}, {Hines}, {Fan}, {Papovich}, {Vanden Berk}, {Trammell},
  {Schneider}, {Vestergaard}, {York}, {Jester}, {Anderson}, {Budav{\'a}ri}, \&
  {Szalay}}]{Richards2006}
{Richards}, G.~T., {Lacy}, M., {Storrie-Lombardi}, L.~J., {et~al.} 2006, \apjs,
  166, 470

\bibitem[{{Richards} {et~al.}(2002){Richards}, {Vanden Berk}, {Reichard},
  {Hall}, {Schneider}, {SubbaRao}, {Thakar}, \& {York}}]{Richards2002}
{Richards}, G.~T., {Vanden Berk}, D.~E., {Reichard}, T.~A., {et~al.} 2002, \aj,
  124, 1

\bibitem[{{Risaliti} {et~al.}(2011){Risaliti}, {Salvati}, \&
  {Marconi}}]{Risaliti2011}
{Risaliti}, G., {Salvati}, M., \& {Marconi}, A. 2011, \mnras, 411, 2223

\bibitem[{{Runnoe} {et~al.}(2012){Runnoe}, {Brotherton}, \&
  {Shang}}]{Runnoe2012}
{Runnoe}, J.~C., {Brotherton}, M.~S., \& {Shang}, Z. 2012, \mnras, 427, 1800

\bibitem[{{Rupke} \& {Veilleux}(2015)}]{Rupke2015}
{Rupke}, D.~S.~N. \& {Veilleux}, S. 2015, \apj, 801, 126

\bibitem[{{Shen}(2016)}]{Shen2016}
{Shen}, Y. 2016, \apj, 817, 55

\bibitem[{{Shen} \& {Ho}(2014)}]{Shen2014}
{Shen}, Y. \& {Ho}, L.~C. 2014, \nat, 513, 210

\bibitem[{{Shen} \& {Liu}(2012)}]{Shen2012}
{Shen}, Y. \& {Liu}, X. 2012, \apj, 753, 125

\bibitem[{{Shen} {et~al.}(2011){Shen}, {Richards}, {Strauss}, {Hall},
  {Schneider}, {Snedden}, {Bizyaev}, {Brewington}, {Malanushenko},
  {Malanushenko}, {Oravetz}, {Pan}, \& {Simmons}}]{Shen2011}
{Shen}, Y., {Richards}, G.~T., {Strauss}, M.~A., {et~al.} 2011, \apjs, 194, 45

\bibitem[{{Silk} \& {Rees}(1998)}]{Silk1998}
{Silk}, J. \& {Rees}, M.~J. 1998, \aap, 331, L1

\bibitem[{{Skrutskie} {et~al.}(2006){Skrutskie}, {Cutri}, {Stiening},
  {Weinberg}, {Schneider}, {Carpenter}, {Beichman}, {Capps}, {Chester},
  {Elias}, {Huchra}, {Liebert}, {Lonsdale}, {Monet}, {Price}, {Seitzer},
  {Jarrett}, {Kirkpatrick}, {Gizis}, {Howard}, {Evans}, {Fowler}, {Fullmer},
  {Hurt}, {Light}, {Kopan}, {Marsh}, {McCallon}, {Tam}, {Van Dyk}, \&
  {Wheelock}}]{Skrutskie2006}
{Skrutskie}, M.~F., {Cutri}, R.~M., {Stiening}, R., {et~al.} 2006, \aj, 131,
  1163

\bibitem[{{Spoon} {et~al.}(2013){Spoon}, {Farrah}, {Lebouteiller},
  {Gonz{\'a}lez-Alfonso}, {Bernard-Salas}, {Urrutia}, {Rigopoulou},
  {Westmoquette}, {Smith}, {Afonso}, {Pearson}, {Cormier}, {Efstathiou},
  {Borys}, {Verma}, {Etxaluze}, \& {Clements}}]{Spoon2013}
{Spoon}, H.~W.~W., {Farrah}, D., {Lebouteiller}, V., {et~al.} 2013, \apj, 775,
  127

\bibitem[{{Steffen} {et~al.}(2006){Steffen}, {Strateva}, {Brandt}, {Alexander},
  {Koekemoer}, {Lehmer}, {Schneider}, \& {Vignali}}]{Steffen2006}
{Steffen}, A.~T., {Strateva}, I., {Brandt}, W.~N., {et~al.} 2006, \aj, 131,
  2826

\bibitem[{{Sulentic} {et~al.}(2007){Sulentic}, {Bachev}, {Marziani}, {Negrete},
  \& {Dultzin}}]{Sulentic2007}
{Sulentic}, J.~W., {Bachev}, R., {Marziani}, P., {Negrete}, C.~A., \&
  {Dultzin}, D. 2007, \apj, 666, 757

\bibitem[{{Sulentic} {et~al.}(2017){Sulentic}, {del Olmo}, {Marziani},
  {Mart{\'{\i}}nez-Carballo}, {D'Onofrio}, {Dultzin}, {Perea},
  {Mart{\'{\i}}nez-Aldama}, {Negrete}, {Stirpe}, \& {Zamfir}}]{Sulentic2017}
{Sulentic}, J.~W., {del Olmo}, A., {Marziani}, P., {et~al.} 2017, \aap, 608,
  A122

\bibitem[{{Sulentic} {et~al.}(2000){Sulentic}, {Marziani}, \&
  {Dultzin-Hacyan}}]{Sulentic2000}
{Sulentic}, J.~W., {Marziani}, P., \& {Dultzin-Hacyan}, D. 2000, \araa, 38, 521

\bibitem[{{Sulentic} {et~al.}(2006){Sulentic}, {Repetto}, {Stirpe}, {Marziani},
  {Dultzin-Hacyan}, \& {Calvani}}]{Sulentic2006}
{Sulentic}, J.~W., {Repetto}, P., {Stirpe}, G.~M., {et~al.} 2006, \aap, 456,
  929

\bibitem[{{Sulentic} {et~al.}(2004){Sulentic}, {Stirpe}, {Marziani}, {Zamanov},
  {Calvani}, \& {Braito}}]{Sulentic2004}
{Sulentic}, J.~W., {Stirpe}, G.~M., {Marziani}, P., {et~al.} 2004, \aap, 423,
  121

\bibitem[{{Tang} {et~al.}(2012){Tang}, {Shang}, {Gu}, {Brotherton}, \&
  {Runnoe}}]{Tang2012}
{Tang}, B., {Shang}, Z., {Gu}, Q., {Brotherton}, M.~S., \& {Runnoe}, J.~C.
  2012, \apjs, 201, 38

\bibitem[{{Tombesi} {et~al.}(2012){Tombesi}, {Cappi}, {Reeves}, \&
  {Braito}}]{Tombesi2012}
{Tombesi}, F., {Cappi}, M., {Reeves}, J.~N., \& {Braito}, V. 2012, \mnras, 422,
  L1

\bibitem[{{Tombesi} {et~al.}(2013{\natexlab{a}}){Tombesi}, {Cappi}, {Reeves},
  {Nemmen}, {Braito}, {Gaspari}, \& {Reynolds}}]{TombesiCappi2013}
{Tombesi}, F., {Cappi}, M., {Reeves}, J.~N., {et~al.} 2013{\natexlab{a}},
  \mnras, 430, 1102

\bibitem[{{Tombesi} {et~al.}(2013{\natexlab{b}}){Tombesi}, {Cappi}, {Reeves},
  {Nemmen}, {Braito}, {Gaspari}, \& {Reynolds}}]{Tombesi2013}
{Tombesi}, F., {Cappi}, M., {Reeves}, J.~N., {et~al.} 2013{\natexlab{b}},
  \mnras, 430, 1102

\bibitem[{{Tombesi} {et~al.}(2010){Tombesi}, {Cappi}, {Reeves}, {Palumbo},
  {Yaqoob}, {Braito}, \& {Dadina}}]{Tombesi2010}
{Tombesi}, F., {Cappi}, M., {Reeves}, J.~N., {et~al.} 2010, \aap, 521, A57

\bibitem[{{Tombesi} {et~al.}(2015){Tombesi}, {Mel{\'e}ndez}, {Veilleux},
  {Reeves}, {Gonz{\'a}lez-Alfonso}, \& {Reynolds}}]{Tombesi2015}
{Tombesi}, F., {Mel{\'e}ndez}, M., {Veilleux}, S., {et~al.} 2015, \nat, 519,
  436

\bibitem[{{Tombesi} {et~al.}(2014){Tombesi}, {Tazaki}, {Mushotzky}, {Ueda},
  {Cappi}, {Gofford}, {Reeves}, \& {Guainazzi}}]{Tombesi2014}
{Tombesi}, F., {Tazaki}, F., {Mushotzky}, R.~F., {et~al.} 2014, \mnras, 443,
  2154

\bibitem[{{Trakhtenbrot} \& {Netzer}(2012)}]{Trakhtenbrot2012}
{Trakhtenbrot}, B. \& {Netzer}, H. 2012, \mnras, 427, 3081

\bibitem[{{Tsuzuki} {et~al.}(2006){Tsuzuki}, {Kawara}, {Yoshii}, {Oyabu},
  {Tanab{\'e}}, \& {Matsuoka}}]{Tsuzuki2006}
{Tsuzuki}, Y., {Kawara}, K., {Yoshii}, Y., {et~al.} 2006, \apj, 650, 57

\bibitem[{{Vacca} {et~al.}(2003){Vacca}, {Cushing}, \& {Rayner}}]{Vacca2003}
{Vacca}, W.~D., {Cushing}, M.~C., \& {Rayner}, J.~T. 2003, \pasp, 115, 389

\bibitem[{{Veilleux} {et~al.}(2017){Veilleux}, {Bolatto}, {Tombesi},
  {Mel{\'e}ndez}, {Sturm}, {Gonz{\'a}lez-Alfonso}, {Fischer}, \&
  {Rupke}}]{Veilleux2017}
{Veilleux}, S., {Bolatto}, A., {Tombesi}, F., {et~al.} 2017, \apj, 843, 18

\bibitem[{{V{\'e}ron-Cetty} {et~al.}(2004){V{\'e}ron-Cetty}, {Joly}, \&
  {V{\'e}ron}}]{Veron2004}
{V{\'e}ron-Cetty}, M.-P., {Joly}, M., \& {V{\'e}ron}, P. 2004, \aap, 417, 515

\bibitem[{{Vestergaard} \& {Wilkes}(2001)}]{Vestergaard2001}
{Vestergaard}, M. \& {Wilkes}, B.~J. 2001, \apjs, 134, 1

\bibitem[{{Vignali} {et~al.}(2003){Vignali}, {Brandt}, \&
  {Schneider}}]{Vignali2003}
{Vignali}, C., {Brandt}, W.~N., \& {Schneider}, D.~P. 2003, \aj, 125, 433

\bibitem[{{Weymann} {et~al.}(1991){Weymann}, {Morris}, {Foltz}, \&
  {Hewett}}]{Weymann1991}
{Weymann}, R.~J., {Morris}, S.~L., {Foltz}, C.~B., \& {Hewett}, P.~C. 1991,
  \apj, 373, 23

\bibitem[{{Wu} {et~al.}(2011){Wu}, {Brandt}, {Hall}, {Gibson}, {Richards},
  {Schneider}, {Shemmer}, {Just}, \& {Schmidt}}]{Wu2011}
{Wu}, J., {Brandt}, W.~N., {Hall}, P.~B., {et~al.} 2011, \apj, 736, 28

\bibitem[{{Wu} {et~al.}(2009){Wu}, {Vanden Berk}, {Brandt}, {Schneider},
  {Gibson}, \& {Wu}}]{Wu2009}
{Wu}, J., {Vanden Berk}, D.~E., {Brandt}, W.~N., {et~al.} 2009, \apj, 702, 767

\bibitem[{{Young} {et~al.}(2007){Young}, {Axon}, {Robinson}, {Hough}, \&
  {Smith}}]{Young2007}
{Young}, S., {Axon}, D.~J., {Robinson}, A., {Hough}, J.~H., \& {Smith}, J.~E.
  2007, \nat, 450, 74

\bibitem[{{Zubovas} \& {King}(2012)}]{Zubovas2012}
{Zubovas}, K. \& {King}, A. 2012, \apjl, 745, L34

\end{thebibliography}

\appendix

\section{SDSS J1538+0855}\label{sec:app_1538}

We present here a detailed analysis of the source J1538+0855 since is the only object exhibiting both NLR and BLR winds.

As reported in Sect. \ref{sec:ekin} we measured the CIV outflow properties based on the so-called outflow component. We estimate the gas mass of the outflow M$\rm_{ion}$ = 590 M$\rm_{\odot}$, the maximum velocity of the outflow as representative of the bulk velocity of the outflow v$\rm_{CIV}^{max}$ = 13,400 \kms , the ionized gas mass rate $\rm\dot{M}\rm_{ion}$ = 24 M$\rm_{\odot}$ yr$^{-1}$ and the kinetic power $\rm\dot{E}\rm_{kin}$ = 1.4 $\times$ 10$^{45}$ \ergs .
We have also measured the properties of the CIV outflow considering the entire CIV profile, with the \vciv\ as representative of the outflow velocity. We estimate M$\rm_{ion}$ = 1210 M$\rm_{\odot}$, a ionized gas mass rate $\rm\dot{M}\rm_{ion}$ = 8 M$\rm_{\odot}$ yr$^{-1}$ and a kinetic power $\rm\dot{E}\rm_{kin}$ = 12 $\times$ 10$^{42}$ \ergs .

As mentioned in Sect. \ref{sec:analysis}, we found a broad ($\sim$2,000 \kms) blueshifted (4987 \AA) [OIII] component, indicative of outflow.
In order to characterize the ionized outflowing gas, we calculated the mass rate $\rm \dot{M}$ and the kinetic power $\rm \dot{E}_{\rm kin}$, assuming the outflow model discussed in Paper I and using the [OIII] emission line. According to the equations (2), (5) and (6) in Paper I, we derived   $\rm {M}_{ion}$ = 4.2 $\times$ 10$^{8}$ M$_{\odot}$,  $\rm \dot{M}_{ion}$ = 530 M$_\odot$ yr$^{-1}$ and $\rm \dot{E}_{\rm kin}$ = 1.4 $\times$ 10$^{45}$ \ergs\ of the ionized outflow, assuming an electron density of n$_{e}$ = 200 cm$^{-3}$, a solar metallicity and an outflow velocity defined as v$_{max}$ = $\rm|\Delta v|$ + 2$\sigma^{broad}_{[OIII]}$, where $\rm\Delta$v is the velocity shift between the broad [OIII] component and the systemic [OIII] assumed at 5007 \AA\ and $\sigma^{broad}_{[OIII]}$ is the velocity dispersion of the broad [OIII] component (v$_{max}$ = 2,900 \kms). We refer to Paper I for more details about the assumptions used in the calculation of the outflow parameters.

We investigated how the values of  $\rm \dot{M}_{ion}^{[OIII]}$ and $\rm \dot{E}_{\rm kin}^{[OIII]}$ are sensitive to different assumed parameters. As in paper I, the upper bound corresponds to the assumption of n$_e$ = 80 cm$^{-3}$, (as in \citealt{Genzel2014}) and we determined the $\rm \dot{M}_{ion}^{[OIII]}$ = 1330 M$_\odot$ yr$^{-1}$ and $\rm \dot{E}_{\rm kin}^{[OIII]}$ = 3.5 $\times$ 10$^{45}$ \ergs , while the lower bound corresponds to n$_e$ = 1000 cm$^{-3}$ (typical value for the NLR, \citealt{Peterson1997}) and velocity of W$_{80}$/1.3 (as in \citealt{Harrison2014}), where W$_{80}$ is the velocity width of the line at the 80\% of the line flux. We derived a lower bound of $\rm \dot{M}_{ion}^{[OIII]}$  = 60 M$_\odot$ yr$^{-1}$ and  $\rm \dot{E}_{\rm kin}^{[OIII]}$ = 5 $\times$ 10$^{43}$ \ergs .


\clearpage
\section{Optical rest-frame spectra of 13 WISSH QSOs}\label{sec:app_lbt}
We present the LBT/LUCI and SINFONI optical rest-frame spectra in the wavelength range of the H$\beta$-[OIII] emission lines for the 13 WISSH QSOs analyzed in the present paper (see Table \ref{tab:data1}). 
The red lines show the best-fit of the spectra. Green curves refer to the core component associated with the NLR emission of H$\beta$ emission line. If present, blue curves refer to the broad blueshifted emission of [OIII]$\lambda\lambda$4959,5007 \AA , indicative of outflow.
Gold curves indicate the broad component of H$\beta$ associated with BLR emission. FeII emission is indicated in magenta. 
Grey band indicates the region excluded from the fit because of the presence of sky line residual.

\begin{minipage}{0.5\textwidth}
\includegraphics[width=1\columnwidth]{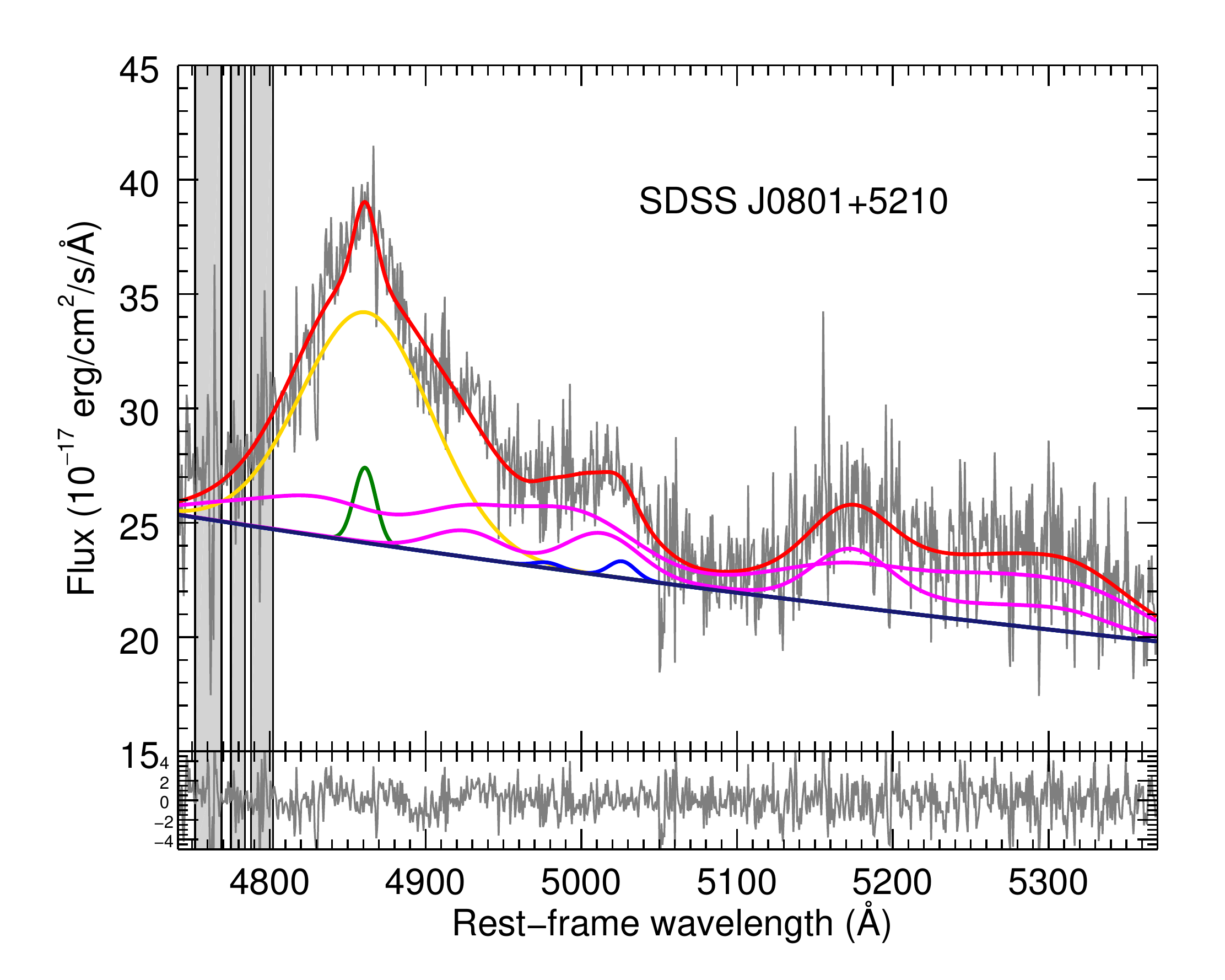}
\includegraphics[width=1\columnwidth]{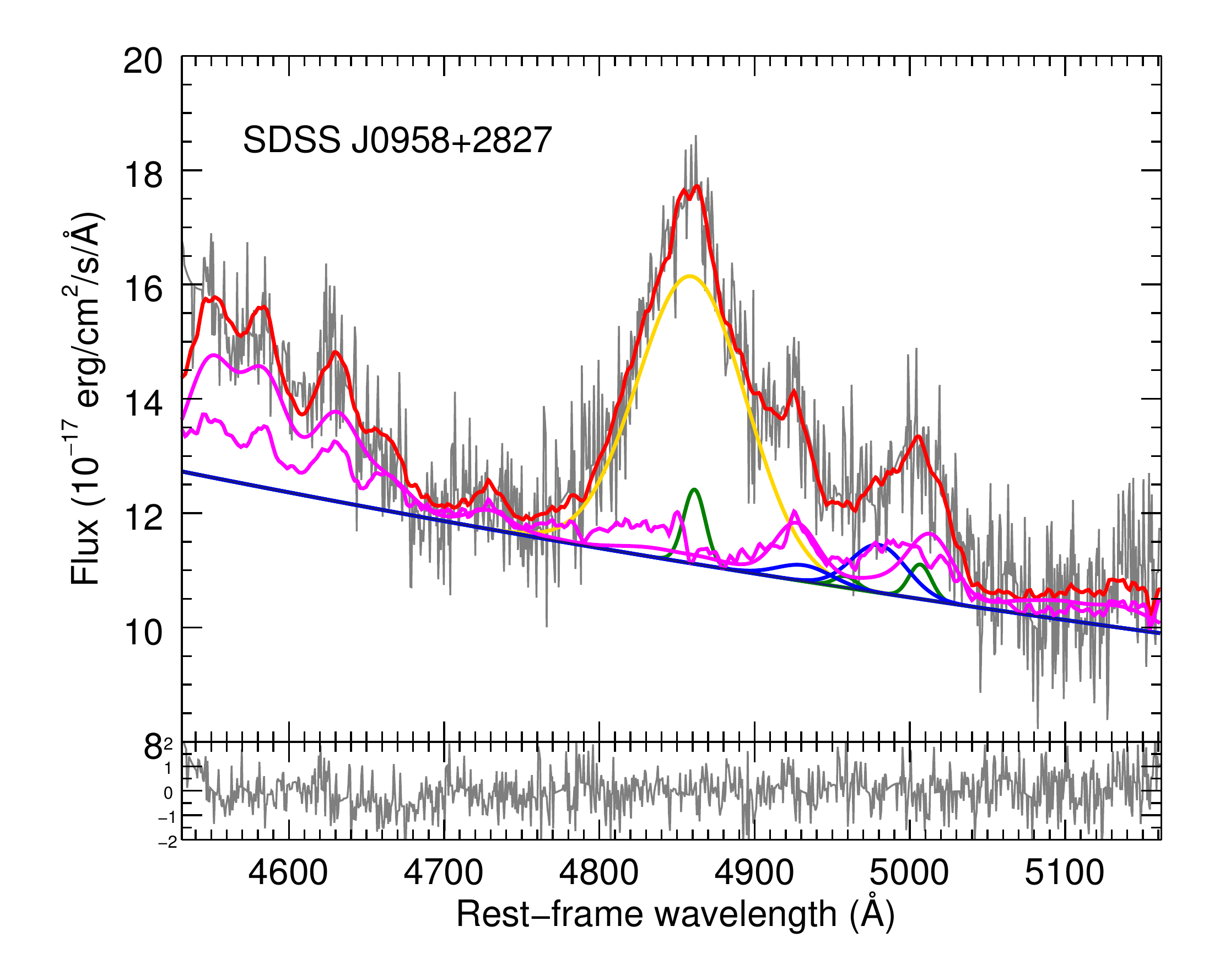}
\end{minipage}

\begin{minipage}{0.5\textwidth}
\includegraphics[width=1\columnwidth]{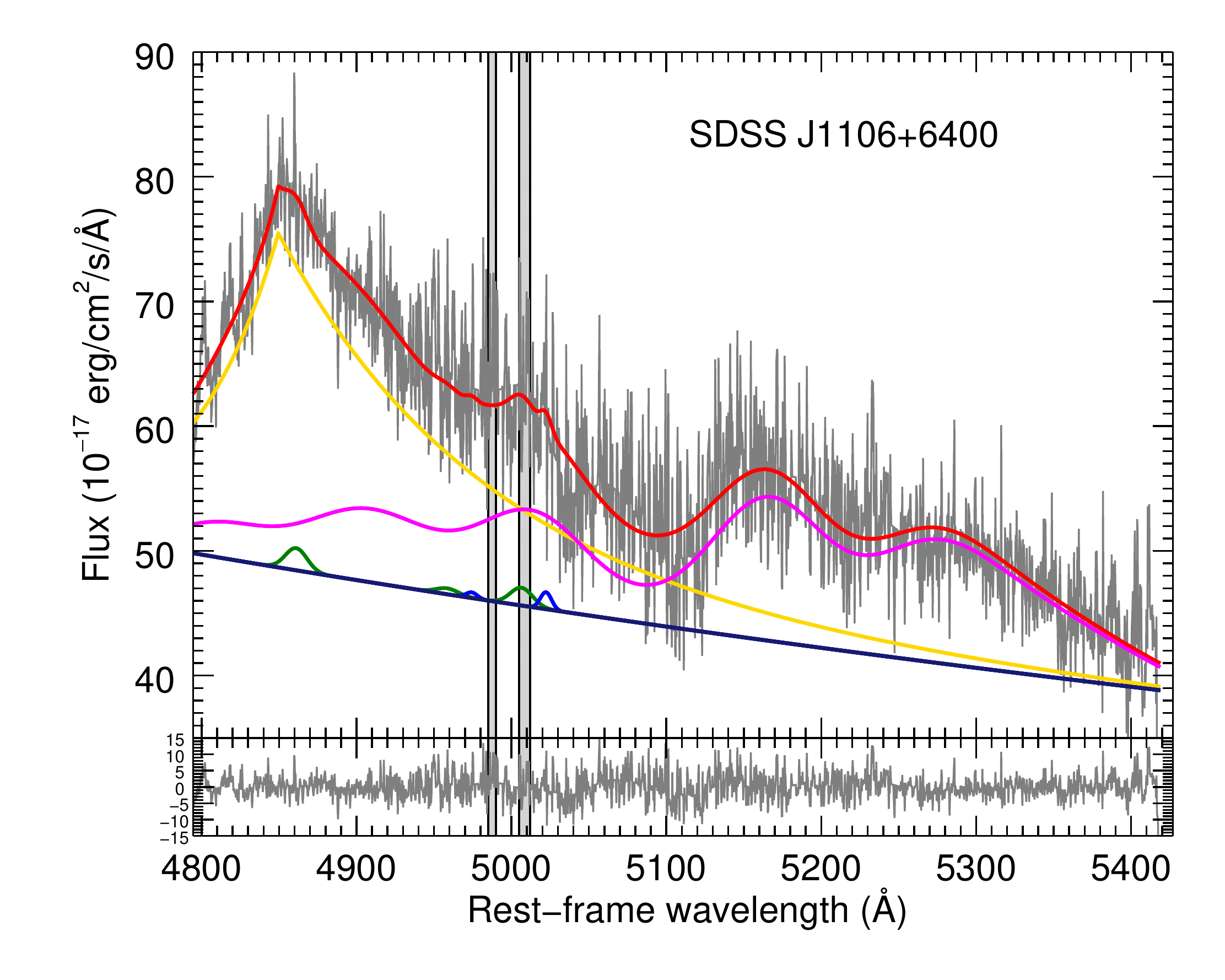}
\includegraphics[width=1\columnwidth]{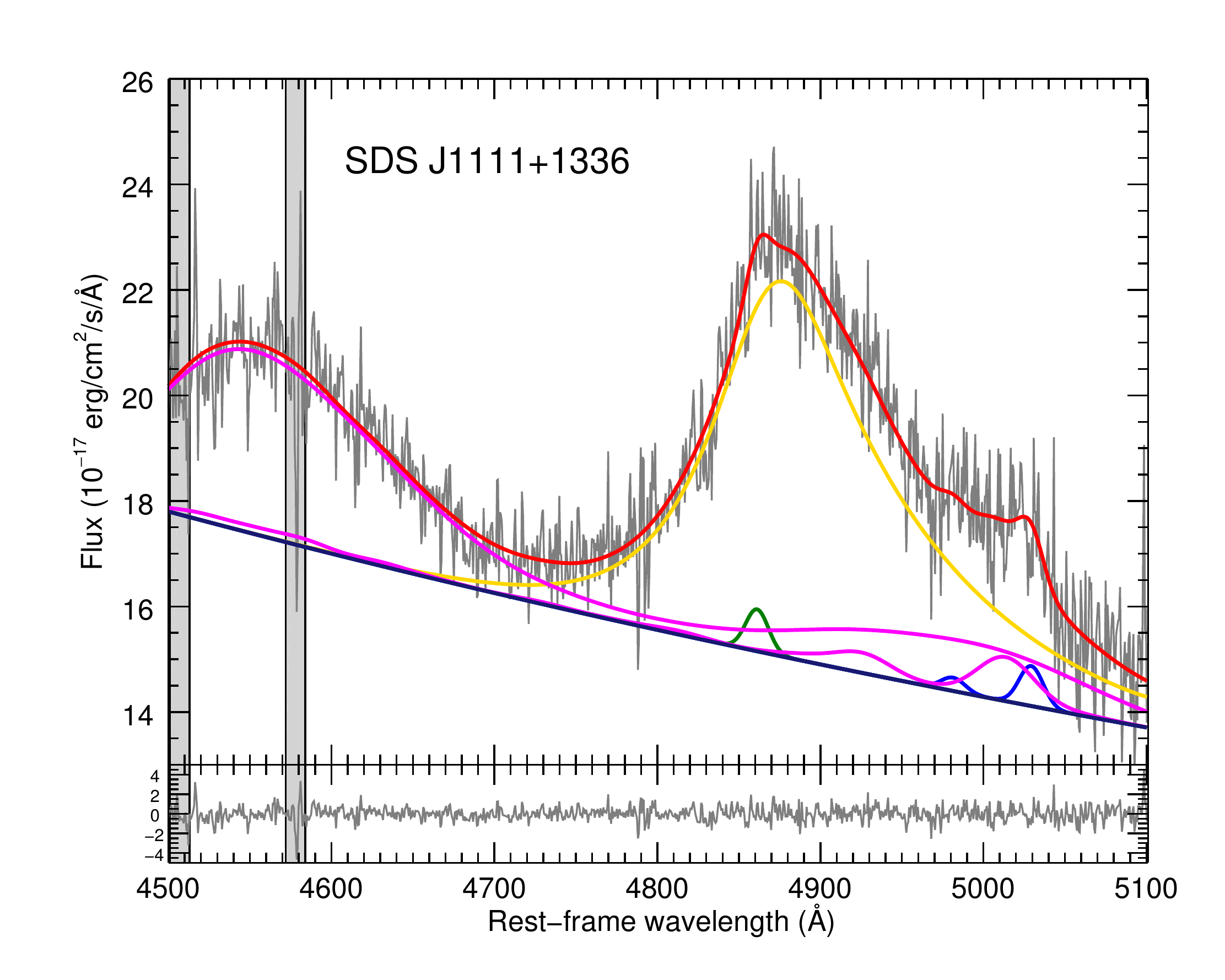}
	\includegraphics[width=1\columnwidth]{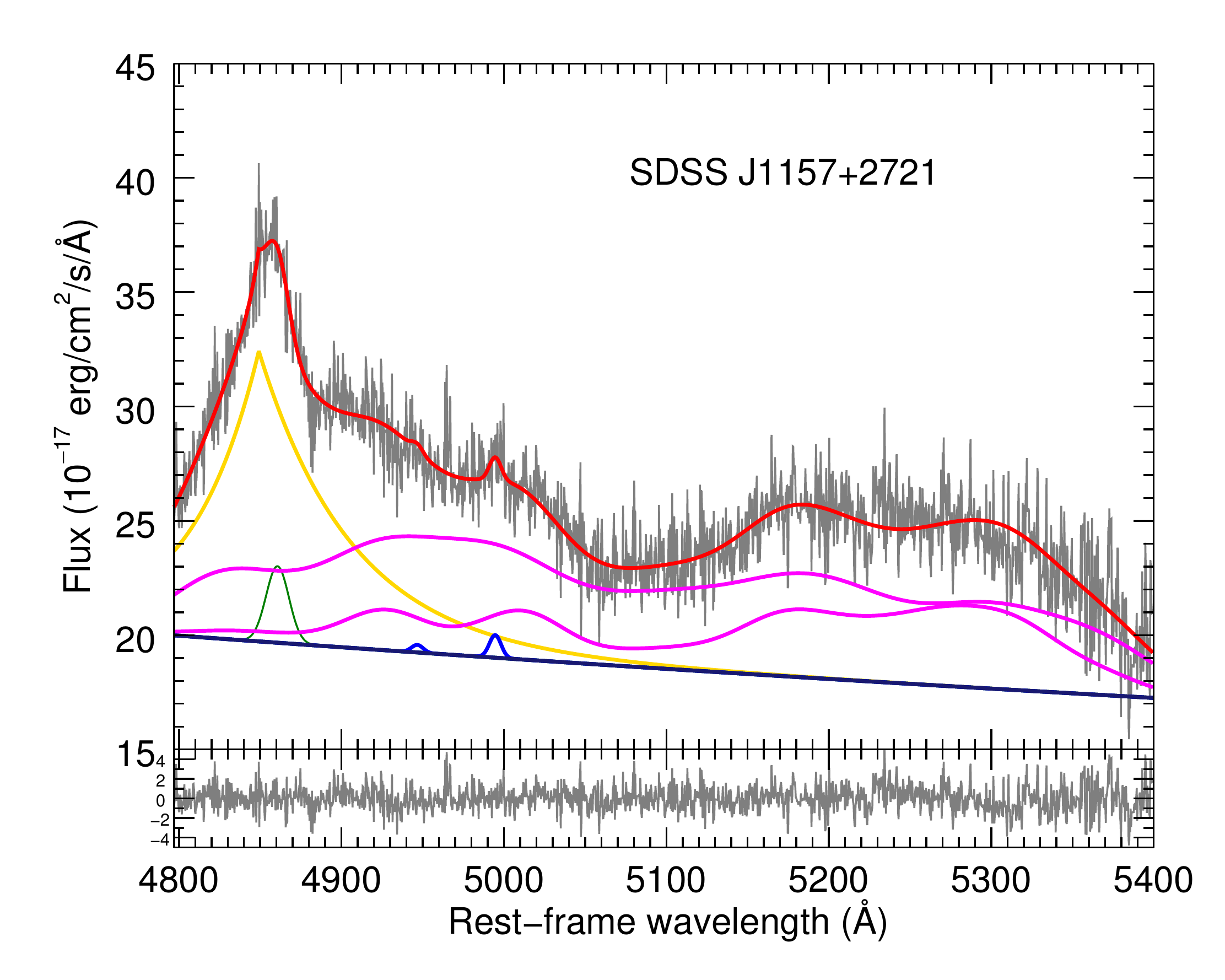}
\end{minipage}

\begin{figure}[]
		\includegraphics[width=1\columnwidth]{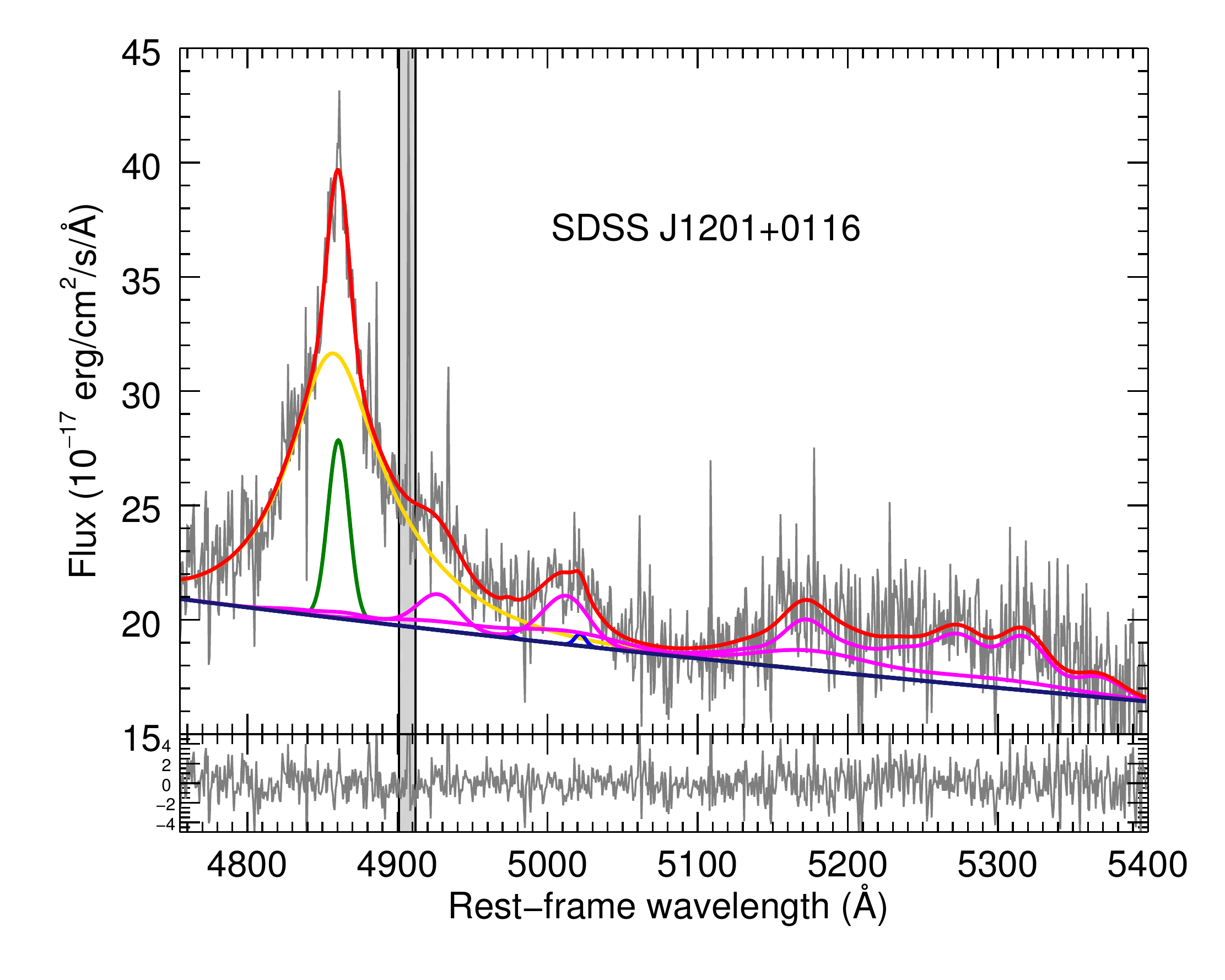}
    \includegraphics[width=1\columnwidth]{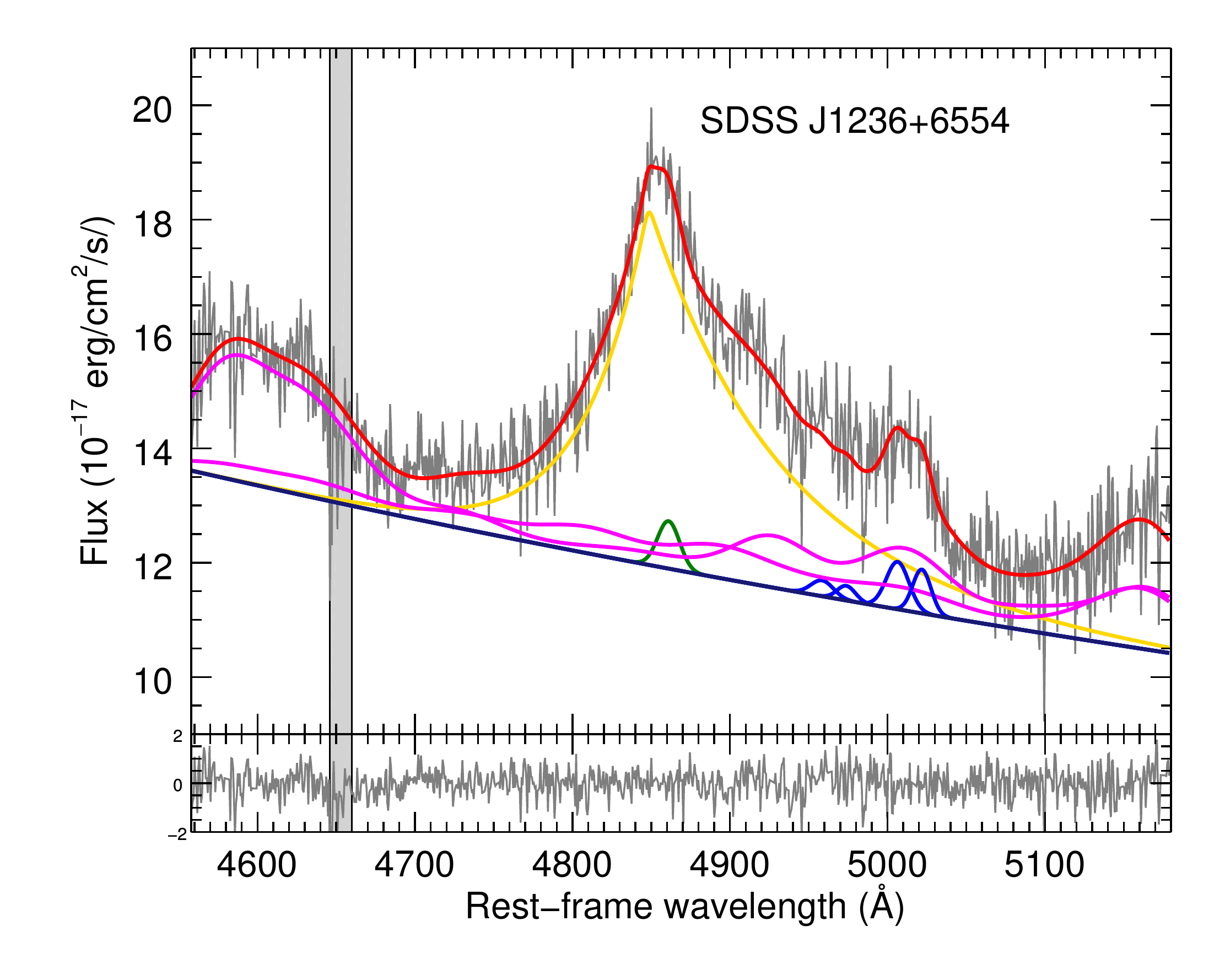}	
	\includegraphics[width=1\columnwidth]{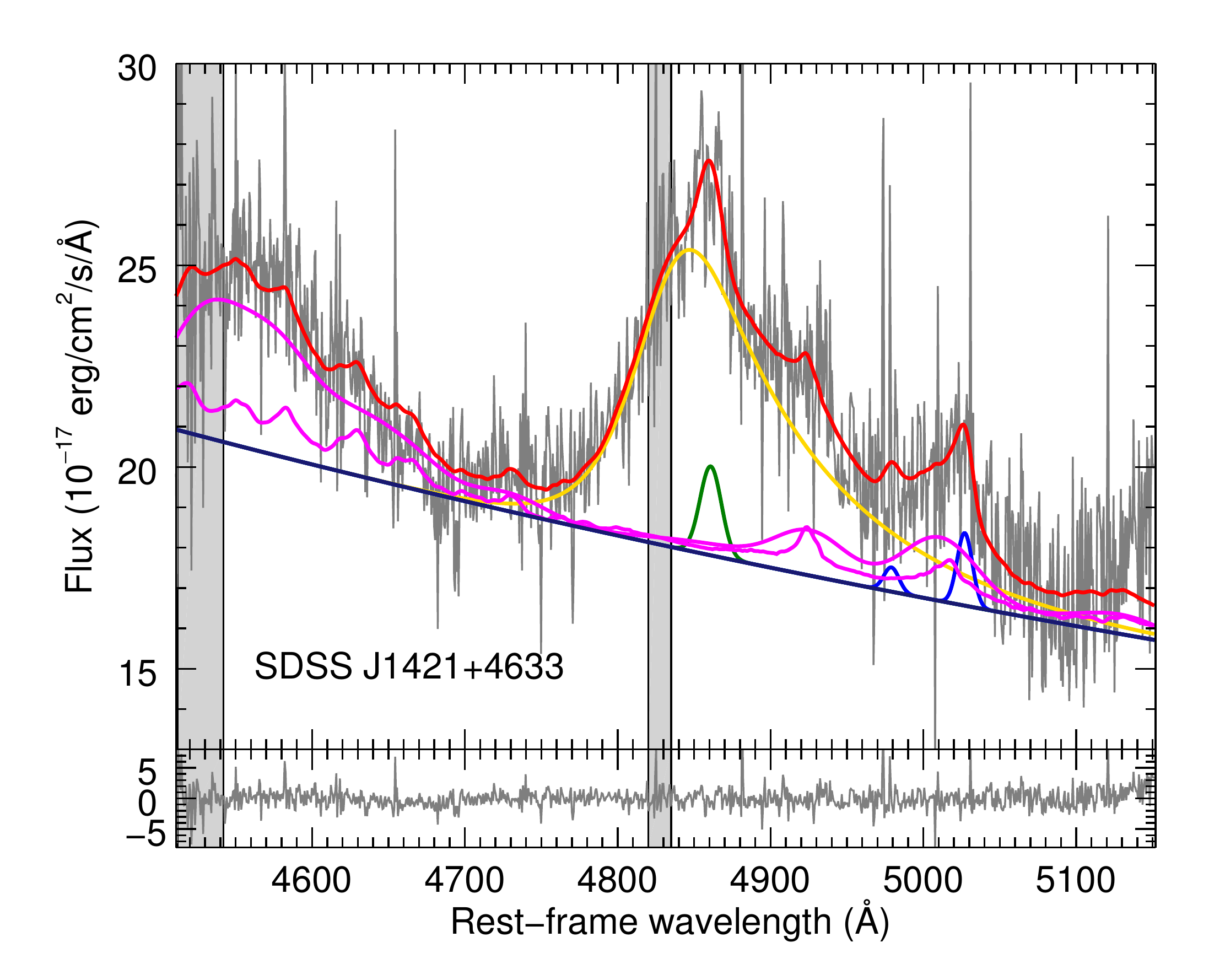}
	
		\end{figure}
		
		\begin{figure}[]
			\includegraphics[width=1\columnwidth]{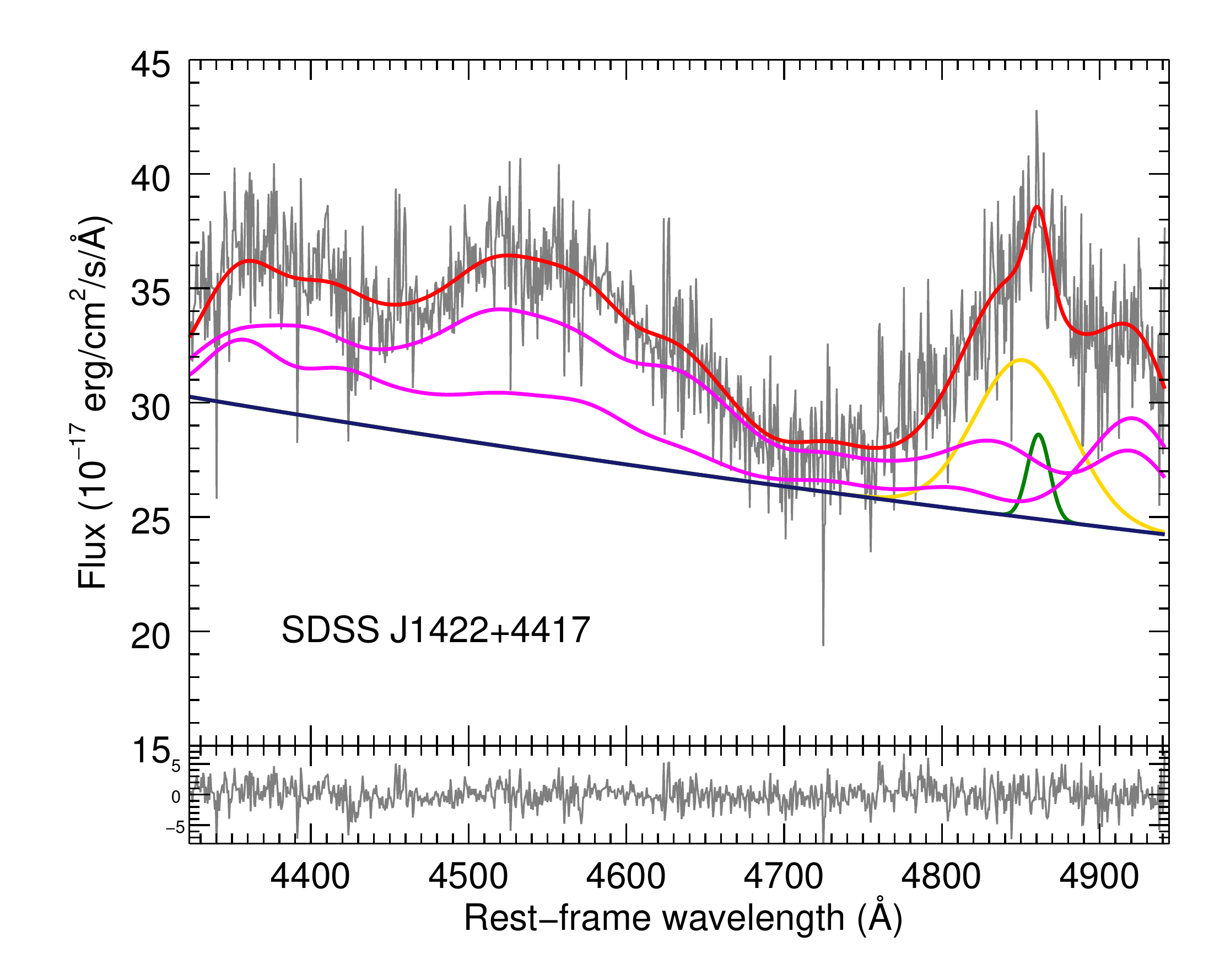}
		\includegraphics[width=1\columnwidth]{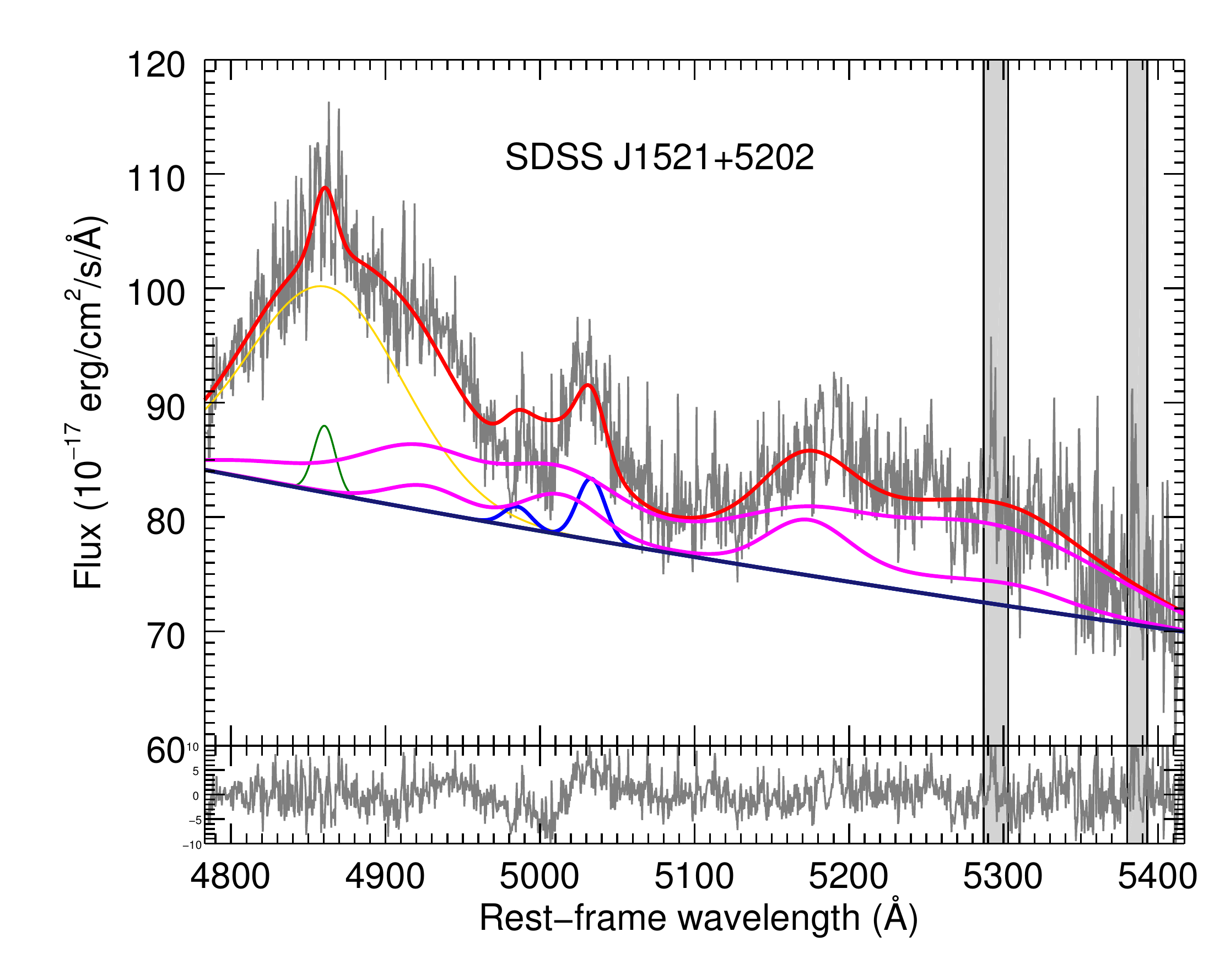}
		\includegraphics[width=1\columnwidth]{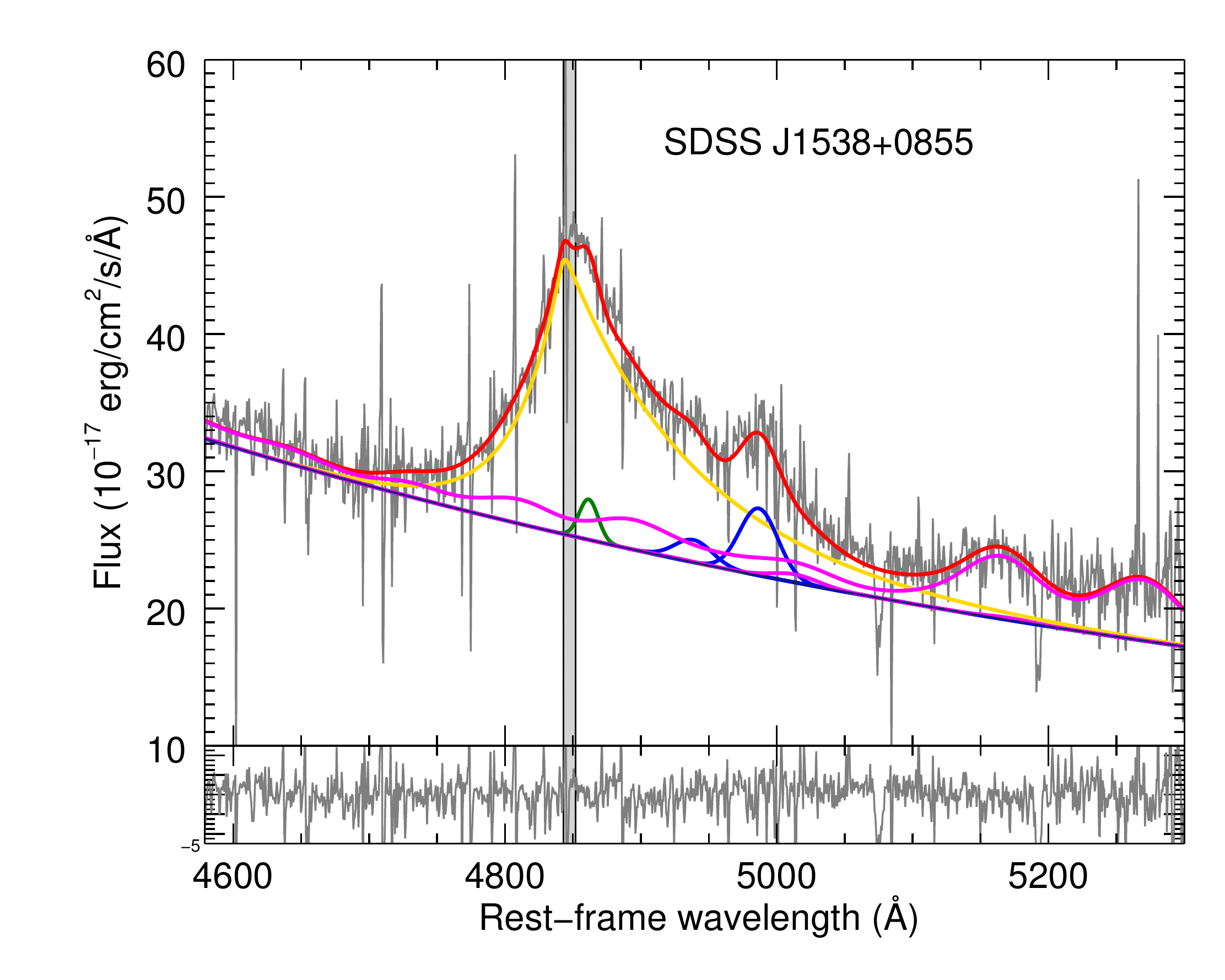}
\end{figure}

\begin{figure}	
		\includegraphics[width=1\columnwidth]{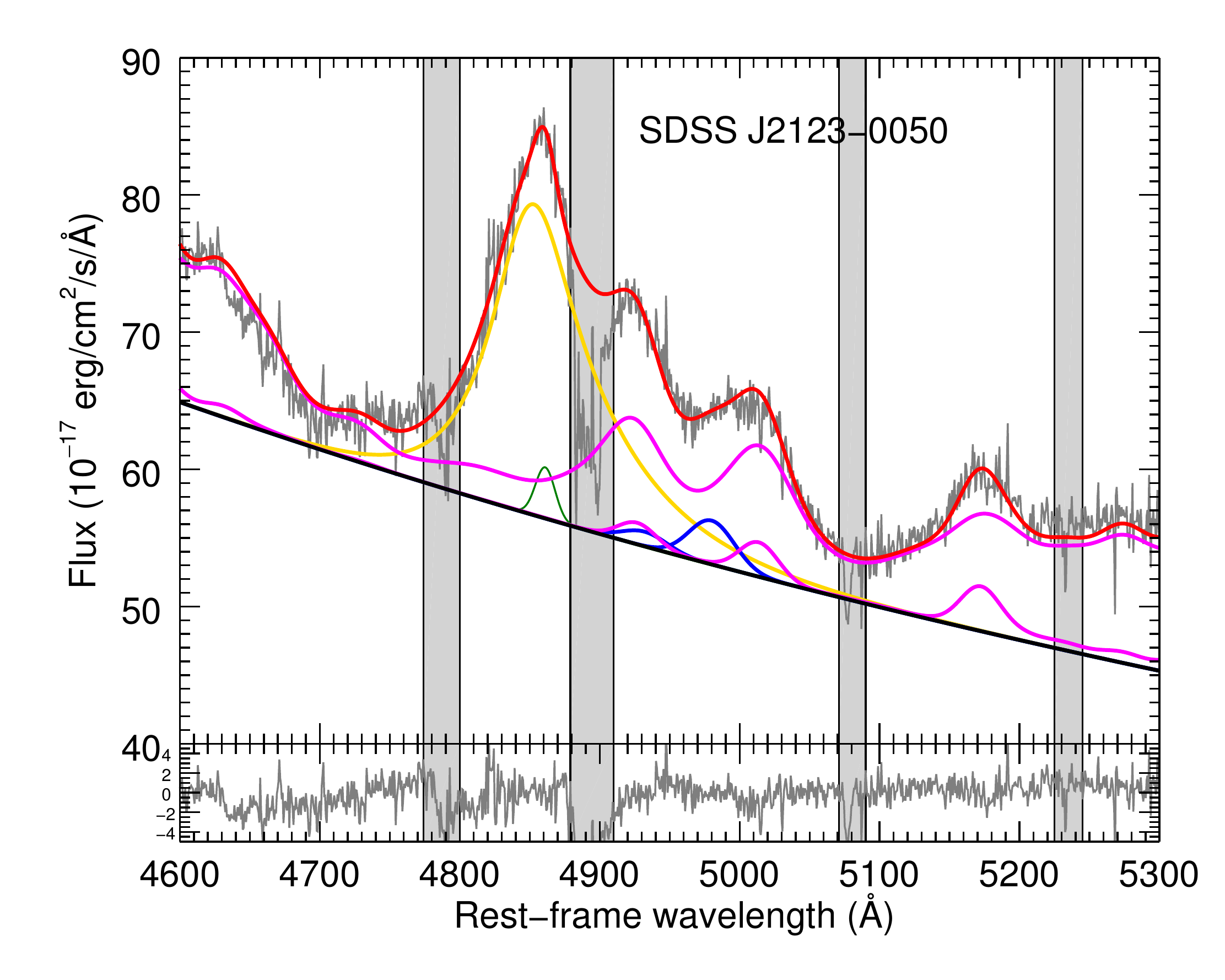}

	\includegraphics[width=1\columnwidth]{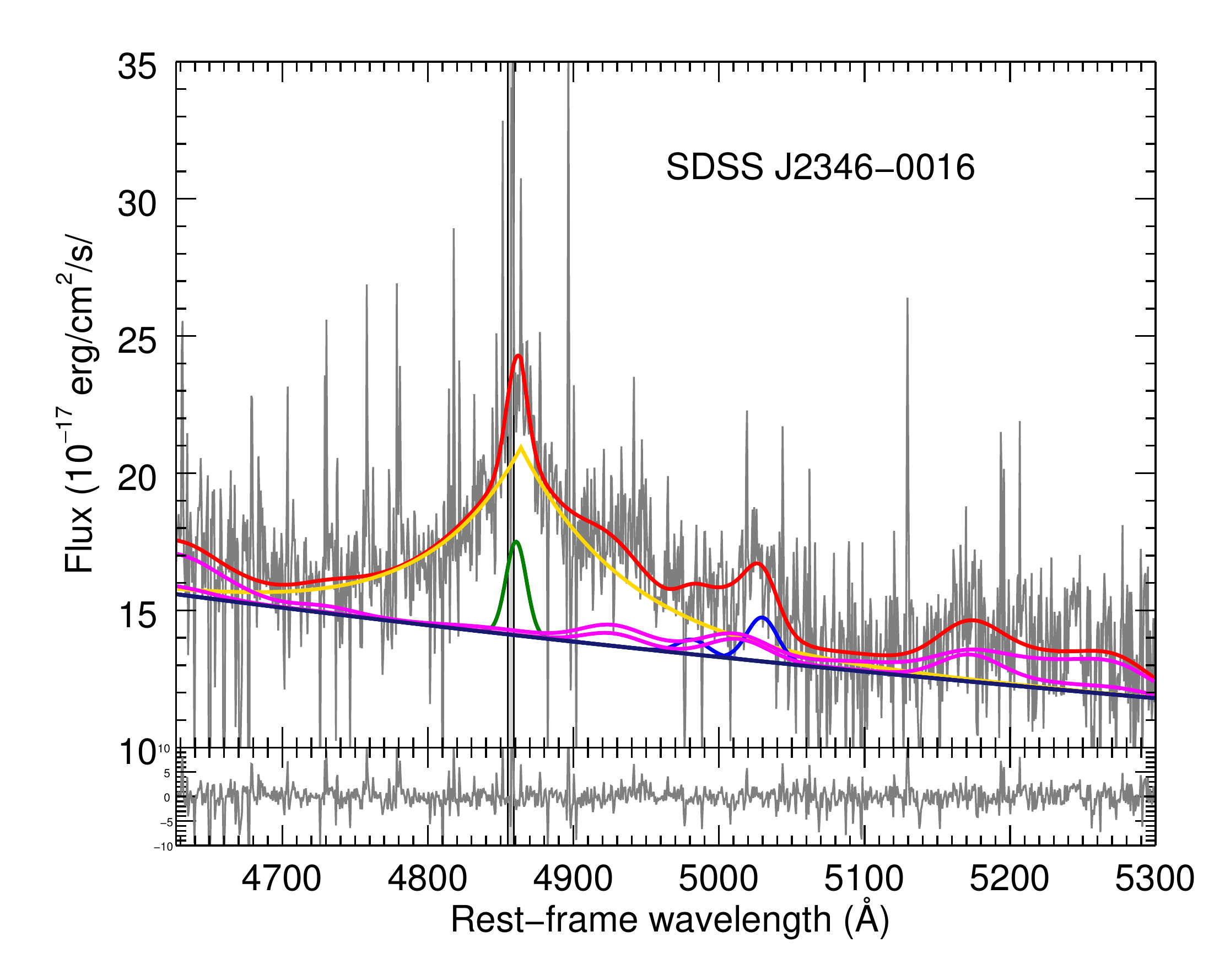}
	\caption{Fig 1. }\label{B13}

\end{figure}

\clearpage
	\section{SDSS Optical/UV rest-frame spectra of 18 WISSH QSOs}\label{sec:app_sdss}
We present the SDSS DR10 UV rest-frame spectra of the 18 WISSH QSOs analyzed in the present paper. 
The red lines show the best-fit of the spectra. If present, green curves refer to the core component likely associated with the NLR emission of the CIV emission. Blue curves refer to the broad blueshifted emission of CIV emission line, indicative of outflow.
Gold curves indicate the broad virialized component associated with the BLR emission. FeII emission is indicated in magenta. Grey bands indicate the region excluded from the fit because of the presence of absorption lines. The dashed lines denote the CIV $\lambda$1549 \AA\ rest-frame wavelength.
Lower panels show the fit residuals.

\begin{figure}[h]

\includegraphics[width=1\columnwidth]{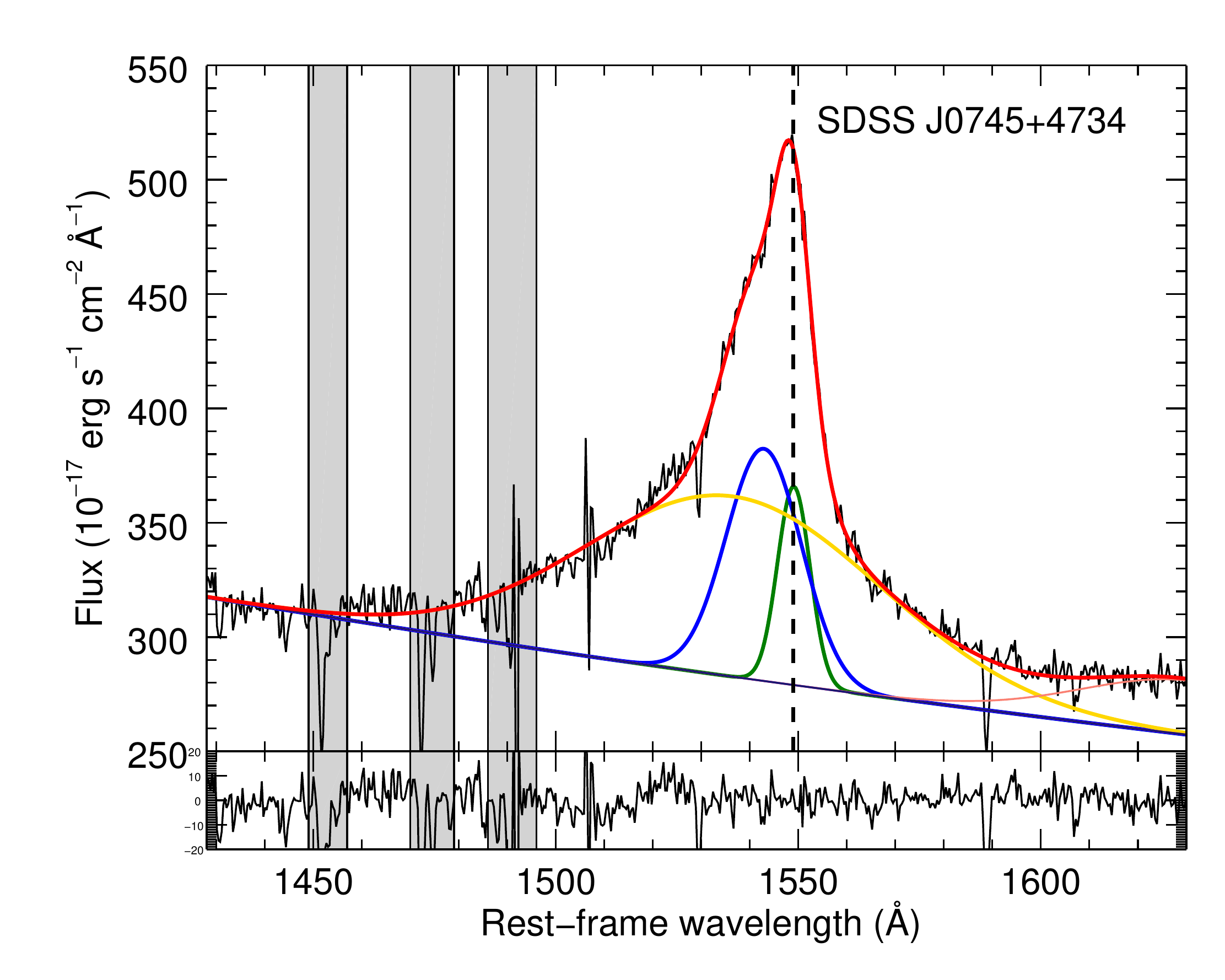}
\caption{}\label{C1}
	\end{figure}
	
	\begin{figure}[h]
	\includegraphics[width=1\columnwidth]{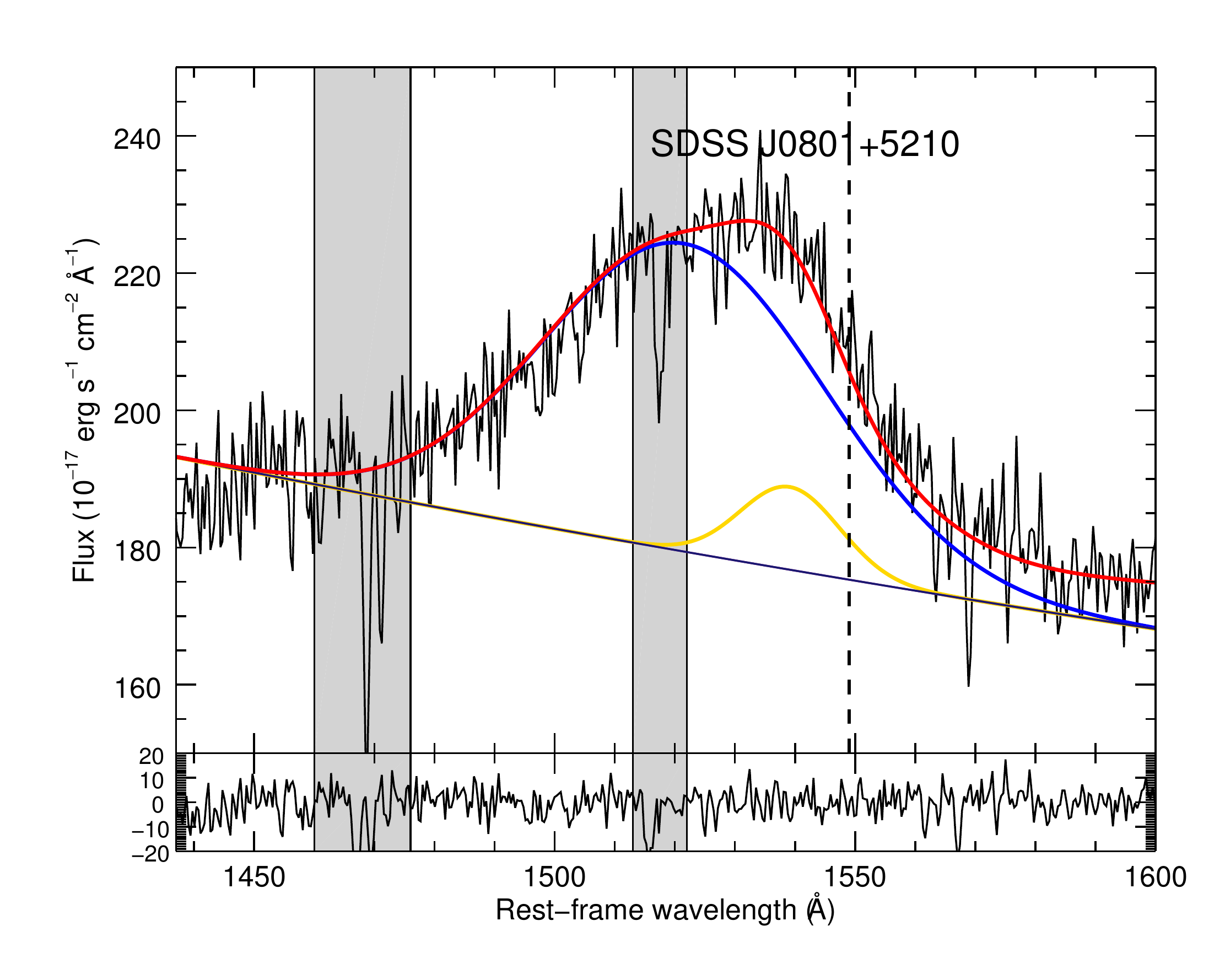}
	\caption{}\label{C2}
		\end{figure}
		
	\begin{figure}[h]
\includegraphics[width=1\columnwidth]{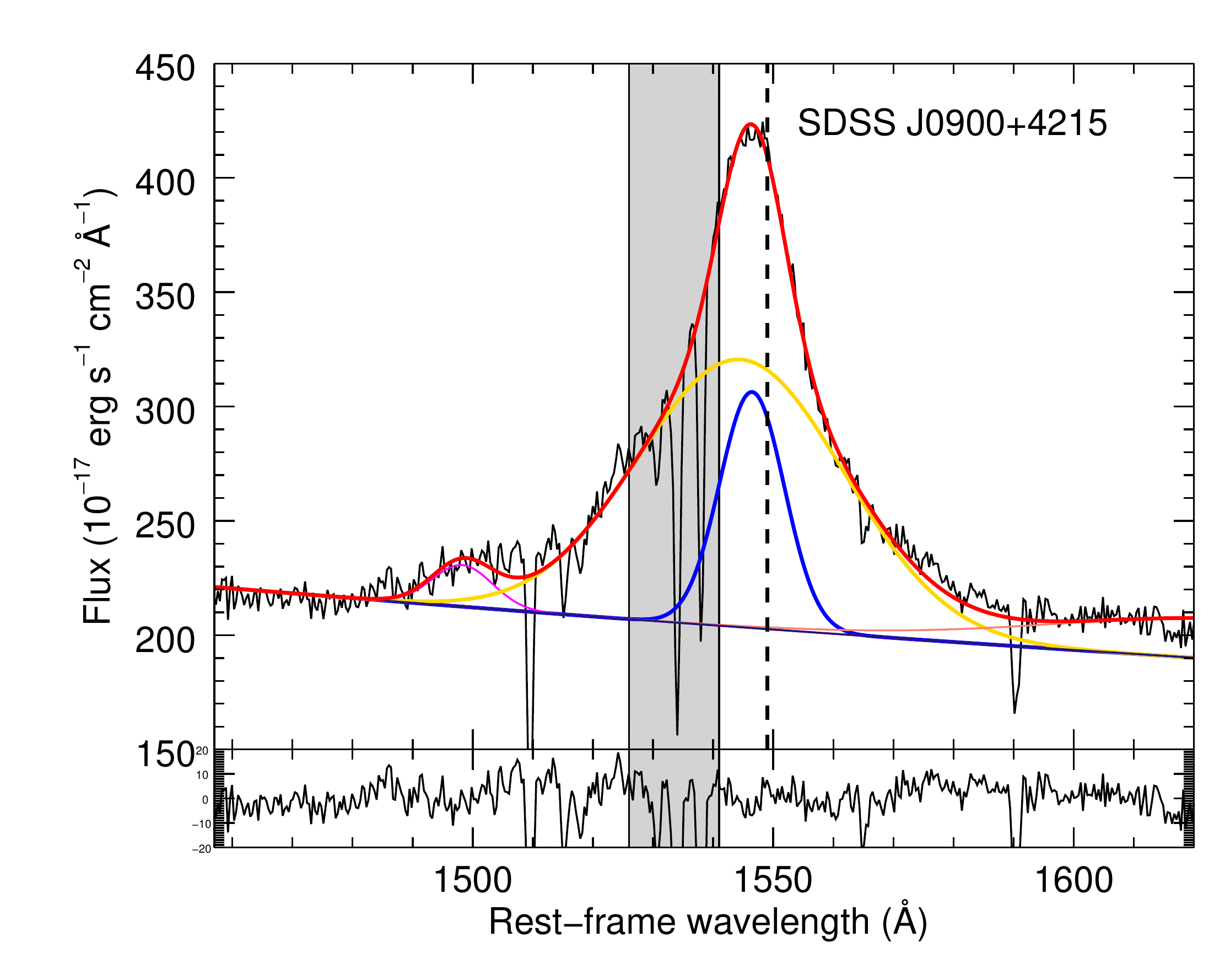}
\caption{}\label{C3}
	\end{figure}

	\begin{figure}[]

	\includegraphics[width=1\columnwidth]{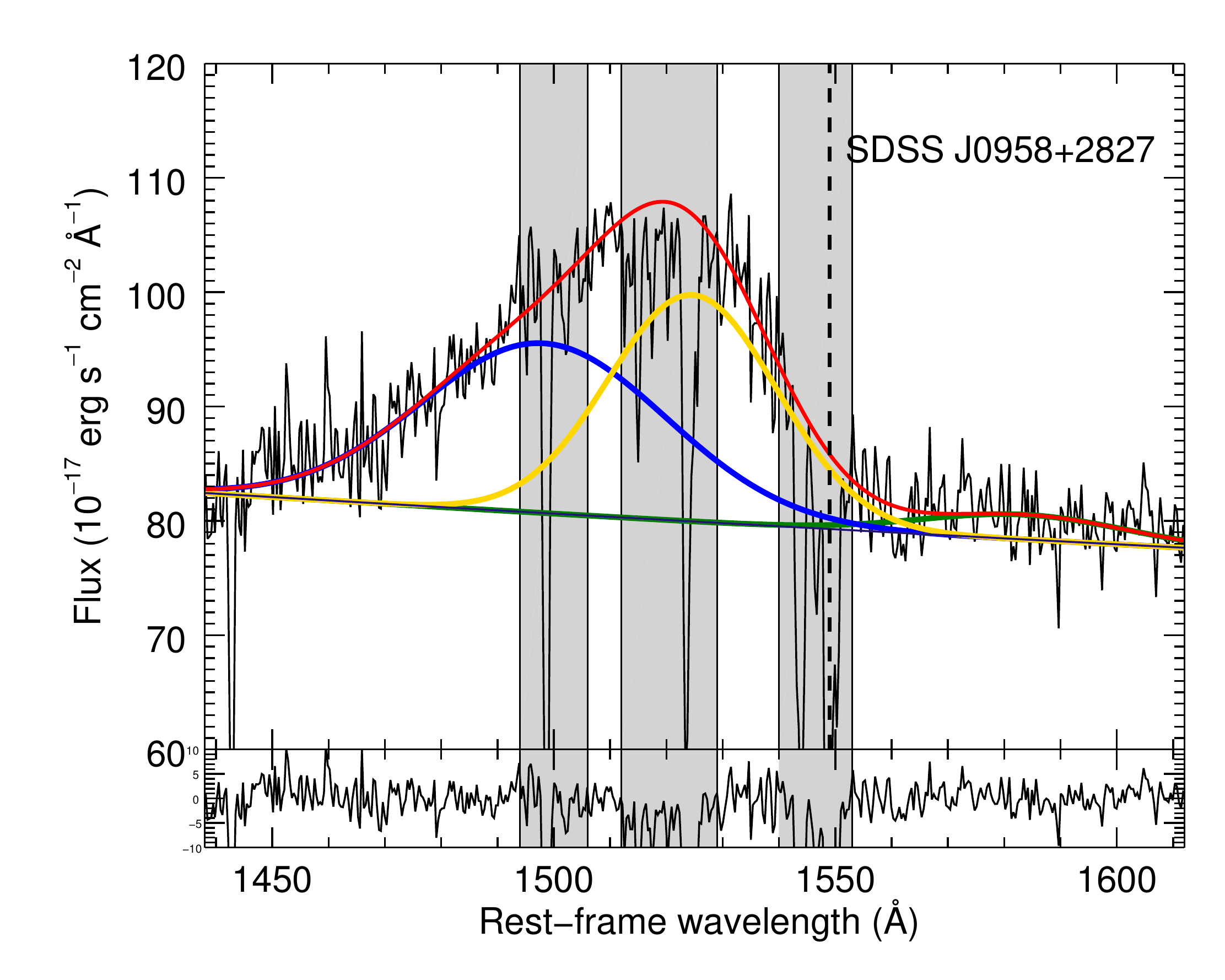}
\caption{}\label{C4}
\end{figure}
\begin{figure}[]
		\includegraphics[width=1\columnwidth]{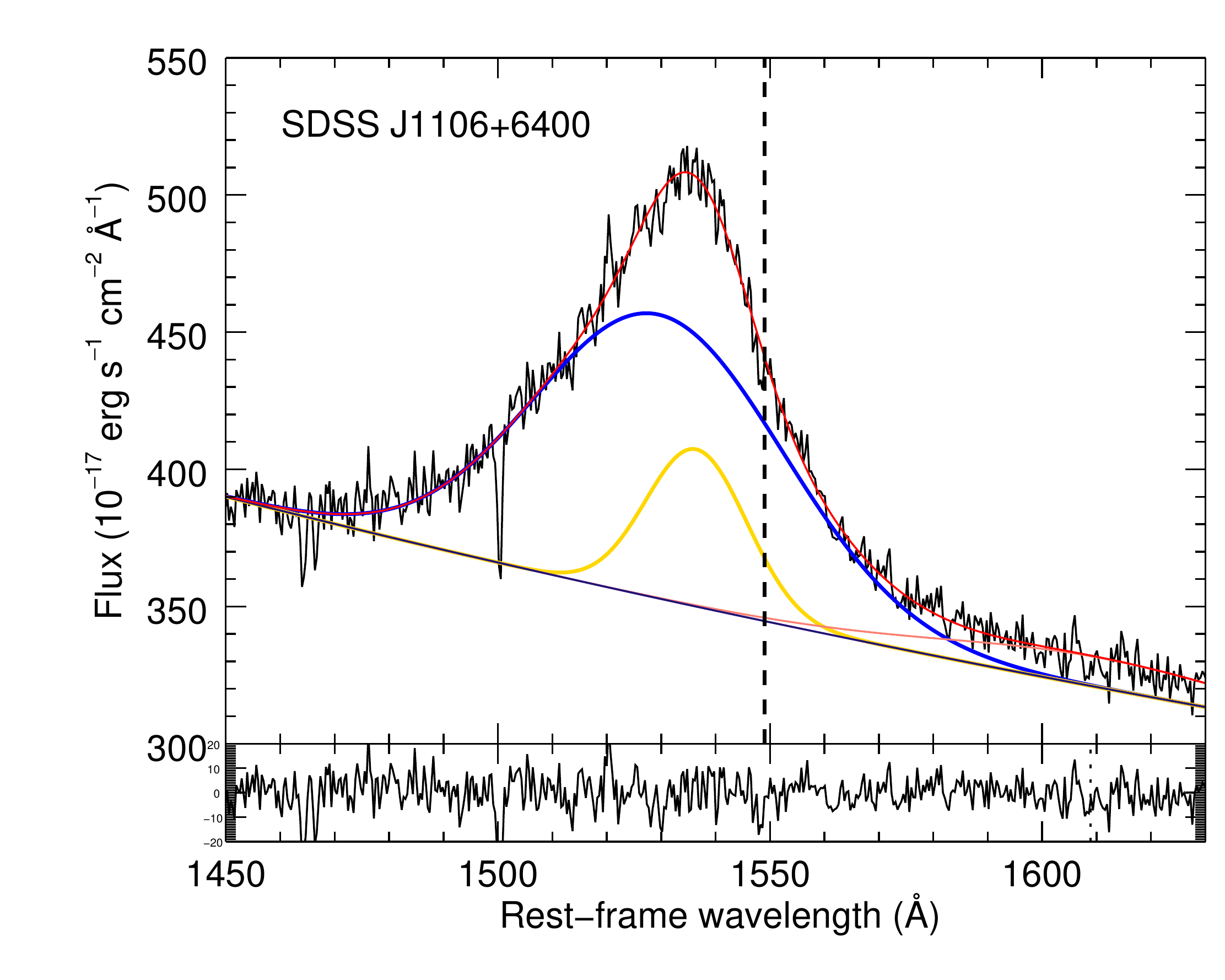}
		\caption{}\label{C5}
\end{figure}

\begin{figure}[]
		\includegraphics[width=1\columnwidth]{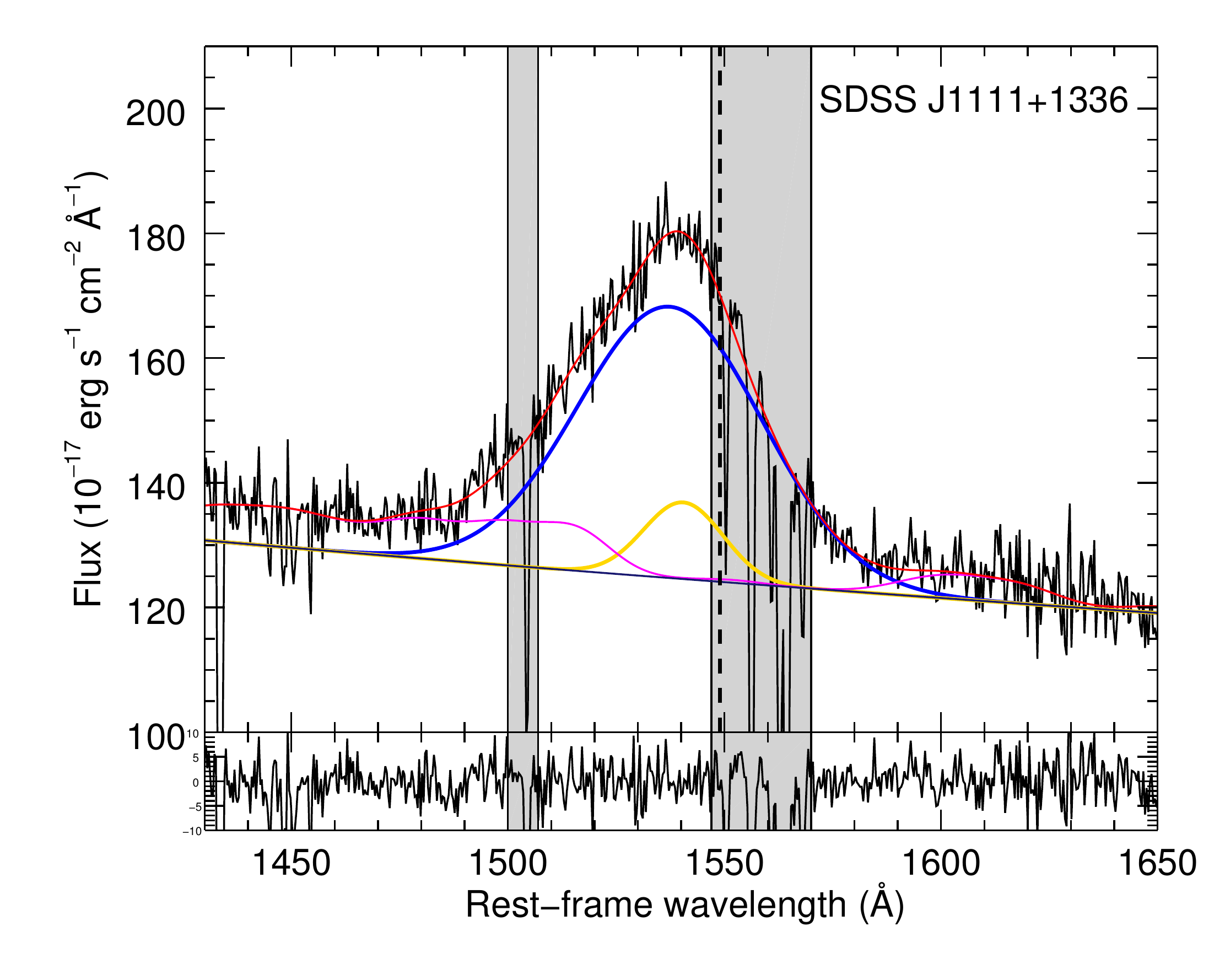}
		\caption{}\label{C6}
			\includegraphics[width=1\columnwidth]{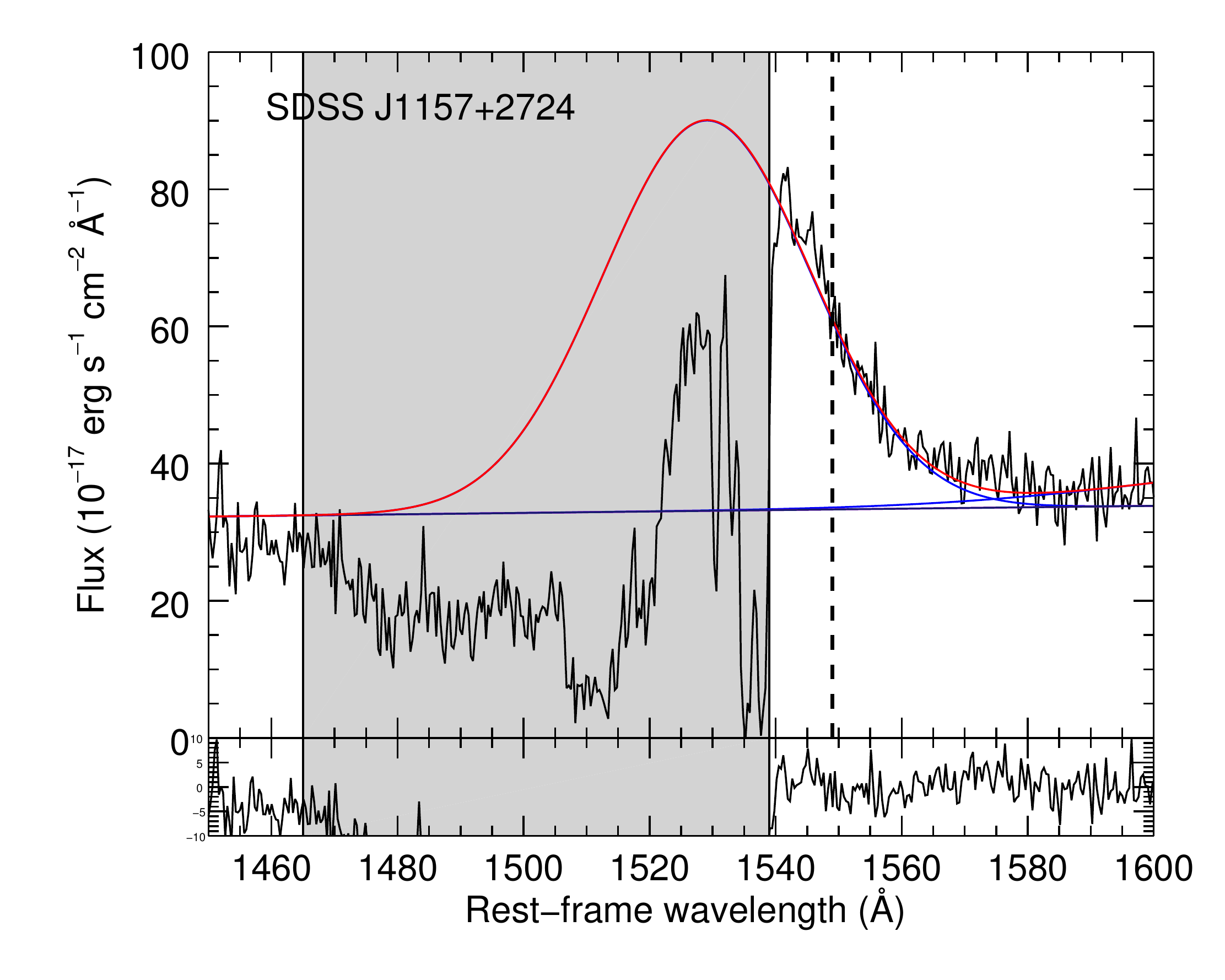}
		\caption{}\label{C7}
		\includegraphics[width=1\columnwidth]{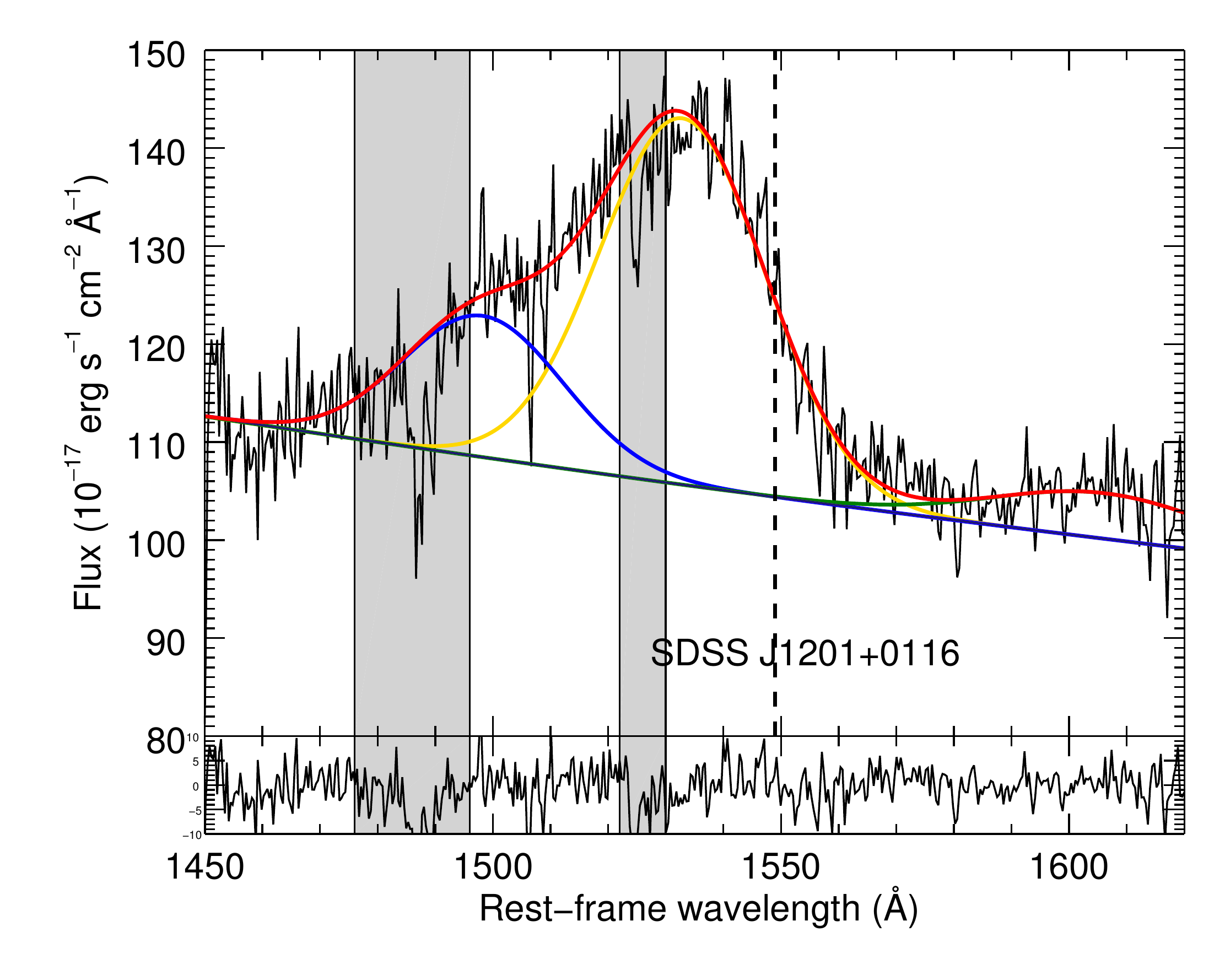}
		\caption{}\label{C8}
	\end{figure}

	\begin{figure}[]

\includegraphics[width=1\columnwidth]{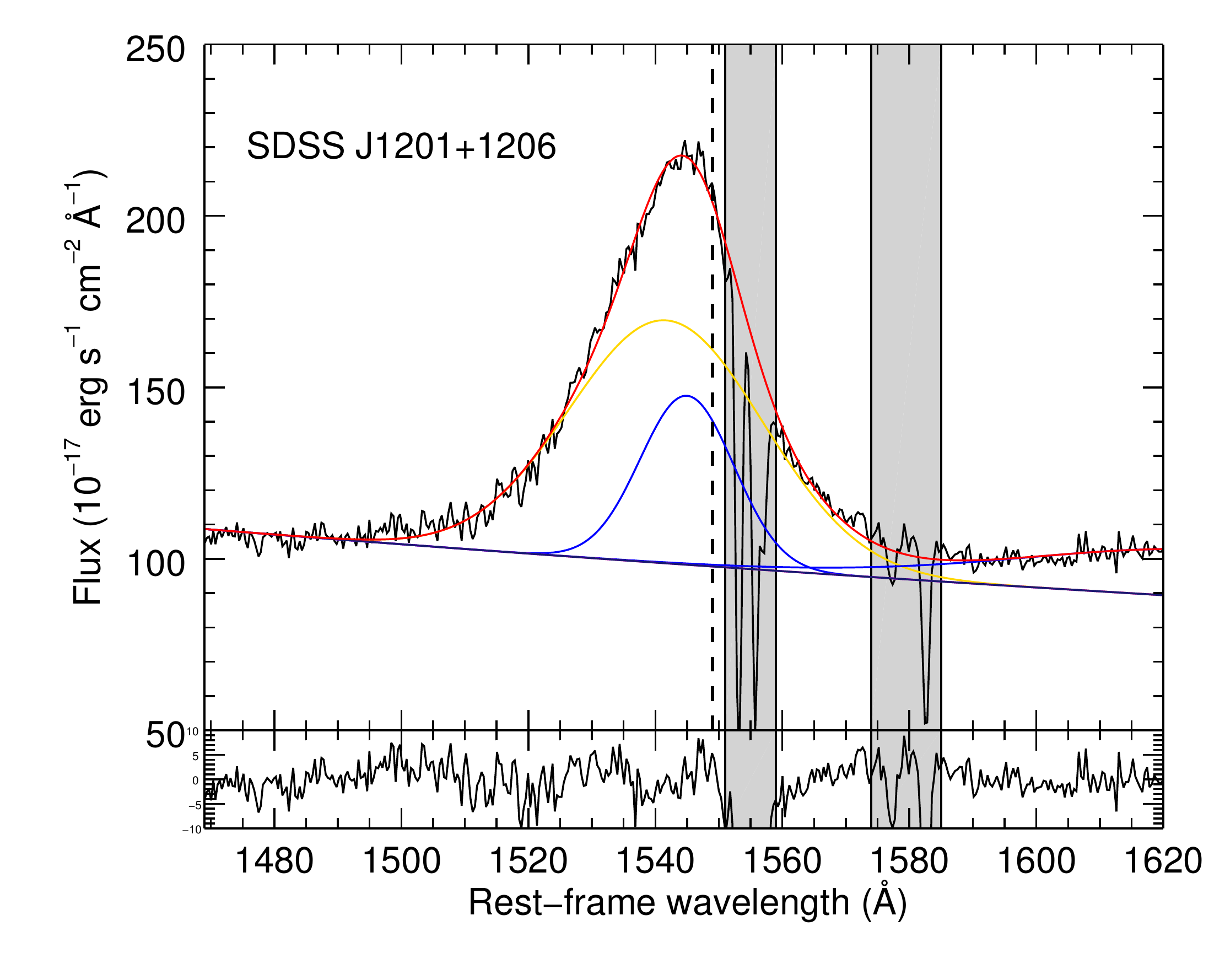}
\caption{}\label{C9}
	\includegraphics[width=1\columnwidth]{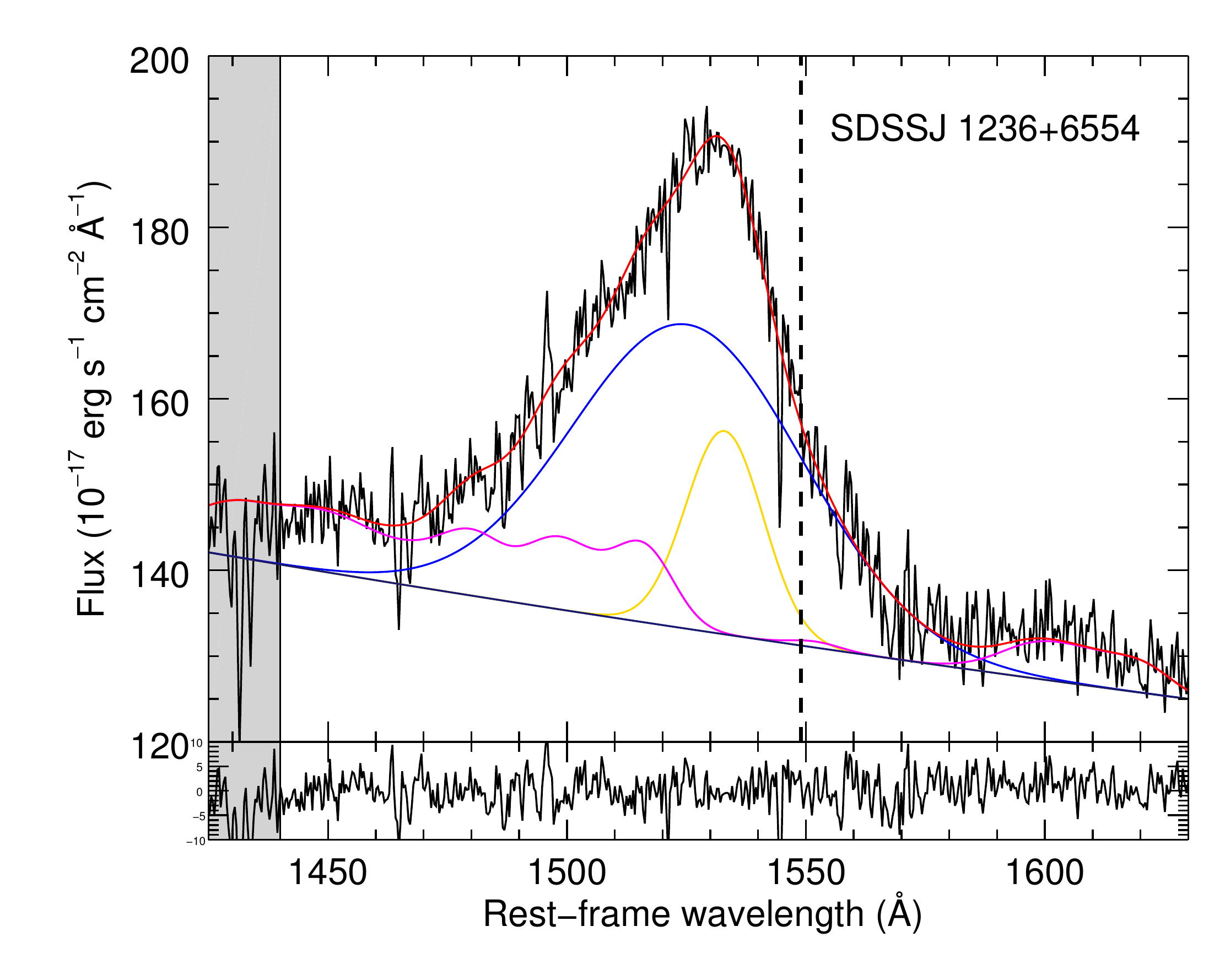}
\caption{}\label{C10}
		\includegraphics[width=1\columnwidth]{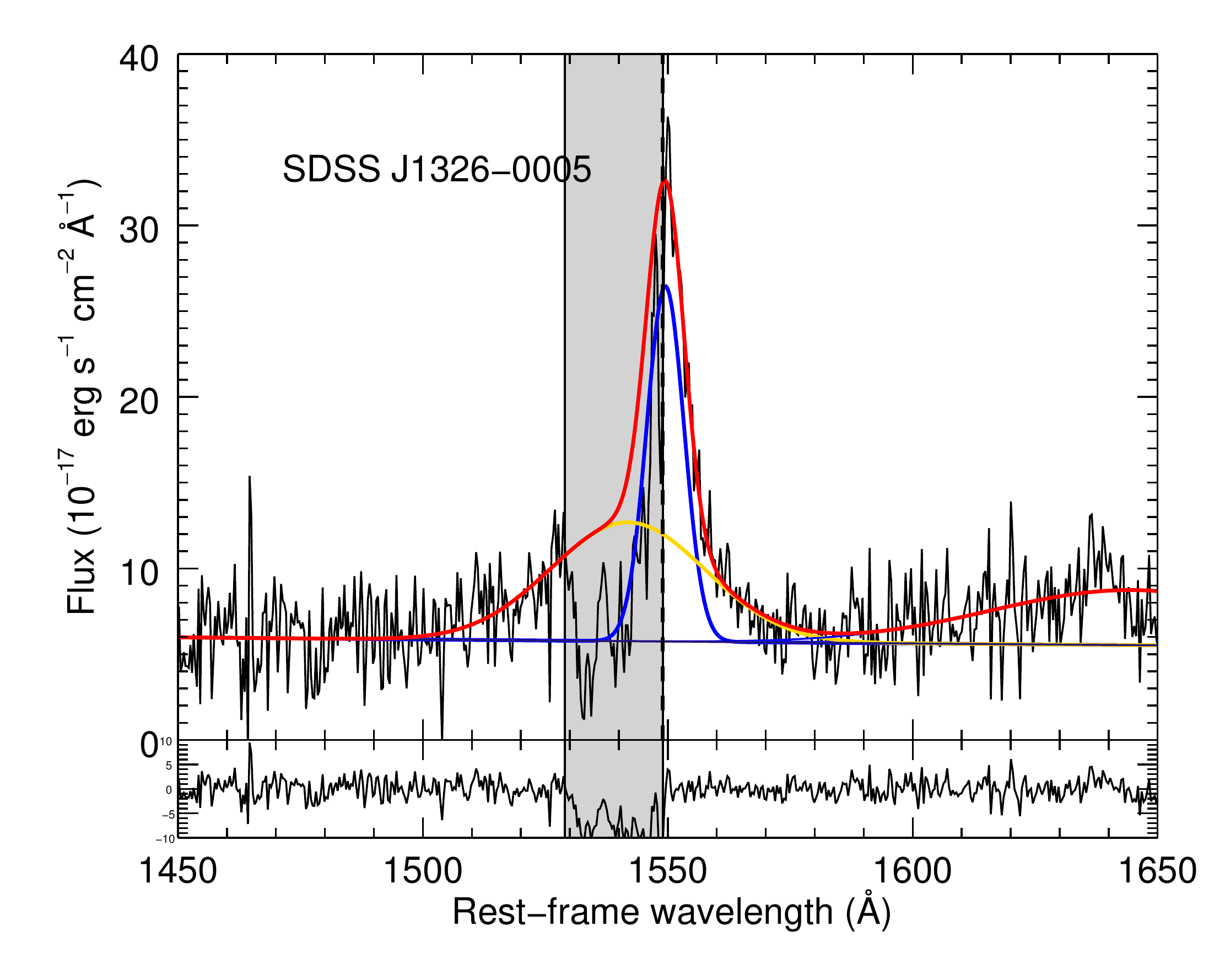}
		\caption{}\label{C11}
	\end{figure}

	\begin{figure}[]
		\includegraphics[width=1\columnwidth]{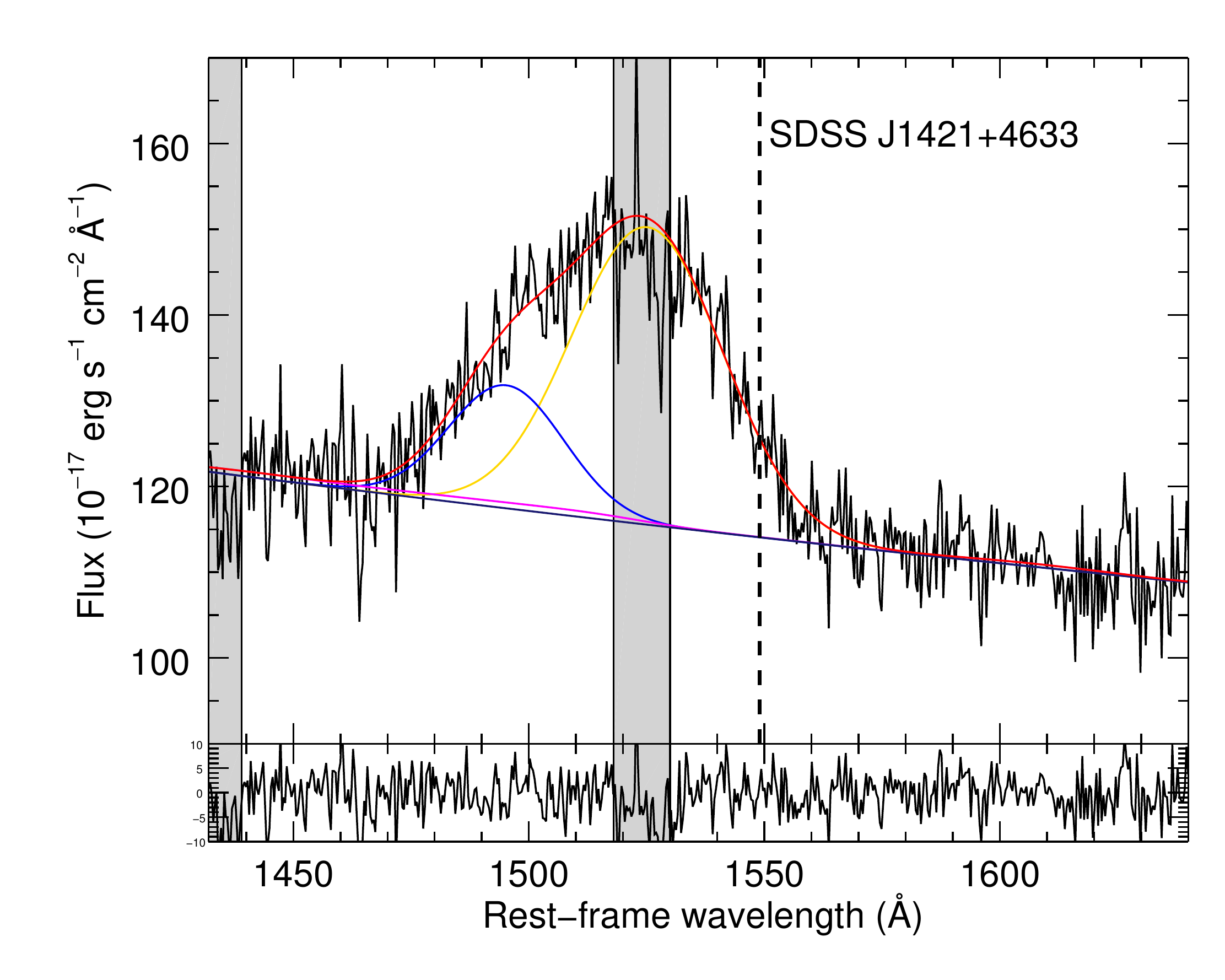}
		\caption{}\label{C12}
		\includegraphics[width=1\columnwidth]{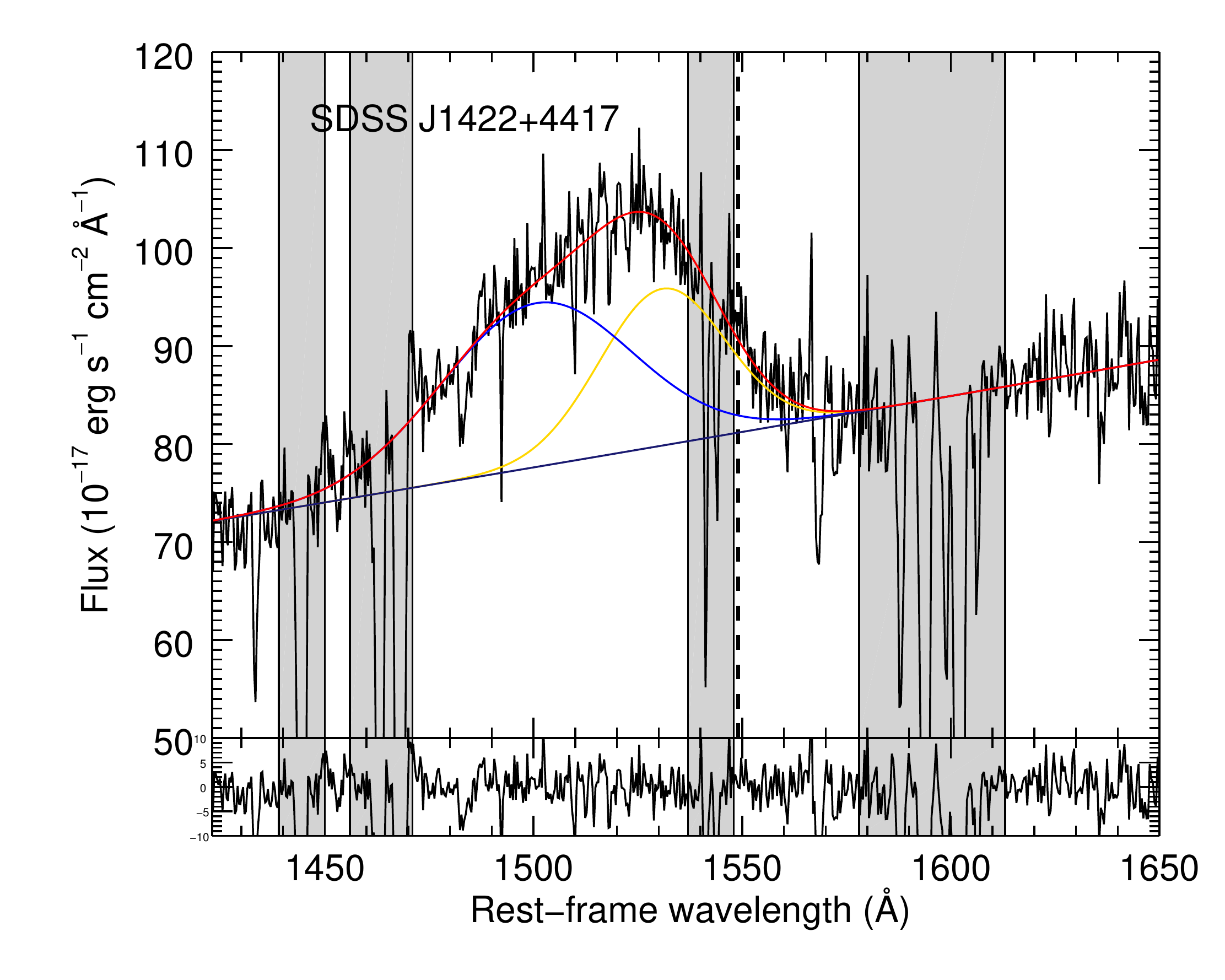}
		\caption{}\label{C13}
		\includegraphics[width=1\columnwidth]{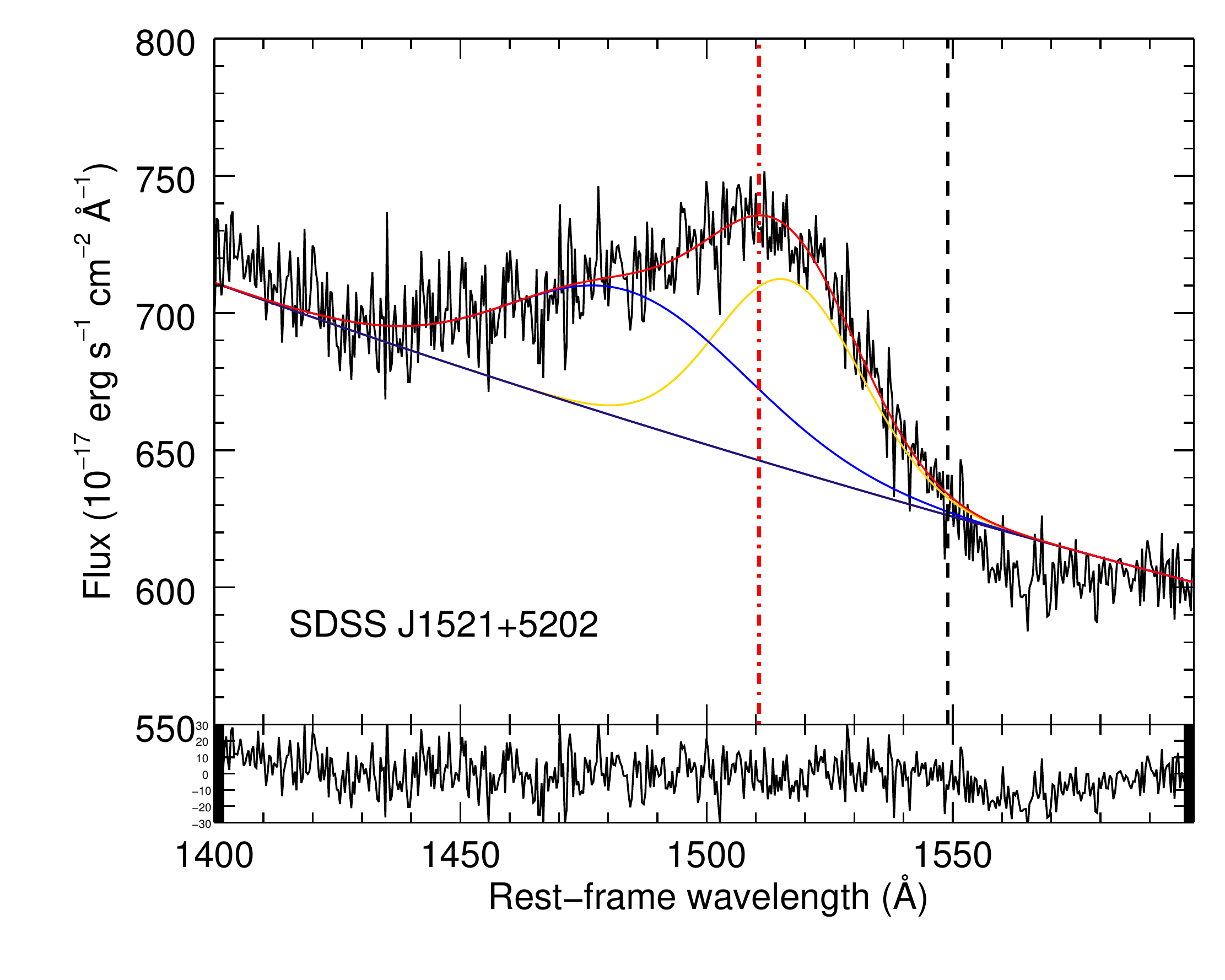}
		\caption{}\label{C14}
	\end{figure}

	\begin{figure}[]
		\includegraphics[width=1\columnwidth]{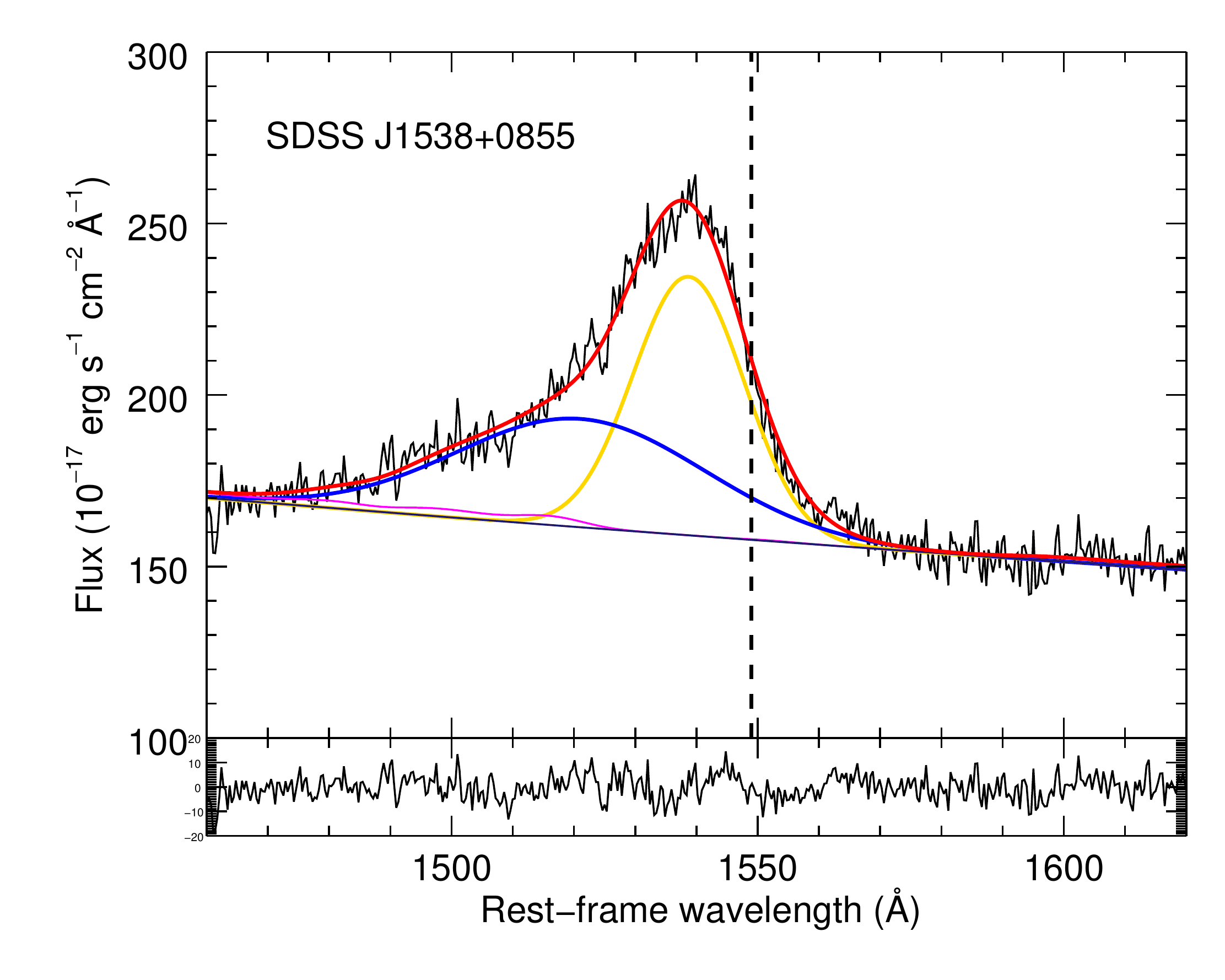}
		\caption{}\label{C15}
		\includegraphics[width=1\columnwidth]{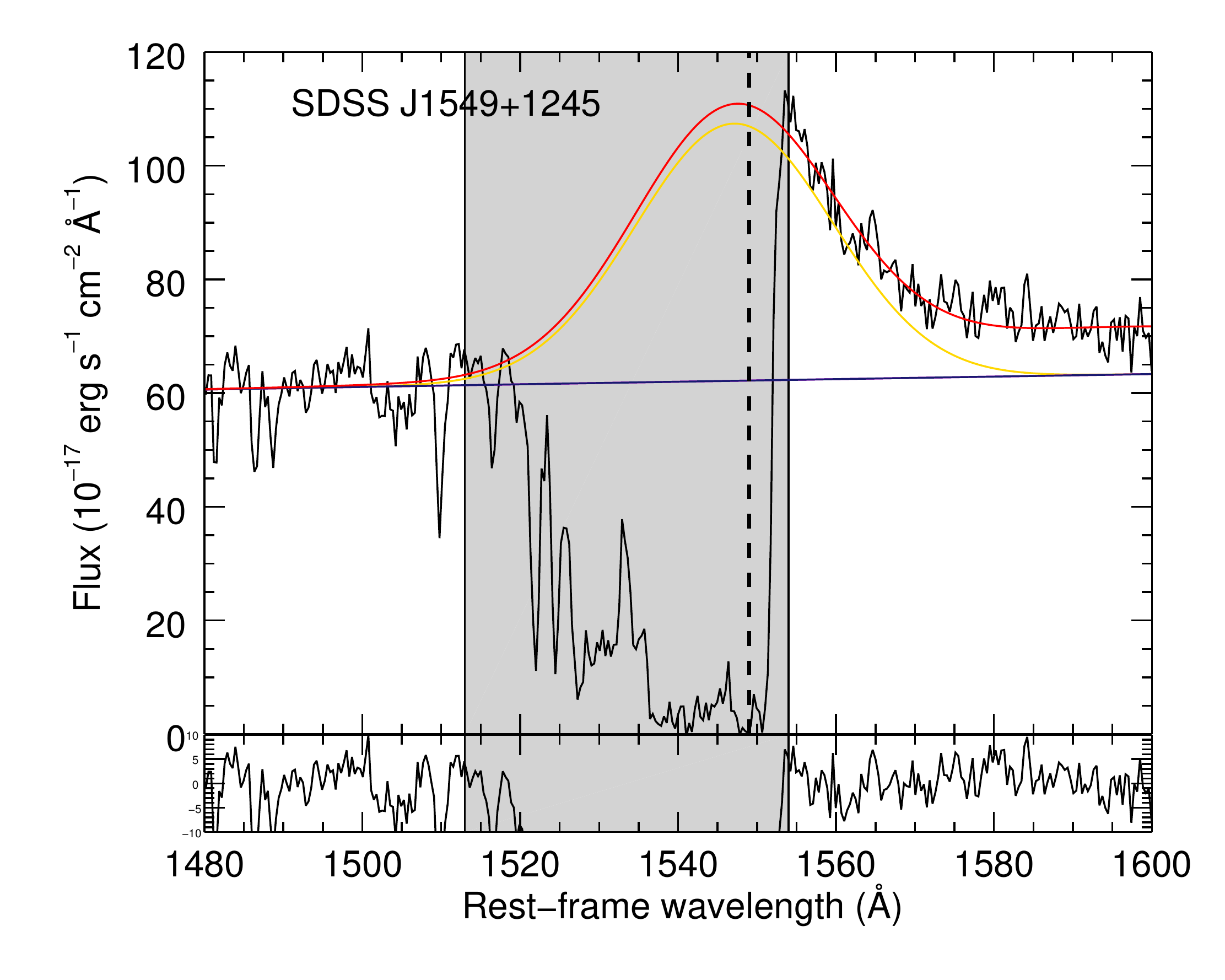}
		\caption{}\label{C16}
	\includegraphics[width=1\columnwidth]{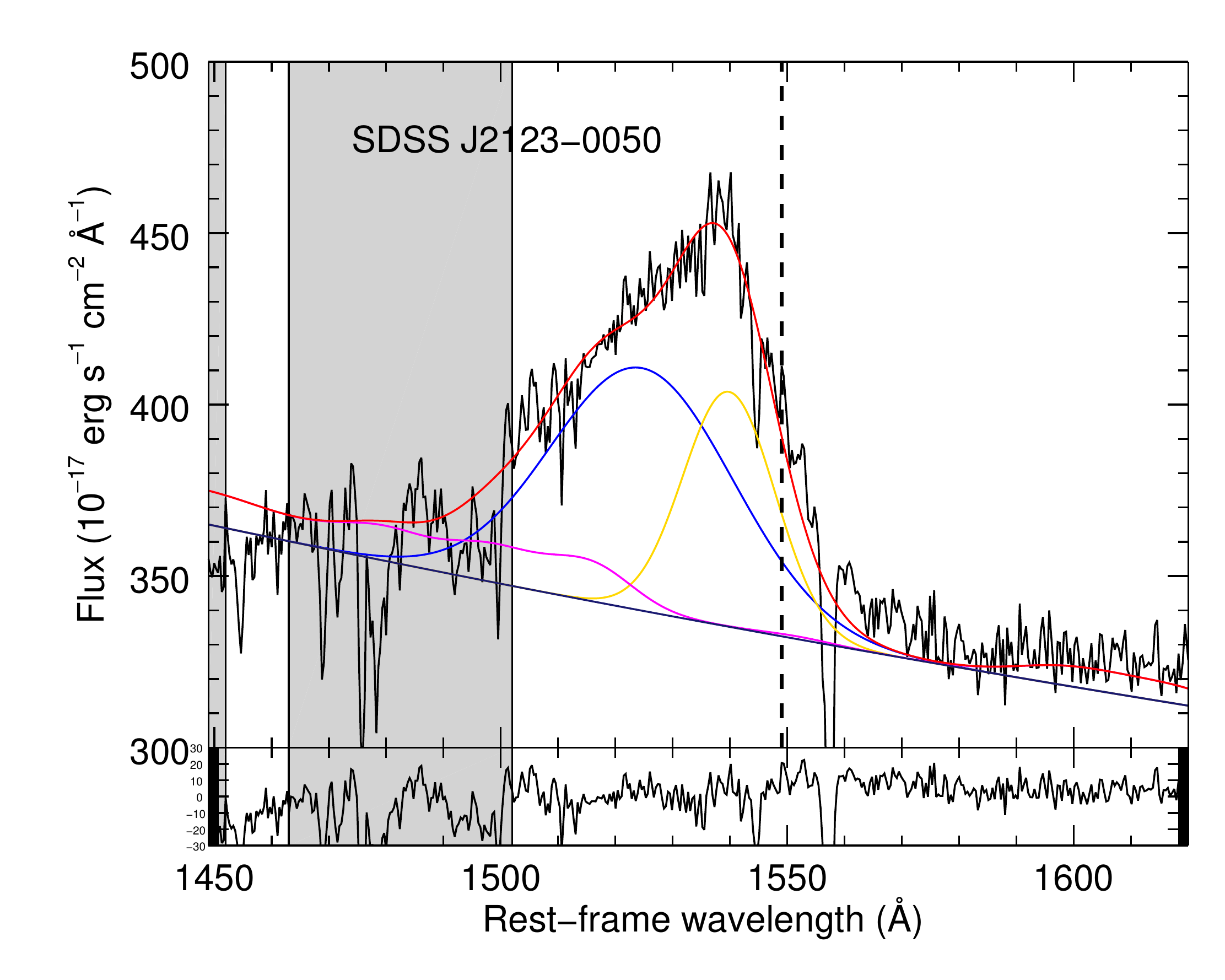}
	\caption{}\label{C17}
  \end{figure}

	\begin{figure}[]

		\includegraphics[width=1\columnwidth]{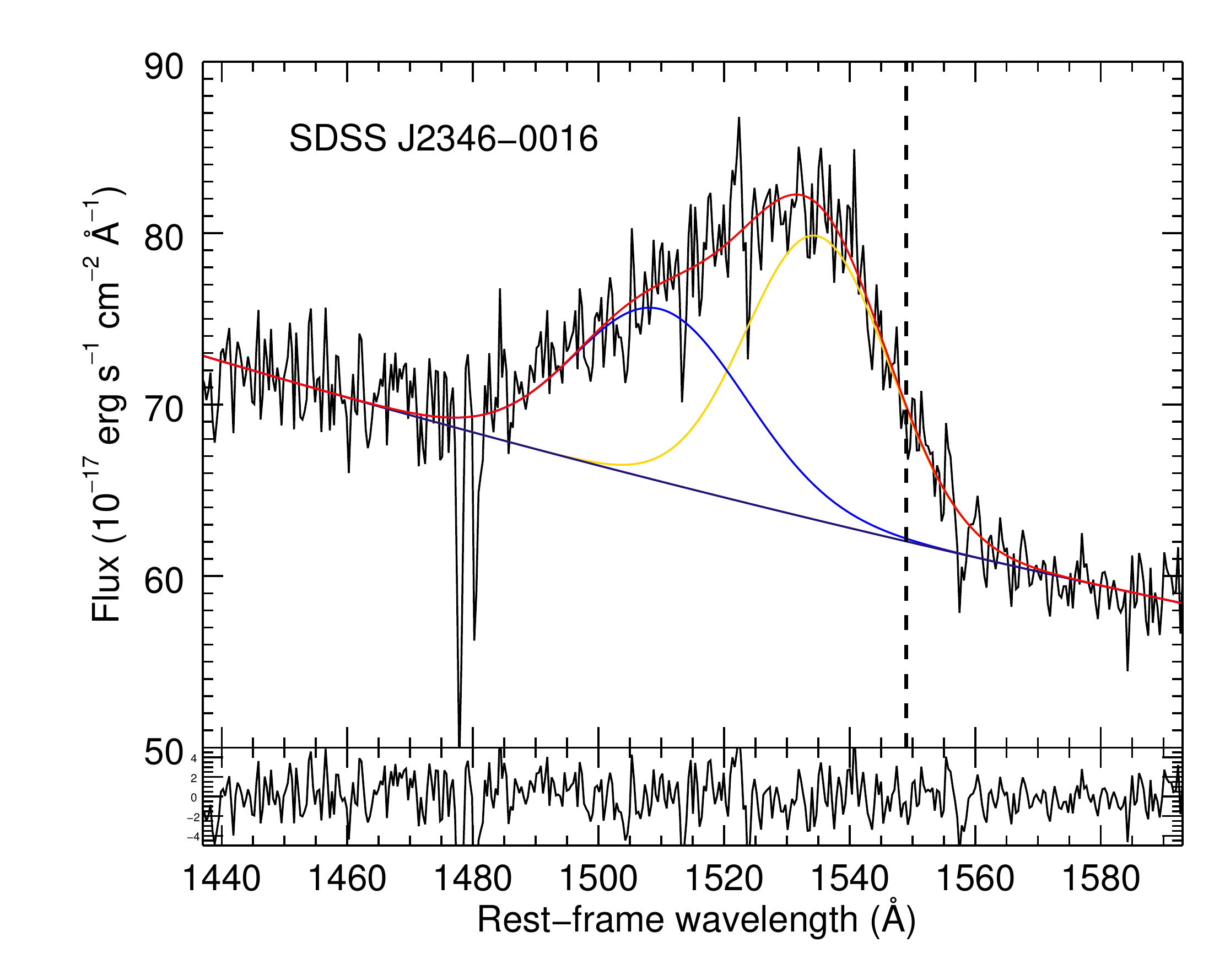}
		\caption{}\label{C18}
		  \end{figure}


 %

%
 
%





%
%

\end{document}